\documentclass[aps,10pt,prl,twocolumn,letterpaper,superscriptaddress]{revtex4-1}

\usepackage{amsmath}
\usepackage{epsfig}
\usepackage{graphicx, caption}
\usepackage{latexsym}
\usepackage{amssymb}
\usepackage{color}
\usepackage{amsfonts}
\usepackage{bm}
\usepackage{upgreek}
\usepackage{hyperref}

\usepackage{blindtext}

\usepackage{adjustbox}
\begin{document}
\title{Three interaction energy scales in single-layer high-T$_C$ cuprate HgBa$_2$CuO$_{4+\delta}$}

\author{Sudheer Anand Sreedhar}
\affiliation{Department of Physics, University of California, Davis, CA 95616, USA}

\author{Antonio Rossi}
\affiliation{Department of Physics, University of California, Davis, CA 95616, USA}
\affiliation{Advanced Light Source, Lawrence Berkeley National Lab, Berkeley, 94720, USA}

\author{J. Nayak}
\altaffiliation[Current address: ]{Department of Physics, Indian Institute of Technology Kanpur, Kanpur - 208016, India}
\affiliation{Department of Physics, University of California, Davis, CA 95616, USA}

\author{Zach Anderson}
\affiliation{University of Minnesota, School of Physics and Astronomy, Minneapolis, Minnesota 55455, USA}

\author{Yang Tang}
\affiliation{University of Minnesota, School of Physics and Astronomy, Minneapolis, Minnesota 55455, USA}

\author{Benjamin Gregory}
\affiliation{Cornell University, Department of Physics, Ithaca, NY 14850, USA}

\author{M.~Hashimoto}
\affiliation{Stanford Synchrotron Radiation Lightsource, SLAC National Accelerator Laboratory, Menlo Park, CA 94025, USA}

\author{D.-H. Lu}
\affiliation{Stanford Synchrotron Radiation Lightsource, SLAC National Accelerator Laboratory, Menlo Park, CA 94025, USA}

\author{Eli Rotenberg}
\affiliation{Advanced Light Source, Lawrence Berkeley National Lab, Berkeley, 94720, USA}

\author{R. J.~Birgeneau}
\affiliation{Department of Physics, University of California, Berkeley, California 94720, USA}

\author{M. Greven}
\affiliation{University of Minnesota, School of Physics and Astronomy, Minneapolis, Minnesota 55455, USA}

\author{M.~Yi}
\affiliation{Department of Physics and Astronomy, Rice University, Houston, Texas 77005, USA}
\affiliation{Department of Physics, University of California, Berkeley, California 94720, USA}

\author{I. M. Vishik}
\email[]{ivishik@ucdavis.edu}
\affiliation{Department of Physics, University of California, Davis, CA 95616, USA}

\begin{abstract}
The lamellar cuprate superconductors exhibit the highest ambient-pressure superconducting transition temperatures (T$_C$) and, after more than three decades of extraordinary research activity, continue to pose formidable scientific challenges. A major experimental obstacle has been to distinguish universal phenomena from materials- or technique-dependent ones. Angle-resolved photoemission spectroscopy (ARPES) measures momentum-dependent single-particle electronic excitations and has been invaluable in the endeavor to determine the anisotropic momentum-space properties of the cuprates. HgBa$_2$CuO$_{4+\delta}$ (Hg1201) is a single-CuO$_2$-layer cuprate with a particularly high optimal T$_C$ and a simple crystal structure; yet there exists little information from ARPES about the electronic properties of this model system. Here we present an ARPES study of doping-, temperature-, and momentum-dependent systematics of near-nodal dispersion anomalies in Hg1201. The data reveal a hierarchy of three distinct energy scales —a sub-gap low-energy kink, an intermediate-energy kink near 55 meV, and a peak-dip-hump structure. The first two features are attributed to the coupling of electrons to Ba-derived optical phonons and in-plane bond-stretching phonons, respectively. The nodal peak-dip-hump structure appears to have a common doping-dependence in several single-layer cuprates, and is interpreted as a manifestation of pseudogap physics at the node. These results establish several universal phenomena, both in terms of connecting multiple experimental techniques for a single material, and in terms of connecting comparable spectral features in multiple structurally similar cuprates. 
\end{abstract}
\maketitle

\section{Introduction}
Cuprate high-temperature superconductors have a rich phase diagram with multiple competing and coexisting phenomena and interactions, which are characterized by strong anisotropy both in real space and momentum space \cite{Howald:inhomogeneities,Lang:ImagingGranular,Cuk:MomentumDepKinkBi2212,Hashimoto:EnergyGaps,Anzai:Kinks}, the latter of which is the focus of this work. Many of these emergent phases, notably superconductivity and the normal state pseudogap, remain without an accepted explanation, and one promising approach is to identify associated collective excitations to which electrons strongly couple. Doing so often requires multiple experimental probes, but different compounds are often favored by different techniques, muddling this correspondence.

Cuprates share a common structural unit of the CuO$_2$ plane, but the hundreds of known cuprate compounds differ from one another in the number of adjacent CuO$_2$ planes (single-layer vs multiple-layer) and in the chemistry of the charge reservoir layers separating CuO$_2$ plane(s) from one another.  While a common framework is the goal,  disparities among compounds can also guide an understanding of the mechanism of producing and/or enhancing $T_c$.  

Nowhere is this materials-dependence more striking than among single-layer cuprates.  These materials are structurally similar, but some have maximum $T_c$ values ($T_{c,max}$)  near 35-40 K, whereas others have $T_{c,max}$ near 100 K \cite{Eisaki:ChemicalInhomogeneity,Sakakibara:OriginMaterialDependenceTc}.  Notably, the momentum-dependent many-body interactions are relatively well characterized in several of the lower-$T_c$ compounds \cite{Ronning:NaCCOC_nodal_PDH,Hashimoto:DopingEvolutionElectronicStructureSingleLayer,Yoshida:LSCO_ARPES}, helping to guide interpretation of bulk phenomena in terms of momentum-dependent microscopic electronic properties, but no comparable systematic ARPES study yet exists for any of the higher-$T_c$ members.

HgBa$_2$CuO$_{4+\delta}$ (Hg1201) has $T_{c,max} = 98$ K and is considered to be a model cuprate because of its single-layer simple-tetragonal crystal structure and relatively minor (point) disorder effects \cite{Eisaki:ChemicalInhomogeneity,Barisic:DemonstratingModelNature}. It is also a member of the family of Hg-cuprates which achieve the maximum $T_c$ among all cuprates in the triple-layer version \cite{Schilling:TripleLayerHgTc}.  There exist extensive experimental results for Hg1201 including charge transport \cite{Barisic:PNAS_phaseDiagram,DoironLeyraud:HallSeebeckNernst}, torque magnetometry \cite{Murayama:DiagonalNematicity}, optical spectroscopy \cite{vanHeumen:OpticalDetermination}, microwave \cite{Grbic:MicrowaveSCFluct}, as well as neutron \cite{Li:UnusualMagneticOrder,Chan:DopingDepMagResMode2016,Chan:Tdep_mag_res_mode2016}, X-ray \cite{Lu:ChargeTransferExcitations,Tabis:ChargeOrderFLtransport,Tabis:SynchrotronXrayStudyCDW}, and Raman \cite{Li:DopingDepPhotonScattering} scattering. However, ARPES measurements have been limited owing to the lack of a neutral cleavage plane and the high sensitivity to photoemission matrix elements \cite{Vishik:Hg1201_PRB2014}.

Here we present a systematic study (doping-, temperature-, momentum-dependence) of near-nodal dispersions and lineshapes that highlights three different energy scales in the ARPES spectra of Hg1201, and establishes key points of universality and deviations therefrom among single-layer cuprates.  First, we observe a low-energy kink (at $\omega\approx 11$ meV), which has so far been reported only for Bi-based cuprates, and associated with the interaction of electrons with acoustic or optical phonons \cite{Rameau:LEKink,Vishik:LowEnergyKink,Kondo:LEKink,Plumb:LEKink,Johnston:ForwardScattering}.  Second, the intermediate-energy kink ($\omega\approx 55$ meV) shows a temperature, doping, and momentum-dependent phenomenology consistent with coupling to in-plane bond-stretching phonons.  While this appears to be common to all single-layer cuprates, the stronger coupling in Hg1201 suggests a possible connection to $T_c$ enhancement.  Finally, a peak-dip-hump (PDH) lineshape is observed at the \textit{node}, the momentum point where the superconducting gap is identically zero.  We find that it also has a doping dependence similar to the that of the pseudogap, the anomalous state existing above $T_c$ in hole doped cuprates.  The nodal PDH appears to be ubiquitous in single-layer cuprates, and may be connected to significant electronic changes at the hole-doping level $p\approx0.2$, favoring explanations of pseudogap phenomena which do not vanish at the node.  Hg1201 has yielded important insights from multiple experimental techniques, and the comprehensive nature of the present ARPES work makes it a starting point for establishing a cohesive multi-technique narrative about this prototype material.

\section{Experiments}
Hg1201 single crystals were grown by a two-step flux method \cite{Zhao:GrowthCharacterizationModelHg1201} and were subsequently annealed to achieve the desired $T_c$ \cite{Barisic:DemonstratingModelNature}.  For ARPES experiments, care was taken to ensure mechanical adhesion and electrical conductance between the samples and the sample holder, because Hg1201 is mechanically robust and reacts with commonly used silver epoxies.  Samples were glued onto copper sample post using Torr seal. Silver paint (DuPont 4922N-100) was used to ground the samples and was cured at room temperature. ARPES experiments were performed at Stanford Synchrotron Radiation Lightsource (SSRL) beamline 5-4 using 19.4 eV photon energy, which was previously shown to optimize valence band spectral quality near the node \cite{Vishik:Hg1201_PRB2014}.  The beam-spot size was approximately 100 microns.  Samples were cleaved \textit{in situ} at the lowest measurement temperature (30 K).  Roughness of the cleaved surface can contribute to extrinsic broadening of ARPES lineshapes, and the quality of the cleave was assessed from the presence of a quasiparticle peak and the Lorentzian width of the momentum distribution curve (MDC) at $E_F$, which was as small as 0.09 \AA$^{-1}$.  Additional X-ray photoelectron spectroscopy (XPS) measurements were performed at beamline 7.0.2 of the Advanced Lightsource (ALS) to ascertain surface termination, and these data are shown in supplementary materials (SM) \cite{supplementaryMaterials}.

\section{Results}

\begin{figure}[ht]
\includegraphics[width=1.0\columnwidth]{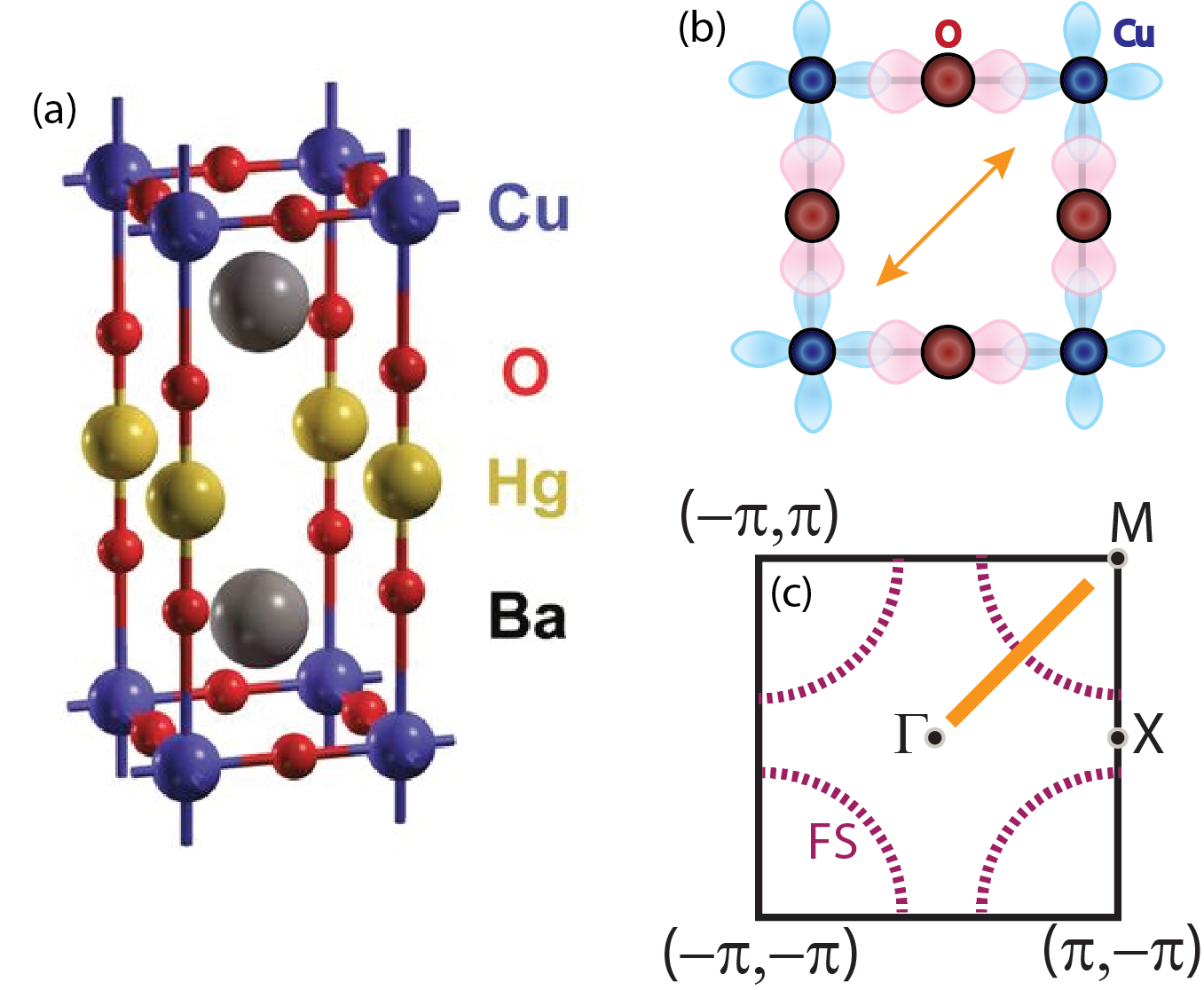}
\caption{Crystal structure and Brillouin zone of Hg1201. a) Schematic crystal structure of Hg1201
for simplicity drawn without oxygen interstitials in the Hg layers.  Image from Ref. \onlinecite{Barisic:PNAS_phaseDiagram}. b) portion of CuO$_2$ plane, with bond-diagonal (nodal) direction indicated by orange arrow c) Two-dimensional projection of tetragonal Brillouin zone with high-symmetry points labeled. Schematic of Fermi surface (FS) is shown by purple dashed line, with nodal cut indicated by thick orange line. }
\label{fig:fig0}
\end{figure}

Figure \ref{fig:fig0} introduces the structure of Hg1201 in real and momentum space.  The unit cell (Fig. \ref{fig:fig0}(a)) has a single CuO$_2$ plane.  Our XPS measurements suggest that cleavage happens along the O-Hg-O barbells \cite{supplementaryMaterials}, which are located relatively far from CuO$_2$ planes and yield an overall neutral cleaved surface, because the bond is equally likely to be broken above (Hg$^{2+}$ termination) and below (O$^{2-}$ termination) the Hg atoms.  The nodal cuts used for the majority of this paper correspond to $45^\circ$ from the Cu-O bond directions in the CuO$_2$ plane (Fig. \ref{fig:fig0}(b)) or along the diagonal of the Brillouin zone (Fig. \ref{fig:fig0}(c)).  The terminology originates from the zero of the superconducting gap, but it is used even in the absence of superconductivity to refer to the aforementioned trajectory.

\begin{figure}[ht]
\includegraphics[width=1.0\columnwidth]{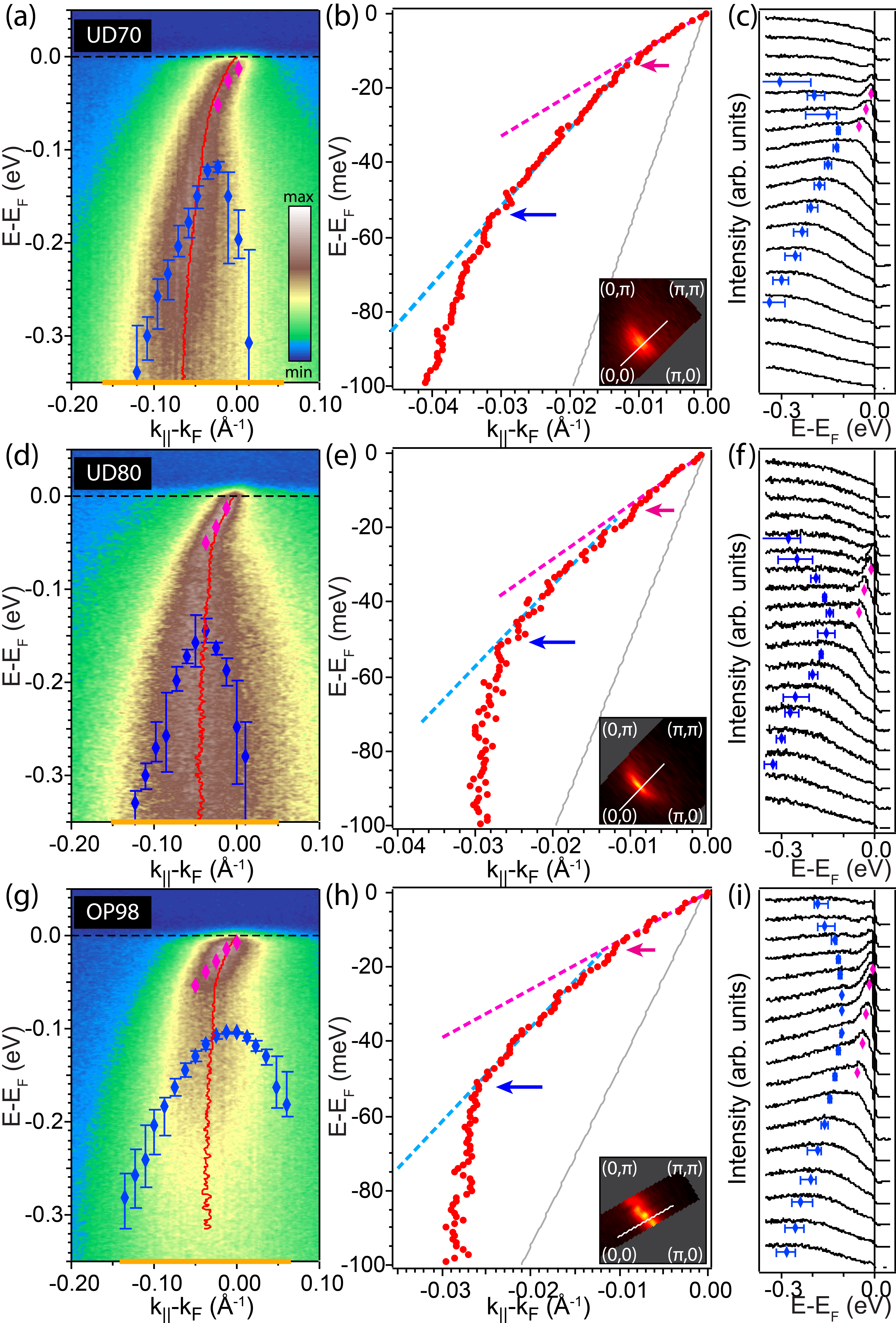}
\caption{Nodal dispersions and lineshapes. a) Nodal cut for UD70. Red line: MDC-fitted dispersion. Magenta and blue symbols: `peak' and `hump' positions (local maxima). Incoherent background, determined from EDC far from the dispersion, has been subtracted. b) MDC-derived dispersion for UD70. Thin grey line: tight-binding band dispersion, used as bare band. Magenta and cyan dotted lines: linear fits between [0, -10] meV and [-25, -45] meV, respectively. Magenta and blue arrows point to the low-energy and intermediate-energy kinks, respectively.  Inset: measured Fermi surface and cut angle (white). c) EDCs for UD70 along momentum region marked by yellow bar in (a), with peak and hump positions indicated. (d)-(f) Same data for UD80. (g)-(i) Same data for OP98. All measurements are taken at 30 K.  Error bars in hump position reflect variation in eight adjacent EDCs, which were averaged to give each EDC in (c),(f),(i).
}
\label{fig:fig1}
\end{figure}

Figure~\ref{fig:fig1} shows dispersions and lineshapes along nodal cuts for the three doping levels investigated in this study: UD70 (underdoped, $T_c=70K$, $p\approx0.095$), UD80 ($p\approx0.11$), and OP98 (optimal doping, $p\approx0.16$), with doping levels estimated from Ref. \onlinecite{Yamamoto2000}. All data in this figure were obtained at 30K, well below $T_c$.  An incoherent background was subtracted to obtain image plots (Fig.~\ref{fig:fig1}(a),(d),(g)).   Dispersions are quantified by fitting MDCs to Lorentzians plus a constant background, and the results, with a focus on the lower-energy anomalies, are shown in Fig.~\ref{fig:fig1}(b),(e),(h). The ubiquitous dispersion kink is observed around 50-70 meV, and its estimated energy is indicated by arrows.  Another bend in the dispersion, the low-energy kink, is observed at $\approx 11$ meV.  An assumed  tight-binding bare band \cite{Das:Hg1201_TB_params,TB:params} is also shown in panels (b),(e), and (h).  Energy distribution curves (EDCs) along the momentum range marked by yellow bars in panels (a),(d),(g) are shown in Fig.~\ref{fig:fig1}(c),(f),(i).  Quasiparticle peaks are observed near $k_F$ at all three doping levels (magenta), which attests to the quality of the spectra.  An additional `hump' feature is observed at higher binding energy, lending a PDH lineshape to \textit{nodal} cuts even at optimal doping. The peak and hump energies were determined from local maxima of smoothed data.  The `hump' feature first disperses toward the Fermi energy ($E_F$), and then disperses away.  The two nodal kinks and the nodal PDH structure provide evidence for strong interactions at three distinct characteristic energy scales.

\begin{figure}[ht]
 	\includegraphics[width=1.0\columnwidth]{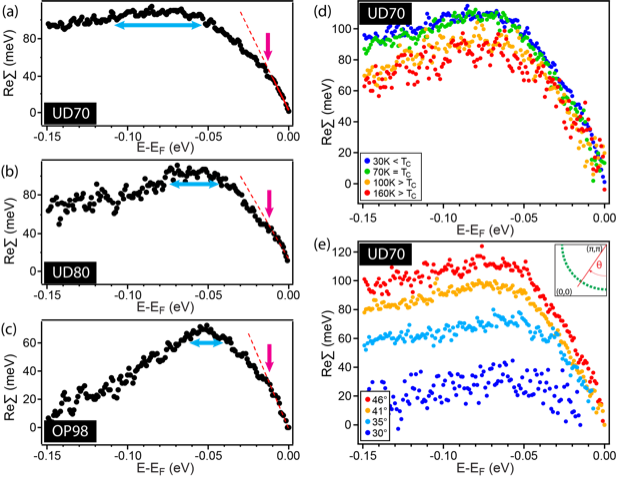}
 	\caption{Real part of the self energy, Re$\Sigma$, obtained by subtracting an assumed tight-binding bare band. (a) Re$\Sigma$ at 30 K at the node for UD70. Blue arrow indicates the estimated energy range of the intermediate-energy kink, chosen to correspond to a 15$\%$ decrease of Re$\Sigma$ from the maximum value. Dashed red line: linear fit to the low-energy [0,-10] meV portion of Re$\Sigma$. Magenta arrow: deviation from this linear fit indicates the low-energy kink. (b)-(c) Same for UD80 and OP98. (d) Temperature dependence of Re$\Sigma$ at the node for UD70. (c) Momentum dependence of Re$\Sigma$ for UD70, with angle as indicated in inset.}
 	\label{fig:fig2}
\end{figure}

Figure~\ref{fig:fig2} shows the evolution of the intermediate-energy nodal kink with doping, temperature, and momentum.  The real part of the self energy ($Re \Sigma$) is estimated by subtracting an assumed tight-binding bare band from the MDC dispersion, a standard technique to quantify the effects of many-body interactions on band position  \cite{johnson2007photoemission, He:Bi2212KinkAcrossTc, Lee:TempEvolutionKinkBi2212, Park:BrokenRelationship}.  The momentum difference at each energy is multiplied by the slope of the bare band, which is nearly linear in the energy region of interest, to recover units of energy. The kink energy is estimated from the broad peak of $Re \Sigma$, and slightly disperses to lower energy with increasing doping (Fig. \ref{fig:fig2} (a), (b), (c)).  Additionally, a subkink feature is observed around 11 meV, an energy scale consistent with the low energy kink previously only reported in Bi-based cuprates\cite{Vishik:LowEnergyKink,Rameau:LEKink,Plumb:LEKink,Kondo:LEKink,Peng:Bi2201_nodalPDH,Anzai:Kinks}.  Temperature dependence of self energy is shown for UD70 in Fig. \ref{fig:fig2}(d), from well below $T_c$ to well above $T_c$. The pseudogap temperature ($T^*\approx 300K$ \cite{Chan:Tdep_mag_res_mode2016}), could not be accessed because of sample aging above 200K.  $Re \Sigma$ is relatively insensitive to temperature, decreasing by less than 20$\%$ at its maximum between 30K and 160K. Figure~\ref{fig:fig2}(e) shows the momentum dependence of $Re \Sigma$ at 30K for UD70 starting at the node (red) and moving approximately 1/3 of the way to the antinode (dark blue).  The momentum dependence is more substantial than the temperature dependence: the magnitude of $Re \Sigma$ decreases by at least 75$\%$ over the aforementioned momentum range.

\begin{figure}[ht]
 	\includegraphics[width=0.8\columnwidth]{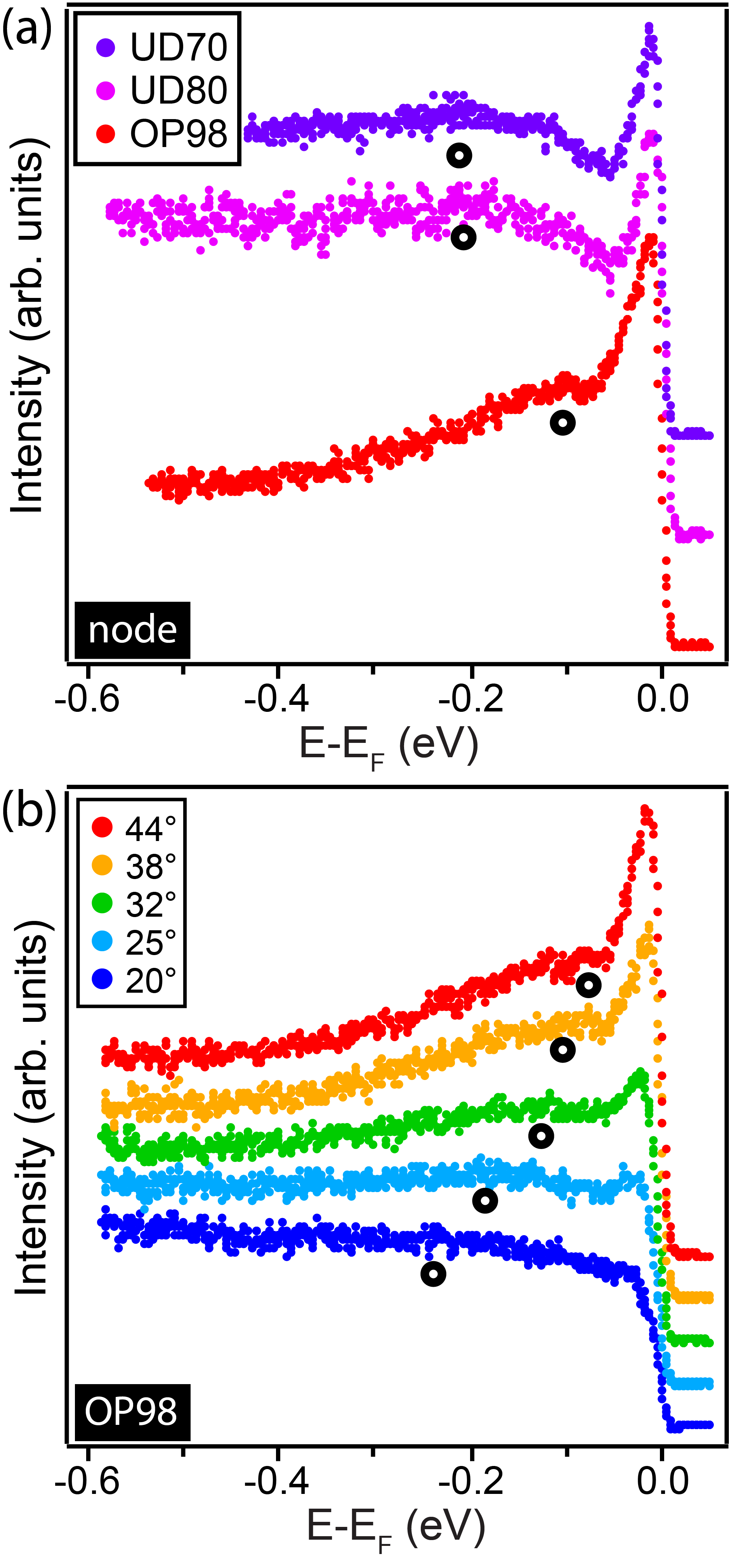}
 	\caption{Nodal peak-dip hump. (a) Doping dependence at the node. (b) Momentum dependence for OP98, showing EDCs at $k_F$. Markers point to the hump energy as defined in Fig. \ref{fig:fig1}. Data in (a) are normalized by maximum height, whereas data in (b) are normalized at -0.55 eV. EDCs offset for clarity.}
 	\label{fig:fig3}
\end{figure}

Figure~\ref{fig:fig3} explores the doping and momentum dependence of the nodal PDH structure.  Figure \ref{fig:fig3}(a) shows the EDCs at the nodal $k_F$ for the three doping levels. The nodal PDH persists to at least optimal doping. Figure \ref{fig:fig3}(b) shows the momentum dependence for OP98, indicating that the hump energy scale disperses to higher energy moving away from the node.

\begin{figure}[ht]
\includegraphics[width=0.8\columnwidth]{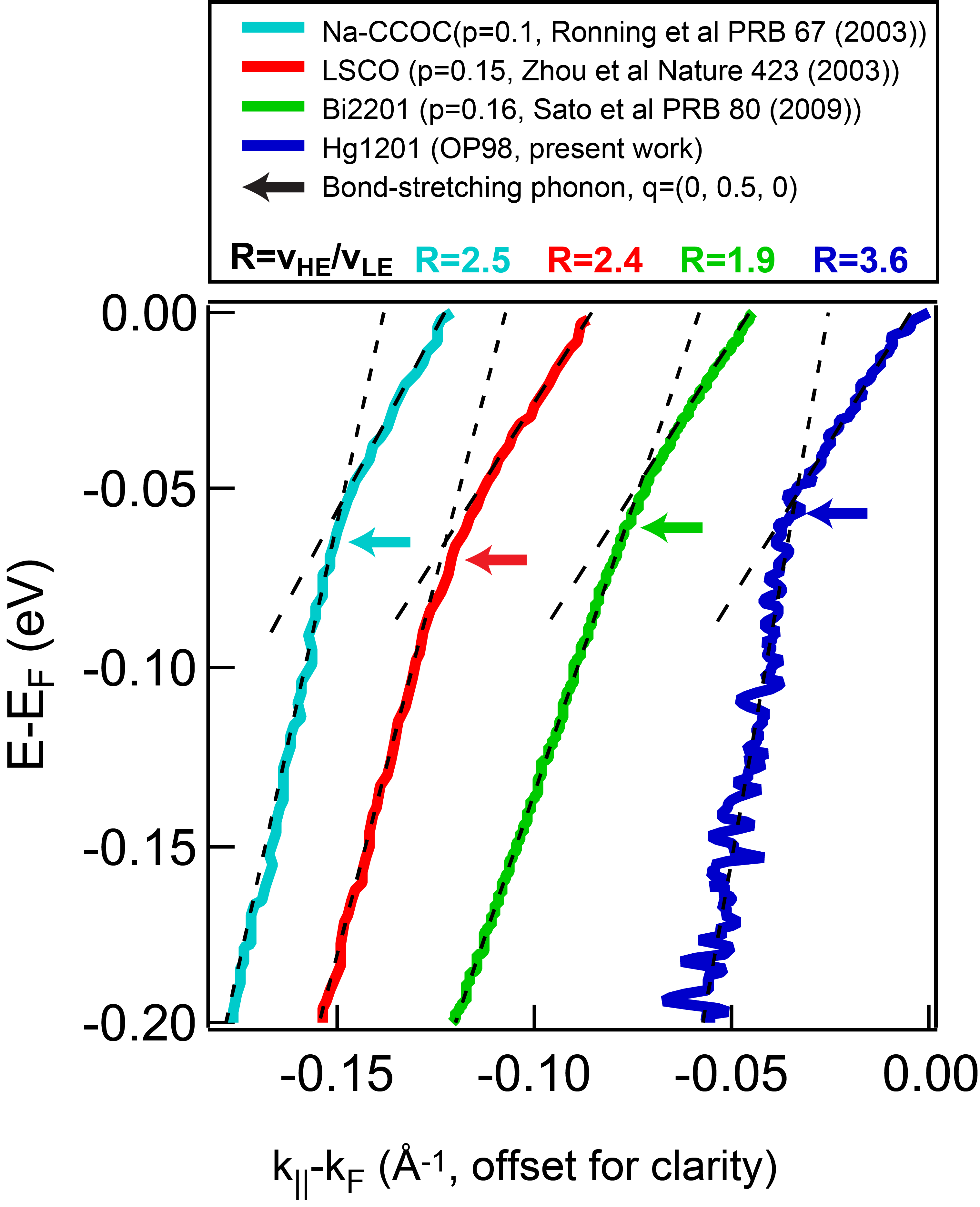}
\caption{Comparison of the intermediate-energy nodal kink for four single-layer cuprates near optimal doping \cite{Ronning:NaCCOC_nodal_PDH, Zhou:UniveralNodalFermiVelocity, Sato:Bi2201_2009} with corresponding values of the bond-stretching phonon at \textbf{q}=(0.5,0,0) \cite{dAstuto:phonon_NaCCOC, Pintschovius:BondStretchDopingDep, graf:momentumDepKink_Bi2201_OP,Uchiyama:PhononsH1201_2004}. Dashed lines: fits to $v_{LE}$ (low-energy velocity, [20,45] meV) and $v_{HE}$ (high-energy velocity, [80,200] meV). The renormalization across the intermediate-energy kink, $R \equiv v_{HE}/v_{LE}$, is indicated. }
\label{fig:fig4a}
\end{figure}

Figure \ref{fig:fig4a} compares the intermediate-energy nodal kink near optimal doping for several single-layer cuprates \textemdash Hg1201, Ca$_{2-x}$Na$_x$CuO$_2$Cl$_2$ (Na-CCOC), La$_{2-x}$Sr$_x$CuO$_4$ (LSCO), and Bi$_2$Sr$_2$CuO$_{6+\delta}$ (Bi2201).  Also shown are the values for the bond-stretching (half-breathing) phonon at \textbf{q}$=$(0.5,0,0) near optimal doping \cite{dAstuto:phonon_NaCCOC,Pintschovius:BondStretchDopingDep,graf:momentumDepKink_Bi2201_OP,Uchiyama:PhononsH1201_2004}).
The velocity renormalization across the intermediate-energy kink at optimal doping, $R=v_{HE}/v_{LE}$, varies among single-layer cuprates from 1.9 (Bi2201) to 3.6 (Hg1201); here, the low-energy velocity, $v_{LE}$, is taken as the slope of the MDC dispersion in the [20,45] meV range, and the high-energy velocity, $v_{HE}$, is the slope in the [80,200] meV range.

\begin{figure}[ht]
\includegraphics[width=0.8\columnwidth]{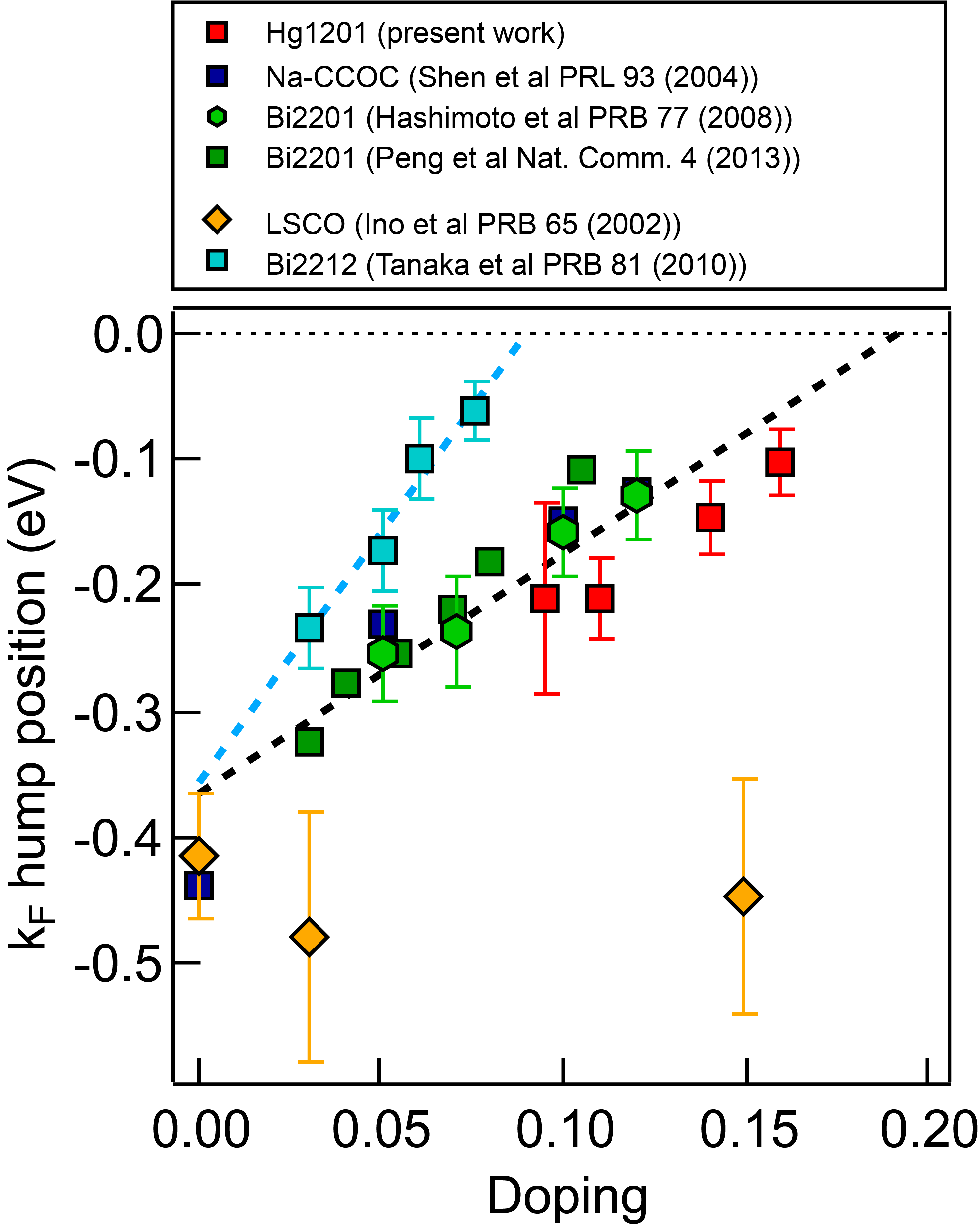}
\caption{Comparison of `hump' energy scales at nodal $k_F$ for five cuprates. Additional datum for Hg1201 at $p=0.14$  from Ref. \onlinecite{Vishik:Hg1201_PRB2014}. Blue dashed line: linear fit to Bi2212 data \cite{Tanaka:EvolutionElectronicStructureBi2212}. Black dashed line: linear fit for Bi2201 \cite{Hashimoto:DopingEvolutionElectronicStructureSingleLayer,Peng:Bi2201_nodalPDH}, Na-CCOC \cite{Shen:DopingDepCCOC}, and Hg1201.  LSCO data from Ref. \onlinecite{Ino:DopingDependenceLSCO}. }
\label{fig:fig4b}
\end{figure}

The doping and materials dependence of the hump energy scale are shown in Fig. \ref{fig:fig4b}, where we compare our result for Hg1201 with data from LSCO \cite{Ino:DopingDependenceLSCO}, Na-CCOC \cite{Shen:DopingDepCCOC}, Bi2201 \cite{Hashimoto:DopingEvolutionElectronicStructureSingleLayer,Peng:Bi2201_nodalPDH}, Bi$_2$Sr$_2$CaCu$_2$O$_{8+\delta}$ (Bi2212) \cite{Tanaka:EvolutionElectronicStructureBi2212}.  We note that while similar spectral features are identified for all the materials, they are not necessarily interpreted the same way in the existing literature.  Three systematic trends are observed:  LSCO shows no doping dependence of the nodal hump feature.  Double-layer Bi2212 exhibits this feature only at low doping, with an energy scale that extrapolates to zero at $p\approx0.09$.  Within error, the remaining materials, all single-layer cuprates, exhibit the same hump energy scale, which extrapolates to zero at $p\approx0.2$.

\section{Discussion}
We first discuss the low-energy kink.  A similar feature was previously observed and studied in Bi-based cuprates.  It was found that the velocity renormalization across this kink increases with underdoping below a certain material-dependent hole concentration ($p\approx0.13$ in Bi2212 \cite{Vishik:LowEnergyKink} and $p\approx0.2$ in Bi2201 \cite{Kondo:LEKink}).  In Hg1201, we obtain a velocity renormalization, $v_{mid}/v_F \approx 2$, for all studied doping levels, where $v_F$ is the slope of the MDC dispersion 0-10 meV and $v_{mid}$ is the slope in the 20-45 meV range. These energy intervals are chosen to avoid low-energy and intermediate-energy kinks.  This suggests that doping dependence, if present in Hg1201, occurs at lower hole concentrations than accessed by this study; alternately the mechanism of the low-energy kink in this system may be different from Bi-based cuprates.

The low-energy kink in Bi-based cuprates has been attributed to coupling to either acoustic \cite{Johnston:ForwardScattering} or optical phonons \cite{Rameau:LEKink}, with electron-phonon coupling peaked at small \textbf{q} for both cases. Our assignment for Hg1201 is guided by phonon dispersions measured near optimal doping \cite{dAstuto:Hg1201_phonon2003, Uchiyama:PhononsH1201_2004}, and by the kink energy that we observe.  A kink at 11 meV is consistent with a zone-center $c$-axis optical branch, and with in-plane ($E_u$) modes at large momentum transfer, but not with acoustic phonons, which reach a maximum of only $\approx$ 7 meV at the zone boundary \cite{dAstuto:Hg1201_phonon2003}. The calculated phonon density of states shows contributions primarily from Ba in the energy range of interest \cite{dAstuto:Hg1201_phonon2003}.   It has been shown that only $A_{1g}$-symmetry optical phonons can have coupling peaked at $\textbf{q} = 0$ for nodal Fermions, and the closest candidate in Hg1201 is a Ba mode with energy 20-26 meV \cite{Zhou:RamanHgCuprate1996,dAstuto:Hg1201_phonon2003}.  This is inconsistent with the low-energy kink energy, unless the phonon frequency softens considerably near the sample surface.  Thus, electron-phonon coupling peaked at larger \textbf{q} needs to be considered instead. This points to Ba-derived optical modes with large reduced momentum. As discussed below, the lack of a gap shift can be achieved by scattering from near-node to near-node.  This mechanism is different from most explanations of the low-energy kink in Bi-based cuprates, and may be related to the different doping dependence of this feature in Hg1201.

The intermediate-energy nodal kink, observed between 50-80 meV in all cuprates \cite{Lanzara:EvidenceUbiquitous,Johnston:MaterialDopingDepkink,Anzai:Kinks}, is a ubiquitous spectral feature with various purported origins. It has been attributed to a coupling to sharp bosonic excitations of lattice, magnetic, or dual origin. There also exist proposals that involve coupling to a gapped continuum of excitations \cite{Chubukov:DispersionAnomalyS-shape} and proposals that do not involve electron-boson coupling \cite{Matsuyama:OriginKinksNoMode}.  Notably, in the present data, the MDC-derived dispersions for nodal cuts show a slight `S' shape, most visible in Fig. \ref{fig:fig4a}, the characteristic of coupling to a sharp bosonic mode \cite{Chubukov:DispersionAnomalyS-shape}. 

In Hg1201, several previously reported excitations have energy scales similar to the intermediate-energy kink: a dispersionless Ising-like excitation \cite{Li:IsingHg1201},
the maximum in the magnetic susceptibility at the antiferromagnetic wavevector \cite{Chan:DopingDepMagResMode2016,Chan:Tdep_mag_res_mode2016}, and bond-stretching in-plane phonons \cite{Uchiyama:PhononsH1201_2004} .
A recent X-ray scattering study of an underdoped $T_c = 70$ K Hg1201 sample revealed a broad spectrum of charge excitations in the vicinity of the two-dimensional wavevector $\textbf{q}_{CDW} = (0.3,0)$, where short-range charge-order is observed, but these excitations are peaked well below the kink energy \cite{yu2019unusual}.

The nodal kink is a somewhat broad feature, possibly with multiple (lattice, magnetic, charge) contributions, and in this manuscript we narrow down the likeliest phonon contribution based on observed phenomenology. As noted, in Hg1201, this kink appears to have some doping dependence (see Fig. \ref{fig:fig2}). At optimal doping, its energy is approximately 55 meV, whereas for UD70, its characteristic energy is closer to 70 meV.
This doping dependence is the opposite of that of the maximum in the local magnetic susceptibility in Hg1201 ($\sim 55$ meV and $\sim 60$ meV for samples with $T_c = 71$ K and 88 K, respectively) \cite{Chan:DopingDepMagResMode2016,Chan:Tdep_mag_res_mode2016} and other cuprates \cite{Yu:UniversalRelationMagResDelta}, which decreases with underdoping. The Ising-like excitations have not been observed in cuprates other than Hg1201, and the energy of the higher energy branch exhibits little doping dependence, and strong temperature dependence \cite{Li:IsingHg1201,Li:HiddenMagneticExcitations}. We will therefore not discuss them any further.
On the other hand, the bond-stretching phonon along [100] is generally found to shift to lower energy (soften) approaching the middle of the Brillouin zone (\textbf{q}=(0.5, 0, 0) and equivalent) in doped cuprates, which has been interpreted as evidence of strong coupling to electrons \cite{Pintschovius:AnomalousDisp}.  Whereas  doping-dependence is not well-established for Hg1201, the bond stretching phonon energy at \textbf{q}=(0.5, 0, 0) has been shown to decrease systematically with increasing doping in LSCO, while the energy of this branch near \textbf{q}=(0, 0, 0) has minimal doping dependence \cite{Pintschovius:BondStretchDopingDep}. In Hg1201, both at optimal doping \cite{Uchiyama:PhononsH1201_2004} and for an underdoped sample with $T_c = 55$ K \cite{ahmadova2020phonon}, the measured energy of the bond-stretching phonon at \textbf{q}=(0.5, 0, 0) is consistent with the energy of the intermediate-energy kink in Fig. \ref{fig:fig2}(a)-(c). Further connections will be made when we discuss the momentum dependence.

The second consideration is the minimal or absent temperature dependence of the kink, including the lack of a shift to lower energy across $T_c$ (Fig. \ref{fig:fig2}(d)).
In a \textit{d}-wave superconductor, a kink observed at energy $\omega$ will generally be associated with a phonon of energy $\Omega=\omega-\Delta_0$, where $\Delta_0$ is the maximum of the superconducting gap, if the momentum-dependent electron-phonon coupling connects the momentum of interest to the momentum where the superconducting gap is maximum \cite{Sandvik:EffectsElectronPhononInteration}. The maximum of the superconducting gap of Hg1201 has previously been measured to be around 39 meV at optimal doping \cite{Vishik:Hg1201_PRB2014}. However, there are exceptions to this expectation of 'gap shifting', e.g. when electron-phonon coupling is strongly peaked in the forward direction ($\mathbf{q}\approx 0$) such that fermions at a given momentum are only sensitive to the local gap \cite{Johnston:ForwardScattering}, or if the coupling is strongly peaked at a particular momentum transfer that connects momenta with zero or small gap, such as two opposite nodes. The latter is consistent with the momentum dependence in the present data (Fig. \ref{fig:fig2}(e) and SM \cite{supplementaryMaterials}).  The minimal temperature dependence of the magnitude of $Re\Sigma$  in Fig. \ref{fig:fig2} suggests a lattice origin of the intermediate-energy kink, and the lack of gap-shifting of the peak position of $Re\Sigma$ resstricts this to phonon modes with strongly \textbf{q}-dependent coupling.  

There exists some materials-dependence with regard to the nodal kink phenomenology.  For example, Bi2212 shows a strong temperature dependence of the coupling strength, primarily across $T_c$ \cite{He:Bi2212KinkAcrossTc}, but also extending to higher temperature \cite{Vishik:Thesis}, whereas other single-layer cuprates show weaker or absent temperature dependence similar to Hg1201 \cite{Lanzara:temperatureDepKinkBi2201,Zhao:QuantitativeBi2201_Tdep}.  This is consistent with additional contributions to the nodal kink, such as coupling to the $B_{1g}$ (buckling) mode, in multi-CuO$_2$-plane cuprates \cite{Devereaux:AnisotropicElectronPhononCoupling2004}.

The third aspect of the intermediate-energy kink phenomenology is the pronounced momentum dependence shown in Fig. \ref{fig:fig2}(e), with the kink nearly disappearing for cuts sufficiently far from the node.  This suggests a strong anisotropy in either the mode or the coupling or both.  As previously discussed  \cite{graf:momentumDepKink_Bi2201_OP}, this momentum dependence can be reproduced by electrons coupling primarily to the \textit{soft} portion of the half-breathing mode \textemdash momentum transfer near \textbf{q}=(0.5,0.0) where the phonon energy is decreased from the parent-compound value.  Notably, this and nearby values of \textbf{q} connects near-node to near-node, but is too large to connect antinodal states.  This can also produce a kink whose energy is not gap shifted by the superconducting gap, as the superconducting gap near the nodes is zero or small relative to the bond-stretching mode energy.  At optimal doping, where both ARPES data and the measured phonon dispersion of the bond stretching phonon is available, there is quantitative agreement in the magnitude and momentum-dependence of the intermediate energy kink and the bond-stretching phonon \cite{supplementaryMaterials}.

This identification also yields consistency when comparing to other single-layer cuprates. At optimal doping, the nodal kink energy varies among single-layer cuprates (Fig. \ref{fig:fig4a}), which is mirrored in the variation in the energy of the [100] bond-stretching mode near the zone center \cite{dAstuto:phonon_NaCCOC,Pintschovius:BondStretchDopingDep,graf:momentumDepKink_Bi2201_OP,Uchiyama:PhononsH1201_2004}.  Notably, for all materials near optimal doping, the bond-stretching phonon energy at \textbf{q}=(0.5,0.0) is lower by 10-15 meV than at the zone center, which further constrains the origin of the kink to the soft part of this phonon branch. 

The identity of the phonon contributing to the intermediate-energy kink is somewhat separate from this mode's role in superconductivity.  The bond-stretching mode has been associated with both enhancement \cite{Ishihara:BondStretchEnhance} and suppression \cite{Sakai:Pairing_tJ_model,Johnston:SystematicStudyEPC} of \textit{d}-wave pairing.  The persistence of the kink into the overdoped, non-superconducting regime has also been interpreted in terms of the bond-stretching mode having no relation to superconductivity \cite{Park:BrokenRelationship}.  Hg1201 shows a stronger velocity renormalization across the intermediate-energy kink, an estimate of the coupling strength, than other single-layer cuprates with lower $T_c$ (Fig. \ref{fig:fig4a}).  This suggests that for Hg1201, the single-layer cuprate with the largest $T_{c,max}$, deleterious effects of the bond-stretching mode on superconductivity are minimal or fully compensated by other factors that raise $T_c$.

Finally, we turn to the \textit{nodal} PDH structure. We first discuss general interpretations of PDH lineshapes in cuprates, and then turn specifically to this feature at the \textit{node} in single-layer cuprates.  At the antinode, the PDH lineshape is weak or absent in single-layer cuprates, but ubiquitous in multilayer compounds \cite{Lee:LayerNumberDep,Wei:Bi2201_OP_lineshape,Feng:Bi2223ARPES}.  In the latter, the dip has been attributed to either a magnetic resonance mode \cite{Campuzano:Resonance}, or a feature of electron-phonon coupling to the B$_{1g}$ oxygen bond-buckling phonon \cite{Cuk:B1g_phononAntinode,He:RapidBi2212_Science}.The antinodal hump feature below $T_c$ in multi-CuO$_2$-plane cuprates has been associated with the pseudogap, as it evolves into the broadly peaked lineshape defining the pseudogap above T$_c$ \cite{Hashimoto:PhaseCompetition} or as a feature also related to electron-phonon coupling \cite{Cuk:MomentumDepKinkBi2212}.

Near the node, a hump with similar energy scale and dispersion, both within one cut and around the FS, has been reported in Na-CCOC \cite{Ronning:NaCCOC_nodal_PDH,Shen:DopingDepCCOC}, and discussed in the context of polaron physics \cite{Shen:LatticePolaronFormation}.  In that material, it was only reported up to 12$\%$ hole doping.  A nodal PDH lineshape is also seen in non-superconducting Bi2201, up to $p\approx0.105$ \cite{Peng:Bi2201_nodalPDH}.  At optimal doping, the nodal PDH structure is not reported in Bi2201 and LSCO, with EDCs below the kink energy smoothly evolving into sharper quasiparticles above the kink energy \cite{Yoshida:LSCO_ARPES, Wei:Bi2201_OP_lineshape}. In LSCO a similar feature, albeit located at systematically higher energy, has been associated with the lower Hubbard band.

Our results are inconsistent with an interpretation of the near-nodal dip as a sharp mode, because its energy disperses with momentum along a cut.    Instead, the hump and the peak appear to be the fundamental spectral elements, with the dip energy reflecting their meeting point.  Additionally, the dip persists above $T_c$ \cite{supplementaryMaterials}, unlike multi-CuO$_2$-plane cuprates, where the antinodal dip characteristically vanishes across $T_c$. We also note that in Ref. \onlinecite{Vishik:Hg1201_PRB2014}, a 'kink' was reported in the MDC dispersion at around 200 meV, which is likely another manifestation of the nodal hump structure in Hg1201.  Robust features of strong electronic interactions typically manifest in both the MDC and EDC channels, but are sometimes better visualized in one or the other, and the present feature is better quantified in EDCs.

In Fig. \ref{fig:fig4b} we highlight the universality of the nodal PDH structure, and its relevance to the phase diagram. Although Bi2212 and LSCO show a nodal PDH lineshape in certain regimes, they are outliers in overall phenomenology.  Bi2212 shows this structure only in the deeply underdoped regime, and interestingly, the hump energy extrapolates to zero near a phase boundary (p $\approx$ 0.09) where a non-superconducting gap opens at the nodal momentum, possibly associated with spin-density wave order \cite{Vishik:PNAS,Drachuk:NodalGapSpin}. In LSCO, the nodal hump feature is thought to be doping-independent because doping that compound does not result in a chemical potential shift in the underdoped regime \cite{Ino:ChemicalPotentialShiftLSCO}.   

Other single-layer cuprates (Bi2201, Na-CCOC, Hg1201) exhibit consistent nodal hump energy scales, which extrapolate to zero energy around $p=0.2$, a hole concentration identified with both a phase boundary \cite{Vishik:PNAS,Michon:quantumCriticality} and percolative delocalization of the Mott-localized holes \cite{Pelc:heterogeneousBehavior,pelc2019resistivity}.  This suggests that the nodal hump may also be connected with the pseudogap or charge order $-$ phenomena associated with that characteristic doping. Notably, structural/electronic inhomogeneity \cite{krumhansl1992fine} has been identified as key to explaining many aspects of cuprate phenomenology, including doping and temperature dependence of resistivity \cite{Pelc:heterogeneousBehavior,pelc2019resistivity}, emergence of superconductivity on cooling \cite{Pelc:UniversalPrecursor}, and the doping and temperature evolution of ARPES lineshapes \cite{Zaki:CupratePhaseDiagram}. The broadly-peaked nodal hump feature may be the manifestation of similar strong spatial inhomogeneity, possibly dynamic in nature \cite{Chen:chargeFlux}, which the present experiment's macroscopic spot size averages together.  Local-probe measurements can clarify this issue in Hg1201, as they have in Bi-based cuprates. We also note that optical probes have reported a broad continuum of excitations at energies of several hundred meV, with purported relevance for the pairing mechanism, and the phenomenology of these features is consistent with the nodal PDH \cite{DalConte:DisentanglingElectronicPhononic,vanHeumen:OpticalDetermination}. At optimal doping, the momentum-dependence of the hump energy scale nearly follows a \textit{d}-wave form, and extrapolated to an antinodal energy scale of $\approx 300$ meV \cite{supplementaryMaterials}, which is consistent with the maximum of the broad continuum of excitations derived from optics measurements at the same doping \cite{vanHeumen:OpticalDetermination}. Taken together, the present results and the discussed literature suggest a cooperative interplay between dynamic lattice and electronic effects as well as an important role played by inherent inhomogeneity.

\section{conclusions}
We have identified three energy scales of interactions in Hg1201 with implications for understanding electron-phonon coupling in cuprates and connections to the phase diagram.  The low-energy (11 meV) and intermediate-energy (55 meV) kinks reflect electron-phonon coupling, with the former most consistent with coupling to out-of-plane Ba modes and the latter reflecting coupling to the in-plane oxygen half-breathing mode.  Both interactions appear to involve large momentum transfer. We note the possibility that the overall broad intermediate-energy feature contains additional electronic contributions. At higher energy, a PDH structure is observed at the node, with connections to a previously reported broad continuum of excitations and the phase boundary of the pseudogap and/or charge order.

This work establishes points of universality and points of difference among cuprates.  With this observation of the low-energy kink outside of the Bi-family of cuprates, strong coupling to sub-gap phonons appears to be a universal aspect of cuprate phenomenology, although the identity of the phonon and the nature of the coupling (large vs small \textbf{q}) may differ among different materials.  The intermediate-energy kink, has long been known the be ubiquitous in cuprates.  The present results highlight that the bond-stretching phonon at \textbf{q}$\approx$(0.5, 0, 0) can reproduce the phenomenology of the intermediate-energy kink in all single layer cuprates.  However, the coupling to this mode in Hg1201 appears to be stronger, suggesting that it is not harmful and may be helpful to superconductivity.  Finally, a nodal PDH structure appears to be ubiquitous in single-layer cuprates with a common doping-dependence among several compounds, but with properties distinct from the antinodal PDH structure previously established in multiple-CuO$_2$ plane cuprates.  Nevertheless, the nodal PDH may also be connected to the pseudogap, highlighting that this enigmatic phase may be relevant to nodal and near-nodal physics.

\begin{acknowledgments}
The authors acknowledge helpful discussions Andrey Chubukov, Tom Devereaux, and Zhenglu Li.  Work at UC Davis was supported by AFOSR grant FA9550-18-1-0156.  Work at the University of Minnesota was funded by the Department of Energy through the University of Minnesota Center for Quantum Materials under DE-SC-0016371. Work at UC Berkeley and Lawrence Berkeley Laboratory was supported by the Office of Science, Office of Basic Energy Sciences (BES), Materials Sciences and Engineering Division of the U.S. Department of Energy (DOE) under Contract No. DE-AC02-05-CH1231 within the Quantum Materials Program (KC2202).  Use of the Stanford Synchrotron Radiation Lightsource, SLAC National Accelerator Laboratory, is supported by the U.S. Department of Energy, Office of Science, Office of Basic Energy Sciences under Contract No. DE-AC02-76SF00515. This research used resources of the Advanced Light Source, a DOE Office of Science User Facility under contract no. DE-AC02-05CH11231.
\end{acknowledgments}

\bibliography{Hg1201.bib}

\begin{thebibliography}{87}%
\makeatletter
\providecommand \@ifxundefined [1]{%
 \@ifx{#1\undefined}
}%
\providecommand \@ifnum [1]{%
 \ifnum #1\expandafter \@firstoftwo
 \else \expandafter \@secondoftwo
 \fi
}%
\providecommand \@ifx [1]{%
 \ifx #1\expandafter \@firstoftwo
 \else \expandafter \@secondoftwo
 \fi
}%
\providecommand \natexlab [1]{#1}%
\providecommand \enquote  [1]{``#1''}%
\providecommand \bibnamefont  [1]{#1}%
\providecommand \bibfnamefont [1]{#1}%
\providecommand \citenamefont [1]{#1}%
\providecommand \href@noop [0]{\@secondoftwo}%
\providecommand \href [0]{\begingroup \@sanitize@url \@href}%
\providecommand \@href[1]{\@@startlink{#1}\@@href}%
\providecommand \@@href[1]{\endgroup#1\@@endlink}%
\providecommand \@sanitize@url [0]{\catcode `\\12\catcode `\$12\catcode
  `\&12\catcode `\#12\catcode `\^12\catcode `\_12\catcode `\%12\relax}%
\providecommand \@@startlink[1]{}%
\providecommand \@@endlink[0]{}%
\providecommand \url  [0]{\begingroup\@sanitize@url \@url }%
\providecommand \@url [1]{\endgroup\@href {#1}{\urlprefix }}%
\providecommand \urlprefix  [0]{URL }%
\providecommand \Eprint [0]{\href }%
\providecommand \doibase [0]{https://doi.org/}%
\providecommand \selectlanguage [0]{\@gobble}%
\providecommand \bibinfo  [0]{\@secondoftwo}%
\providecommand \bibfield  [0]{\@secondoftwo}%
\providecommand \translation [1]{[#1]}%
\providecommand \BibitemOpen [0]{}%
\providecommand \bibitemStop [0]{}%
\providecommand \bibitemNoStop [0]{.\EOS\space}%
\providecommand \EOS [0]{\spacefactor3000\relax}%
\providecommand \BibitemShut  [1]{\csname bibitem#1\endcsname}%
\let\auto@bib@innerbib\@empty
\bibitem [{\citenamefont {Howald}\ \emph {et~al.}(2001)\citenamefont {Howald},
  \citenamefont {Fournier},\ and\ \citenamefont
  {Kapitulnik}}]{Howald:inhomogeneities}%
  \BibitemOpen
  \bibfield  {author} {\bibinfo {author} {\bibfnamefont {C.}~\bibnamefont
  {Howald}}, \bibinfo {author} {\bibfnamefont {P.}~\bibnamefont {Fournier}},\
  and\ \bibinfo {author} {\bibfnamefont {A.}~\bibnamefont {Kapitulnik}},\
  }\bibfield  {title} {\bibinfo {title} {Inherent inhomogeneities in tunneling
  spectra of \text{Bi}$_2$\text{Sr}$_2$\text{Ca}\text{Cu}$_2$\text{O}$_{8-x}$
  crystals in the superconducting state},\ }\href
  {https://doi.org/10.1103/PhysRevB.64.100504} {\bibfield  {journal} {\bibinfo
  {journal} {Phys. Rev. B}\ }\textbf {\bibinfo {volume} {64}},\ \bibinfo
  {pages} {100504} (\bibinfo {year} {2001})}\BibitemShut {NoStop}%
\bibitem [{\citenamefont {Lang}\ \emph {et~al.}(2002)\citenamefont {Lang},
  \citenamefont {Madhavan}, \citenamefont {Hoffman}, \citenamefont {Hudson},
  \citenamefont {Eisaki}, \citenamefont {Uchida},\ and\ \citenamefont
  {Davis}}]{Lang:ImagingGranular}%
  \BibitemOpen
  \bibfield  {author} {\bibinfo {author} {\bibfnamefont {K.~M.}\ \bibnamefont
  {Lang}}, \bibinfo {author} {\bibfnamefont {V.}~\bibnamefont {Madhavan}},
  \bibinfo {author} {\bibfnamefont {J.~E.}\ \bibnamefont {Hoffman}}, \bibinfo
  {author} {\bibfnamefont {E.~W.}\ \bibnamefont {Hudson}}, \bibinfo {author}
  {\bibfnamefont {H.}~\bibnamefont {Eisaki}}, \bibinfo {author} {\bibfnamefont
  {S.}~\bibnamefont {Uchida}},\ and\ \bibinfo {author} {\bibfnamefont {J.~C.}\
  \bibnamefont {Davis}},\ }\bibfield  {title} {\bibinfo {title} {Imaging the
  granular structure of high-tc superconductivity in underdoped
  \text{Bi}$_2$\text{Sr}$_2$\text{CaCu}$_2$\text{O}$_{8+\delta}$},\ }\href
  {https://doi.org/10.1038/415412a} {\bibfield  {journal} {\bibinfo  {journal}
  {Nature}\ }\textbf {\bibinfo {volume} {415}},\ \bibinfo {pages} {412}
  (\bibinfo {year} {2002})}\BibitemShut {NoStop}%
\bibitem [{\citenamefont {Cuk}\ \emph {et~al.}(2005)\citenamefont {Cuk},
  \citenamefont {Lu}, \citenamefont {Zhou}, \citenamefont {Shen}, \citenamefont
  {Devereaux},\ and\ \citenamefont {Nagaosa}}]{Cuk:MomentumDepKinkBi2212}%
  \BibitemOpen
  \bibfield  {author} {\bibinfo {author} {\bibfnamefont {T.}~\bibnamefont
  {Cuk}}, \bibinfo {author} {\bibfnamefont {D.~H.}\ \bibnamefont {Lu}},
  \bibinfo {author} {\bibfnamefont {X.~J.}\ \bibnamefont {Zhou}}, \bibinfo
  {author} {\bibfnamefont {Z.-X.}\ \bibnamefont {Shen}}, \bibinfo {author}
  {\bibfnamefont {T.~P.}\ \bibnamefont {Devereaux}},\ and\ \bibinfo {author}
  {\bibfnamefont {N.}~\bibnamefont {Nagaosa}},\ }\bibfield  {title} {\bibinfo
  {title} {A review of electron–phonon coupling seen in the high-t$_c$
  superconductors by angle-resolved photoemission studies (\text{ARPES})},\
  }\href {https://doi.org/10.1002/pssb.200404959} {\bibfield  {journal}
  {\bibinfo  {journal} {physica status solidi (b)}\ }\textbf {\bibinfo {volume}
  {242}},\ \bibinfo {pages} {11} (\bibinfo {year} {2005})}\BibitemShut
  {NoStop}%
\bibitem [{\citenamefont {Hashimoto}\ \emph
  {et~al.}(2014{\natexlab{a}})\citenamefont {Hashimoto}, \citenamefont
  {Vishik}, \citenamefont {He}, \citenamefont {Devereaux},\ and\ \citenamefont
  {Shen}}]{Hashimoto:EnergyGaps}%
  \BibitemOpen
  \bibfield  {author} {\bibinfo {author} {\bibfnamefont {M.}~\bibnamefont
  {Hashimoto}}, \bibinfo {author} {\bibfnamefont {I.~M.}\ \bibnamefont
  {Vishik}}, \bibinfo {author} {\bibfnamefont {R.-H.}\ \bibnamefont {He}},
  \bibinfo {author} {\bibfnamefont {T.~P.}\ \bibnamefont {Devereaux}},\ and\
  \bibinfo {author} {\bibfnamefont {Z.-X.}\ \bibnamefont {Shen}},\ }\bibfield
  {title} {\bibinfo {title} {Energy gaps in high-transition-temperature cuprate
  superconductors},\ }\href {https://doi.org/10.1038/nphys3009} {\bibfield
  {journal} {\bibinfo  {journal} {Nature Physics}\ }\textbf {\bibinfo {volume}
  {10}},\ \bibinfo {pages} {483} (\bibinfo {year}
  {2014}{\natexlab{a}})}\BibitemShut {NoStop}%
\bibitem [{\citenamefont {Anzai}\ \emph {et~al.}(2017)\citenamefont {Anzai},
  \citenamefont {Arita}, \citenamefont {Namatame}, \citenamefont {Taniguchi},
  \citenamefont {Ishikado}, \citenamefont {Fujita}, \citenamefont {Ishida},
  \citenamefont {Uchida},\ and\ \citenamefont {Ino}}]{Anzai:Kinks}%
  \BibitemOpen
  \bibfield  {author} {\bibinfo {author} {\bibfnamefont {H.}~\bibnamefont
  {Anzai}}, \bibinfo {author} {\bibfnamefont {M.}~\bibnamefont {Arita}},
  \bibinfo {author} {\bibfnamefont {H.}~\bibnamefont {Namatame}}, \bibinfo
  {author} {\bibfnamefont {M.}~\bibnamefont {Taniguchi}}, \bibinfo {author}
  {\bibfnamefont {M.}~\bibnamefont {Ishikado}}, \bibinfo {author}
  {\bibfnamefont {K.}~\bibnamefont {Fujita}}, \bibinfo {author} {\bibfnamefont
  {S.}~\bibnamefont {Ishida}}, \bibinfo {author} {\bibfnamefont
  {S.}~\bibnamefont {Uchida}},\ and\ \bibinfo {author} {\bibfnamefont
  {A.}~\bibnamefont {Ino}},\ }\bibfield  {title} {\bibinfo {title} {A new
  landscape of multiple dispersion kinks in a high-t$_c$ cuprate
  superconductor},\ }\href {https://doi.org/10.1038/s41598-017-04983-0}
  {\bibfield  {journal} {\bibinfo  {journal} {Scientific Reports}\ }\textbf
  {\bibinfo {volume} {7}},\ \bibinfo {pages} {4830} (\bibinfo {year}
  {2017})}\BibitemShut {NoStop}%
\bibitem [{\citenamefont {Eisaki}\ \emph {et~al.}(2004)\citenamefont {Eisaki},
  \citenamefont {Kaneko}, \citenamefont {Feng}, \citenamefont {Damascelli},
  \citenamefont {Mang}, \citenamefont {Shen}, \citenamefont {Shen},\ and\
  \citenamefont {Greven}}]{Eisaki:ChemicalInhomogeneity}%
  \BibitemOpen
  \bibfield  {author} {\bibinfo {author} {\bibfnamefont {H.}~\bibnamefont
  {Eisaki}}, \bibinfo {author} {\bibfnamefont {N.}~\bibnamefont {Kaneko}},
  \bibinfo {author} {\bibfnamefont {D.~L.}\ \bibnamefont {Feng}}, \bibinfo
  {author} {\bibfnamefont {A.}~\bibnamefont {Damascelli}}, \bibinfo {author}
  {\bibfnamefont {P.~K.}\ \bibnamefont {Mang}}, \bibinfo {author}
  {\bibfnamefont {K.~M.}\ \bibnamefont {Shen}}, \bibinfo {author}
  {\bibfnamefont {Z.-X.}\ \bibnamefont {Shen}},\ and\ \bibinfo {author}
  {\bibfnamefont {M.}~\bibnamefont {Greven}},\ }\bibfield  {title} {\bibinfo
  {title} {Effect of chemical inhomogeneity in bismuth-based copper oxide
  superconductors},\ }\href {https://doi.org/10.1103/PhysRevB.69.064512}
  {\bibfield  {journal} {\bibinfo  {journal} {Phys. Rev. B}\ }\textbf {\bibinfo
  {volume} {69}},\ \bibinfo {pages} {064512} (\bibinfo {year}
  {2004})}\BibitemShut {NoStop}%
\bibitem [{\citenamefont {Sakakibara}\ \emph {et~al.}(2012)\citenamefont
  {Sakakibara}, \citenamefont {Usui}, \citenamefont {Kuroki}, \citenamefont
  {Arita},\ and\ \citenamefont {Aoki}}]{Sakakibara:OriginMaterialDependenceTc}%
  \BibitemOpen
  \bibfield  {author} {\bibinfo {author} {\bibfnamefont {H.}~\bibnamefont
  {Sakakibara}}, \bibinfo {author} {\bibfnamefont {H.}~\bibnamefont {Usui}},
  \bibinfo {author} {\bibfnamefont {K.}~\bibnamefont {Kuroki}}, \bibinfo
  {author} {\bibfnamefont {R.}~\bibnamefont {Arita}},\ and\ \bibinfo {author}
  {\bibfnamefont {H.}~\bibnamefont {Aoki}},\ }\bibfield  {title} {\bibinfo
  {title} {Origin of the material dependence of ${T}_{c}$ in the single-layered
  cuprates},\ }\href {https://doi.org/10.1103/PhysRevB.85.064501} {\bibfield
  {journal} {\bibinfo  {journal} {Phys. Rev. B}\ }\textbf {\bibinfo {volume}
  {85}},\ \bibinfo {pages} {064501} (\bibinfo {year} {2012})}\BibitemShut
  {NoStop}%
\bibitem [{\citenamefont {Ronning}\ \emph {et~al.}(2003)\citenamefont
  {Ronning}, \citenamefont {Sasagawa}, \citenamefont {Kohsaka}, \citenamefont
  {Shen}, \citenamefont {Damascelli}, \citenamefont {Kim}, \citenamefont
  {Yoshida}, \citenamefont {Armitage}, \citenamefont {Lu}, \citenamefont
  {Feng}, \citenamefont {Miller}, \citenamefont {Takagi},\ and\ \citenamefont
  {Shen}}]{Ronning:NaCCOC_nodal_PDH}%
  \BibitemOpen
  \bibfield  {author} {\bibinfo {author} {\bibfnamefont {F.}~\bibnamefont
  {Ronning}}, \bibinfo {author} {\bibfnamefont {T.}~\bibnamefont {Sasagawa}},
  \bibinfo {author} {\bibfnamefont {Y.}~\bibnamefont {Kohsaka}}, \bibinfo
  {author} {\bibfnamefont {K.~M.}\ \bibnamefont {Shen}}, \bibinfo {author}
  {\bibfnamefont {A.}~\bibnamefont {Damascelli}}, \bibinfo {author}
  {\bibfnamefont {C.}~\bibnamefont {Kim}}, \bibinfo {author} {\bibfnamefont
  {T.}~\bibnamefont {Yoshida}}, \bibinfo {author} {\bibfnamefont {N.~P.}\
  \bibnamefont {Armitage}}, \bibinfo {author} {\bibfnamefont {D.~H.}\
  \bibnamefont {Lu}}, \bibinfo {author} {\bibfnamefont {D.~L.}\ \bibnamefont
  {Feng}}, \bibinfo {author} {\bibfnamefont {L.~L.}\ \bibnamefont {Miller}},
  \bibinfo {author} {\bibfnamefont {H.}~\bibnamefont {Takagi}},\ and\ \bibinfo
  {author} {\bibfnamefont {Z.-X.}\ \bibnamefont {Shen}},\ }\bibfield  {title}
  {\bibinfo {title} {Evolution of a metal to insulator transition in
  \text{Ca}$_{2-x}$\text{Na}$_x$\text{CuO}$_2$\text{Cl}$_2$ as seen by
  angle-resolved photoemission},\ }\href
  {https://doi.org/10.1103/PhysRevB.67.165101} {\bibfield  {journal} {\bibinfo
  {journal} {Phys. Rev. B}\ }\textbf {\bibinfo {volume} {67}},\ \bibinfo
  {pages} {165101} (\bibinfo {year} {2003})}\BibitemShut {NoStop}%
\bibitem [{\citenamefont {Hashimoto}\ \emph {et~al.}(2008)\citenamefont
  {Hashimoto}, \citenamefont {Yoshida}, \citenamefont {Yagi}, \citenamefont
  {Takizawa}, \citenamefont {Fujimori}, \citenamefont {Kubota}, \citenamefont
  {Ono}, \citenamefont {Tanaka}, \citenamefont {Lu}, \citenamefont {Shen},
  \citenamefont {Ono},\ and\ \citenamefont
  {Ando}}]{Hashimoto:DopingEvolutionElectronicStructureSingleLayer}%
  \BibitemOpen
  \bibfield  {author} {\bibinfo {author} {\bibfnamefont {M.}~\bibnamefont
  {Hashimoto}}, \bibinfo {author} {\bibfnamefont {T.}~\bibnamefont {Yoshida}},
  \bibinfo {author} {\bibfnamefont {H.}~\bibnamefont {Yagi}}, \bibinfo {author}
  {\bibfnamefont {M.}~\bibnamefont {Takizawa}}, \bibinfo {author}
  {\bibfnamefont {A.}~\bibnamefont {Fujimori}}, \bibinfo {author}
  {\bibfnamefont {M.}~\bibnamefont {Kubota}}, \bibinfo {author} {\bibfnamefont
  {K.}~\bibnamefont {Ono}}, \bibinfo {author} {\bibfnamefont {K.}~\bibnamefont
  {Tanaka}}, \bibinfo {author} {\bibfnamefont {D.~H.}\ \bibnamefont {Lu}},
  \bibinfo {author} {\bibfnamefont {Z.-X.}\ \bibnamefont {Shen}}, \bibinfo
  {author} {\bibfnamefont {S.}~\bibnamefont {Ono}},\ and\ \bibinfo {author}
  {\bibfnamefont {Y.}~\bibnamefont {Ando}},\ }\bibfield  {title} {\bibinfo
  {title} {Doping evolution of the electronic structure in the single-layer
  cuprate
  \text{Bi}$_2$\text{Sr}$_{2-x}$\text{La}$_x$\text{Cu}\text{O}$_{6+\delta}$:
  Comparison with other single-layer cuprates},\ }\href
  {https://doi.org/10.1103/PhysRevB.77.094516} {\bibfield  {journal} {\bibinfo
  {journal} {Phys. Rev. B}\ }\textbf {\bibinfo {volume} {77}},\ \bibinfo
  {pages} {094516} (\bibinfo {year} {2008})}\BibitemShut {NoStop}%
\bibitem [{\citenamefont {Yoshida}\ \emph {et~al.}(2007)\citenamefont
  {Yoshida}, \citenamefont {Zhou}, \citenamefont {Lu}, \citenamefont {Komiya},
  \citenamefont {Ando}, \citenamefont {Eisaki}, \citenamefont {Kakeshita},
  \citenamefont {Uchida}, \citenamefont {Hussain}, \citenamefont {Shen},\ and\
  \citenamefont {Fujimori}}]{Yoshida:LSCO_ARPES}%
  \BibitemOpen
  \bibfield  {author} {\bibinfo {author} {\bibfnamefont {T.}~\bibnamefont
  {Yoshida}}, \bibinfo {author} {\bibfnamefont {X.~J.}\ \bibnamefont {Zhou}},
  \bibinfo {author} {\bibfnamefont {D.~H.}\ \bibnamefont {Lu}}, \bibinfo
  {author} {\bibfnamefont {S.}~\bibnamefont {Komiya}}, \bibinfo {author}
  {\bibfnamefont {Y.}~\bibnamefont {Ando}}, \bibinfo {author} {\bibfnamefont
  {H.}~\bibnamefont {Eisaki}}, \bibinfo {author} {\bibfnamefont
  {T.}~\bibnamefont {Kakeshita}}, \bibinfo {author} {\bibfnamefont
  {S.}~\bibnamefont {Uchida}}, \bibinfo {author} {\bibfnamefont
  {Z.}~\bibnamefont {Hussain}}, \bibinfo {author} {\bibfnamefont {Z.-X.}\
  \bibnamefont {Shen}},\ and\ \bibinfo {author} {\bibfnamefont
  {A.}~\bibnamefont {Fujimori}},\ }\bibfield  {title} {\bibinfo {title}
  {Low-energy electronic structure of the high-\text{Tc} cuprates\text{
  La}$_{2-x}$\text{Sr}$_x$\text{CuO}$_4$ studied by angle-resolved
  photoemission spectroscopy},\ }\href
  {https://doi.org/10.1088/0953-8984/19/12/125209} {\bibfield  {journal}
  {\bibinfo  {journal} {Journal of Physics: Condensed Matter}\ }\textbf
  {\bibinfo {volume} {19}},\ \bibinfo {pages} {125209} (\bibinfo {year}
  {2007})}\BibitemShut {NoStop}%
\bibitem [{\citenamefont {Bari\ifmmode \check{s}\else
  \v{s}\fi{}i\ifmmode~\acute{c}\else \'{c}\fi{}}\ \emph
  {et~al.}(2008)\citenamefont {Bari\ifmmode \check{s}\else
  \v{s}\fi{}i\ifmmode~\acute{c}\else \'{c}\fi{}}, \citenamefont {Li},
  \citenamefont {Zhao}, \citenamefont {Cho}, \citenamefont {Chabot-Couture},
  \citenamefont {Yu},\ and\ \citenamefont
  {Greven}}]{Barisic:DemonstratingModelNature}%
  \BibitemOpen
  \bibfield  {author} {\bibinfo {author} {\bibfnamefont {N.}~\bibnamefont
  {Bari\ifmmode \check{s}\else \v{s}\fi{}i\ifmmode~\acute{c}\else \'{c}\fi{}}},
  \bibinfo {author} {\bibfnamefont {Y.}~\bibnamefont {Li}}, \bibinfo {author}
  {\bibfnamefont {X.}~\bibnamefont {Zhao}}, \bibinfo {author} {\bibfnamefont
  {Y.-C.}\ \bibnamefont {Cho}}, \bibinfo {author} {\bibfnamefont
  {G.}~\bibnamefont {Chabot-Couture}}, \bibinfo {author} {\bibfnamefont
  {G.}~\bibnamefont {Yu}},\ and\ \bibinfo {author} {\bibfnamefont
  {M.}~\bibnamefont {Greven}},\ }\bibfield  {title} {\bibinfo {title}
  {Demonstrating the model nature of the high-temperature superconductor
  \text{Hg}\text{Ba}$_2$\text{Cu}\text{O}$_{4+\delta}$},\ }\href
  {https://doi.org/10.1103/PhysRevB.78.054518} {\bibfield  {journal} {\bibinfo
  {journal} {Phys. Rev. B}\ }\textbf {\bibinfo {volume} {78}},\ \bibinfo
  {pages} {054518} (\bibinfo {year} {2008})}\BibitemShut {NoStop}%
\bibitem [{\citenamefont {Schilling}\ \emph {et~al.}(1993)\citenamefont
  {Schilling}, \citenamefont {Cantoni}, \citenamefont {Guo},\ and\
  \citenamefont {Ott}}]{Schilling:TripleLayerHgTc}%
  \BibitemOpen
  \bibfield  {author} {\bibinfo {author} {\bibfnamefont {A.}~\bibnamefont
  {Schilling}}, \bibinfo {author} {\bibfnamefont {M.}~\bibnamefont {Cantoni}},
  \bibinfo {author} {\bibfnamefont {J.~D.}\ \bibnamefont {Guo}},\ and\ \bibinfo
  {author} {\bibfnamefont {H.~R.}\ \bibnamefont {Ott}},\ }\bibfield  {title}
  {\bibinfo {title} {Superconductivity above 130 k in the
  \text{Hg}–\text{Ba}–\text{Ca}–\text{Cu}–\text{O} system},\ }\href
  {https://doi.org/10.1038/363056a0} {\bibfield  {journal} {\bibinfo  {journal}
  {Nature}\ }\textbf {\bibinfo {volume} {363}},\ \bibinfo {pages} {56}
  (\bibinfo {year} {1993})}\BibitemShut {NoStop}%
\bibitem [{\citenamefont {Bari{\v s}i{\'c}}\ \emph {et~al.}(2013)\citenamefont
  {Bari{\v s}i{\'c}}, \citenamefont {Chan}, \citenamefont {Li}, \citenamefont
  {Yu}, \citenamefont {Zhao}, \citenamefont {Dressel}, \citenamefont
  {Smontara},\ and\ \citenamefont {Greven}}]{Barisic:PNAS_phaseDiagram}%
  \BibitemOpen
  \bibfield  {author} {\bibinfo {author} {\bibfnamefont {N.}~\bibnamefont
  {Bari{\v s}i{\'c}}}, \bibinfo {author} {\bibfnamefont {M.~K.}\ \bibnamefont
  {Chan}}, \bibinfo {author} {\bibfnamefont {Y.}~\bibnamefont {Li}}, \bibinfo
  {author} {\bibfnamefont {G.}~\bibnamefont {Yu}}, \bibinfo {author}
  {\bibfnamefont {X.}~\bibnamefont {Zhao}}, \bibinfo {author} {\bibfnamefont
  {M.}~\bibnamefont {Dressel}}, \bibinfo {author} {\bibfnamefont
  {A.}~\bibnamefont {Smontara}},\ and\ \bibinfo {author} {\bibfnamefont
  {M.}~\bibnamefont {Greven}},\ }\bibfield  {title} {\bibinfo {title}
  {Universal sheet resistance and revised phase diagram of the cuprate
  high-temperature superconductors},\ }\href
  {https://doi.org/10.1073/pnas.1301989110} {\bibfield  {journal} {\bibinfo
  {journal} {Proceedings of the National Academy of Sciences}\ }\textbf
  {\bibinfo {volume} {110}},\ \bibinfo {pages} {12235} (\bibinfo {year}
  {2013})}\BibitemShut {NoStop}%
\bibitem [{\citenamefont {Doiron-Leyraud}\ \emph {et~al.}(2013)\citenamefont
  {Doiron-Leyraud}, \citenamefont {Lepault}, \citenamefont {Cyr-Choinière},
  \citenamefont {Vignolle}, \citenamefont {Grissonnanche}, \citenamefont
  {Laliberté}, \citenamefont {Chang}, \citenamefont {Barišić}, \citenamefont
  {Chan}, \citenamefont {Ji}, \citenamefont {Zhao}, \citenamefont {Li},
  \citenamefont {Greven}, \citenamefont {Proust},\ and\ \citenamefont
  {Taillefer}}]{DoironLeyraud:HallSeebeckNernst}%
  \BibitemOpen
  \bibfield  {author} {\bibinfo {author} {\bibfnamefont {N.}~\bibnamefont
  {Doiron-Leyraud}}, \bibinfo {author} {\bibfnamefont {S.}~\bibnamefont
  {Lepault}}, \bibinfo {author} {\bibfnamefont {O.}~\bibnamefont
  {Cyr-Choinière}}, \bibinfo {author} {\bibfnamefont {B.}~\bibnamefont
  {Vignolle}}, \bibinfo {author} {\bibfnamefont {G.}~\bibnamefont
  {Grissonnanche}}, \bibinfo {author} {\bibfnamefont {F.}~\bibnamefont
  {Laliberté}}, \bibinfo {author} {\bibfnamefont {J.}~\bibnamefont {Chang}},
  \bibinfo {author} {\bibfnamefont {N.}~\bibnamefont {Barišić}}, \bibinfo
  {author} {\bibfnamefont {M.~K.}\ \bibnamefont {Chan}}, \bibinfo {author}
  {\bibfnamefont {L.}~\bibnamefont {Ji}}, \bibinfo {author} {\bibfnamefont
  {X.}~\bibnamefont {Zhao}}, \bibinfo {author} {\bibfnamefont {Y.}~\bibnamefont
  {Li}}, \bibinfo {author} {\bibfnamefont {M.}~\bibnamefont {Greven}}, \bibinfo
  {author} {\bibfnamefont {C.}~\bibnamefont {Proust}},\ and\ \bibinfo {author}
  {\bibfnamefont {L.}~\bibnamefont {Taillefer}},\ }\bibfield  {title} {\bibinfo
  {title} {Hall, seebeck, and nernst coefficients of underdoped
  \text{Hg}\text{Ba}$_2$\text{Cu}\text{O}$_{4+\delta}$: Fermi-surface
  reconstruction in an archetypal cuprate superconductor},\ }\href
  {https://doi.org/10.1103/PhysRevX.3.021019} {\bibfield  {journal} {\bibinfo
  {journal} {Physical Review X}\ }\textbf {\bibinfo {volume} {3}},\ \bibinfo
  {pages} {021019} (\bibinfo {year} {2013})}\BibitemShut {NoStop}%
\bibitem [{\citenamefont {Murayama}\ \emph {et~al.}(2019)\citenamefont
  {Murayama}, \citenamefont {Sato}, \citenamefont {Kurihara}, \citenamefont
  {Kasahara}, \citenamefont {Mizukami}, \citenamefont {Kasahara}, \citenamefont
  {Uchiyama}, \citenamefont {Yamamoto}, \citenamefont {Moon}, \citenamefont
  {Cai}, \citenamefont {Freyermuth}, \citenamefont {Greven}, \citenamefont
  {Shibauchi},\ and\ \citenamefont {Matsuda}}]{Murayama:DiagonalNematicity}%
  \BibitemOpen
  \bibfield  {author} {\bibinfo {author} {\bibfnamefont {H.}~\bibnamefont
  {Murayama}}, \bibinfo {author} {\bibfnamefont {Y.}~\bibnamefont {Sato}},
  \bibinfo {author} {\bibfnamefont {R.}~\bibnamefont {Kurihara}}, \bibinfo
  {author} {\bibfnamefont {S.}~\bibnamefont {Kasahara}}, \bibinfo {author}
  {\bibfnamefont {Y.}~\bibnamefont {Mizukami}}, \bibinfo {author}
  {\bibfnamefont {Y.}~\bibnamefont {Kasahara}}, \bibinfo {author}
  {\bibfnamefont {H.}~\bibnamefont {Uchiyama}}, \bibinfo {author}
  {\bibfnamefont {A.}~\bibnamefont {Yamamoto}}, \bibinfo {author}
  {\bibfnamefont {E.~G.}\ \bibnamefont {Moon}}, \bibinfo {author}
  {\bibfnamefont {J.}~\bibnamefont {Cai}}, \bibinfo {author} {\bibfnamefont
  {J.}~\bibnamefont {Freyermuth}}, \bibinfo {author} {\bibfnamefont
  {M.}~\bibnamefont {Greven}}, \bibinfo {author} {\bibfnamefont
  {T.}~\bibnamefont {Shibauchi}},\ and\ \bibinfo {author} {\bibfnamefont
  {Y.}~\bibnamefont {Matsuda}},\ }\bibfield  {title} {\bibinfo {title}
  {Diagonal nematicity in the pseudogap phase of
  \text{HgBa}$_2$\text{CuO}$_{4+\delta}$},\ }\href
  {https://doi.org/10.1038/s41467-019-11200-1} {\bibfield  {journal} {\bibinfo
  {journal} {Nature Communications}\ }\textbf {\bibinfo {volume} {10}},\
  \bibinfo {pages} {3282} (\bibinfo {year} {2019})}\BibitemShut {NoStop}%
\bibitem [{\citenamefont {van Heumen}\ \emph {et~al.}(2009)\citenamefont {van
  Heumen}, \citenamefont {Muhlethaler}, \citenamefont {Kuzmenko}, \citenamefont
  {Eisaki}, \citenamefont {Meevasana}, \citenamefont {Greven},\ and\
  \citenamefont {van~der Marel}}]{vanHeumen:OpticalDetermination}%
  \BibitemOpen
  \bibfield  {author} {\bibinfo {author} {\bibfnamefont {E.}~\bibnamefont {van
  Heumen}}, \bibinfo {author} {\bibfnamefont {E.}~\bibnamefont {Muhlethaler}},
  \bibinfo {author} {\bibfnamefont {A.~B.}\ \bibnamefont {Kuzmenko}}, \bibinfo
  {author} {\bibfnamefont {H.}~\bibnamefont {Eisaki}}, \bibinfo {author}
  {\bibfnamefont {W.}~\bibnamefont {Meevasana}}, \bibinfo {author}
  {\bibfnamefont {M.}~\bibnamefont {Greven}},\ and\ \bibinfo {author}
  {\bibfnamefont {D.}~\bibnamefont {van~der Marel}},\ }\bibfield  {title}
  {\bibinfo {title} {Optical determination of the relation between the
  electron-boson coupling function and the critical temperature in
  high-${\text{t}}_{c}$ cuprates},\ }\href
  {https://doi.org/10.1103/PhysRevB.79.184512} {\bibfield  {journal} {\bibinfo
  {journal} {Phys. Rev. B}\ }\textbf {\bibinfo {volume} {79}},\ \bibinfo
  {pages} {184512} (\bibinfo {year} {2009})}\BibitemShut {NoStop}%
\bibitem [{\citenamefont {Grbi\ifmmode~\acute{c}\else \'{c}\fi{}}\ \emph
  {et~al.}(2009)\citenamefont {Grbi\ifmmode~\acute{c}\else \'{c}\fi{}},
  \citenamefont {Bari\ifmmode \check{s}\else \v{s}\fi{}i\ifmmode~\acute{c}\else
  \'{c}\fi{}}, \citenamefont {Dul\ifmmode \check{c}\else
  \v{c}\fi{}i\ifmmode~\acute{c}\else \'{c}\fi{}}, \citenamefont {Kup\ifmmode
  \check{c}\else \v{c}\fi{}i\ifmmode~\acute{c}\else \'{c}\fi{}}, \citenamefont
  {Li}, \citenamefont {Zhao}, \citenamefont {Yu}, \citenamefont {Dressel},
  \citenamefont {Greven},\ and\ \citenamefont {Po\ifmmode~\check{z}\else
  \v{z}\fi{}ek}}]{Grbic:MicrowaveSCFluct}%
  \BibitemOpen
  \bibfield  {author} {\bibinfo {author} {\bibfnamefont {M.~S.}\ \bibnamefont
  {Grbi\ifmmode~\acute{c}\else \'{c}\fi{}}}, \bibinfo {author} {\bibfnamefont
  {N.}~\bibnamefont {Bari\ifmmode \check{s}\else
  \v{s}\fi{}i\ifmmode~\acute{c}\else \'{c}\fi{}}}, \bibinfo {author}
  {\bibfnamefont {A.}~\bibnamefont {Dul\ifmmode \check{c}\else
  \v{c}\fi{}i\ifmmode~\acute{c}\else \'{c}\fi{}}}, \bibinfo {author}
  {\bibfnamefont {I.}~\bibnamefont {Kup\ifmmode \check{c}\else
  \v{c}\fi{}i\ifmmode~\acute{c}\else \'{c}\fi{}}}, \bibinfo {author}
  {\bibfnamefont {Y.}~\bibnamefont {Li}}, \bibinfo {author} {\bibfnamefont
  {X.}~\bibnamefont {Zhao}}, \bibinfo {author} {\bibfnamefont {G.}~\bibnamefont
  {Yu}}, \bibinfo {author} {\bibfnamefont {M.}~\bibnamefont {Dressel}},
  \bibinfo {author} {\bibfnamefont {M.}~\bibnamefont {Greven}},\ and\ \bibinfo
  {author} {\bibfnamefont {M.}~\bibnamefont {Po\ifmmode~\check{z}\else
  \v{z}\fi{}ek}},\ }\bibfield  {title} {\bibinfo {title} {Microwave
  measurements of the in-plane and $c$-axis conductivity in
  \text{HgBa}$_{2}$\text{CuO}$_{4+\delta}$: Discriminating between
  superconducting fluctuations and pseudogap effects},\ }\href
  {https://doi.org/10.1103/PhysRevB.80.094511} {\bibfield  {journal} {\bibinfo
  {journal} {Physical Review B}\ }\textbf {\bibinfo {volume} {80}},\ \bibinfo
  {pages} {094511} (\bibinfo {year} {2009})}\BibitemShut {NoStop}%
\bibitem [{\citenamefont {Li}\ \emph {et~al.}(2008)\citenamefont {Li},
  \citenamefont {Balédent}, \citenamefont {Barišić}, \citenamefont {Cho},
  \citenamefont {Fauqué}, \citenamefont {Sidis}, \citenamefont {Yu},
  \citenamefont {Zhao}, \citenamefont {Bourges},\ and\ \citenamefont
  {Greven}}]{Li:UnusualMagneticOrder}%
  \BibitemOpen
  \bibfield  {author} {\bibinfo {author} {\bibfnamefont {Y.}~\bibnamefont
  {Li}}, \bibinfo {author} {\bibfnamefont {V.}~\bibnamefont {Balédent}},
  \bibinfo {author} {\bibfnamefont {N.}~\bibnamefont {Barišić}}, \bibinfo
  {author} {\bibfnamefont {Y.}~\bibnamefont {Cho}}, \bibinfo {author}
  {\bibfnamefont {B.}~\bibnamefont {Fauqué}}, \bibinfo {author} {\bibfnamefont
  {Y.}~\bibnamefont {Sidis}}, \bibinfo {author} {\bibfnamefont
  {G.}~\bibnamefont {Yu}}, \bibinfo {author} {\bibfnamefont {X.}~\bibnamefont
  {Zhao}}, \bibinfo {author} {\bibfnamefont {P.}~\bibnamefont {Bourges}},\ and\
  \bibinfo {author} {\bibfnamefont {M.}~\bibnamefont {Greven}},\ }\bibfield
  {title} {\bibinfo {title} {Unusual magnetic order in the pseudogap region of
  the superconductor \text{HgBa}$_2$\text{CuO}$_{4+\delta}$},\ }\href
  {https://doi.org/10.1038/nature07251} {\bibfield  {journal} {\bibinfo
  {journal} {Nature}\ }\textbf {\bibinfo {volume} {455}},\ \bibinfo {pages}
  {372} (\bibinfo {year} {2008})}\BibitemShut {NoStop}%
\bibitem [{\citenamefont {Chan}\ \emph
  {et~al.}(2016{\natexlab{a}})\citenamefont {Chan}, \citenamefont {Tang},
  \citenamefont {Dorow}, \citenamefont {Jeong}, \citenamefont {Mangin-Thro},
  \citenamefont {Veit}, \citenamefont {Ge}, \citenamefont {Abernathy},
  \citenamefont {Sidis}, \citenamefont {Bourges},\ and\ \citenamefont
  {Greven}}]{Chan:DopingDepMagResMode2016}%
  \BibitemOpen
  \bibfield  {author} {\bibinfo {author} {\bibfnamefont {M.~K.}\ \bibnamefont
  {Chan}}, \bibinfo {author} {\bibfnamefont {Y.}~\bibnamefont {Tang}}, \bibinfo
  {author} {\bibfnamefont {C.~J.}\ \bibnamefont {Dorow}}, \bibinfo {author}
  {\bibfnamefont {J.}~\bibnamefont {Jeong}}, \bibinfo {author} {\bibfnamefont
  {L.}~\bibnamefont {Mangin-Thro}}, \bibinfo {author} {\bibfnamefont {M.~J.}\
  \bibnamefont {Veit}}, \bibinfo {author} {\bibfnamefont {Y.}~\bibnamefont
  {Ge}}, \bibinfo {author} {\bibfnamefont {D.~L.}\ \bibnamefont {Abernathy}},
  \bibinfo {author} {\bibfnamefont {Y.}~\bibnamefont {Sidis}}, \bibinfo
  {author} {\bibfnamefont {P.}~\bibnamefont {Bourges}},\ and\ \bibinfo {author}
  {\bibfnamefont {M.}~\bibnamefont {Greven}},\ }\bibfield  {title} {\bibinfo
  {title} {Hourglass dispersion and resonance of magnetic excitations in the
  superconducting state of the single-layer cuprate
  \text{Hg}\text{Ba}$_2$\text{Cu}\text{O}$_{4+\delta}$ near optimal doping},\
  }\href {https://doi.org/10.1103/PhysRevLett.117.277002} {\bibfield  {journal}
  {\bibinfo  {journal} {Phys. Rev. Lett.}\ }\textbf {\bibinfo {volume} {117}},\
  \bibinfo {pages} {277002} (\bibinfo {year} {2016}{\natexlab{a}})}\BibitemShut
  {NoStop}%
\bibitem [{\citenamefont {Chan}\ \emph
  {et~al.}(2016{\natexlab{b}})\citenamefont {Chan}, \citenamefont {Dorow},
  \citenamefont {Mangin-Thro}, \citenamefont {Tang}, \citenamefont {Ge},
  \citenamefont {Veit}, \citenamefont {Yu}, \citenamefont {Zhao}, \citenamefont
  {Christianson}, \citenamefont {Park}, \citenamefont {Sidis}, \citenamefont
  {Steffens}, \citenamefont {Abernathy}, \citenamefont {Bourges},\ and\
  \citenamefont {Greven}}]{Chan:Tdep_mag_res_mode2016}%
  \BibitemOpen
  \bibfield  {author} {\bibinfo {author} {\bibfnamefont {M.~K.}\ \bibnamefont
  {Chan}}, \bibinfo {author} {\bibfnamefont {C.~J.}\ \bibnamefont {Dorow}},
  \bibinfo {author} {\bibfnamefont {L.}~\bibnamefont {Mangin-Thro}}, \bibinfo
  {author} {\bibfnamefont {Y.}~\bibnamefont {Tang}}, \bibinfo {author}
  {\bibfnamefont {Y.}~\bibnamefont {Ge}}, \bibinfo {author} {\bibfnamefont
  {M.~J.}\ \bibnamefont {Veit}}, \bibinfo {author} {\bibfnamefont
  {G.}~\bibnamefont {Yu}}, \bibinfo {author} {\bibfnamefont {X.}~\bibnamefont
  {Zhao}}, \bibinfo {author} {\bibfnamefont {A.~D.}\ \bibnamefont
  {Christianson}}, \bibinfo {author} {\bibfnamefont {J.~T.}\ \bibnamefont
  {Park}}, \bibinfo {author} {\bibfnamefont {Y.}~\bibnamefont {Sidis}},
  \bibinfo {author} {\bibfnamefont {P.}~\bibnamefont {Steffens}}, \bibinfo
  {author} {\bibfnamefont {D.~L.}\ \bibnamefont {Abernathy}}, \bibinfo {author}
  {\bibfnamefont {P.}~\bibnamefont {Bourges}},\ and\ \bibinfo {author}
  {\bibfnamefont {M.}~\bibnamefont {Greven}},\ }\bibfield  {title} {\bibinfo
  {title} {Commensurate antiferromagnetic excitations as a signature of the
  pseudogap in the tetragonal high-t$_c$ cuprate
  \text{Hg}\text{Ba}$_2$\text{Cu}\text{O}$_{4+\delta}$},\ }\href@noop {}
  {\bibfield  {journal} {\bibinfo  {journal} {Nature communications}\ }\textbf
  {\bibinfo {volume} {7}},\ \bibinfo {pages} {10819} (\bibinfo {year}
  {2016}{\natexlab{b}})}\BibitemShut {NoStop}%
\bibitem [{\citenamefont {Lu}\ \emph {et~al.}(2005)\citenamefont {Lu},
  \citenamefont {Chabot-Couture}, \citenamefont {Zhao}, \citenamefont
  {Hancock}, \citenamefont {Kaneko}, \citenamefont {Vajk}, \citenamefont {Yu},
  \citenamefont {Grenier}, \citenamefont {Kim}, \citenamefont {Casa},
  \citenamefont {Gog},\ and\ \citenamefont
  {Greven}}]{Lu:ChargeTransferExcitations}%
  \BibitemOpen
  \bibfield  {author} {\bibinfo {author} {\bibfnamefont {L.}~\bibnamefont
  {Lu}}, \bibinfo {author} {\bibfnamefont {G.}~\bibnamefont {Chabot-Couture}},
  \bibinfo {author} {\bibfnamefont {X.}~\bibnamefont {Zhao}}, \bibinfo {author}
  {\bibfnamefont {J.~N.}\ \bibnamefont {Hancock}}, \bibinfo {author}
  {\bibfnamefont {N.}~\bibnamefont {Kaneko}}, \bibinfo {author} {\bibfnamefont
  {O.~P.}\ \bibnamefont {Vajk}}, \bibinfo {author} {\bibfnamefont
  {G.}~\bibnamefont {Yu}}, \bibinfo {author} {\bibfnamefont {S.}~\bibnamefont
  {Grenier}}, \bibinfo {author} {\bibfnamefont {Y.~J.}\ \bibnamefont {Kim}},
  \bibinfo {author} {\bibfnamefont {D.}~\bibnamefont {Casa}}, \bibinfo {author}
  {\bibfnamefont {T.}~\bibnamefont {Gog}},\ and\ \bibinfo {author}
  {\bibfnamefont {M.}~\bibnamefont {Greven}},\ }\bibfield  {title} {\bibinfo
  {title} {Charge-transfer excitations in the model superconductor
  \text{HgBa}$_2$\text{CuO}$_{4+\delta}$},\ }\href
  {https://doi.org/10.1103/PhysRevLett.95.217003} {\bibfield  {journal}
  {\bibinfo  {journal} {Physical Review Letters}\ }\textbf {\bibinfo {volume}
  {95}},\ \bibinfo {pages} {217003} (\bibinfo {year} {2005})}\BibitemShut
  {NoStop}%
\bibitem [{\citenamefont {Tabis}\ \emph {et~al.}(2014)\citenamefont {Tabis},
  \citenamefont {Li}, \citenamefont {Tacon}, \citenamefont {Braicovich},
  \citenamefont {Kreyssig}, \citenamefont {Minola}, \citenamefont {Dellea},
  \citenamefont {Weschke}, \citenamefont {Veit}, \citenamefont {Ramazanoglu},
  \citenamefont {Goldman}, \citenamefont {Schmitt}, \citenamefont
  {Ghiringhelli}, \citenamefont {Barišić}, \citenamefont {Chan},
  \citenamefont {Dorow}, \citenamefont {Yu}, \citenamefont {Zhao},
  \citenamefont {Keimer},\ and\ \citenamefont
  {Greven}}]{Tabis:ChargeOrderFLtransport}%
  \BibitemOpen
  \bibfield  {author} {\bibinfo {author} {\bibfnamefont {W.}~\bibnamefont
  {Tabis}}, \bibinfo {author} {\bibfnamefont {Y.}~\bibnamefont {Li}}, \bibinfo
  {author} {\bibfnamefont {M.~L.}\ \bibnamefont {Tacon}}, \bibinfo {author}
  {\bibfnamefont {L.}~\bibnamefont {Braicovich}}, \bibinfo {author}
  {\bibfnamefont {A.}~\bibnamefont {Kreyssig}}, \bibinfo {author}
  {\bibfnamefont {M.}~\bibnamefont {Minola}}, \bibinfo {author} {\bibfnamefont
  {G.}~\bibnamefont {Dellea}}, \bibinfo {author} {\bibfnamefont
  {E.}~\bibnamefont {Weschke}}, \bibinfo {author} {\bibfnamefont {M.~J.}\
  \bibnamefont {Veit}}, \bibinfo {author} {\bibfnamefont {M.}~\bibnamefont
  {Ramazanoglu}}, \bibinfo {author} {\bibfnamefont {A.~I.}\ \bibnamefont
  {Goldman}}, \bibinfo {author} {\bibfnamefont {T.}~\bibnamefont {Schmitt}},
  \bibinfo {author} {\bibfnamefont {G.}~\bibnamefont {Ghiringhelli}}, \bibinfo
  {author} {\bibfnamefont {N.}~\bibnamefont {Barišić}}, \bibinfo {author}
  {\bibfnamefont {M.~K.}\ \bibnamefont {Chan}}, \bibinfo {author}
  {\bibfnamefont {C.~J.}\ \bibnamefont {Dorow}}, \bibinfo {author}
  {\bibfnamefont {G.}~\bibnamefont {Yu}}, \bibinfo {author} {\bibfnamefont
  {X.}~\bibnamefont {Zhao}}, \bibinfo {author} {\bibfnamefont {B.}~\bibnamefont
  {Keimer}},\ and\ \bibinfo {author} {\bibfnamefont {M.}~\bibnamefont
  {Greven}},\ }\bibfield  {title} {\bibinfo {title} {Charge order and its
  connection with fermi-liquid charge transport in a pristine high-tc
  cuprate},\ }\href {https://doi.org/10.1038/ncomms6875} {\bibfield  {journal}
  {\bibinfo  {journal} {Nature Communications}\ }\textbf {\bibinfo {volume}
  {5}},\ \bibinfo {pages} {5875} (\bibinfo {year} {2014})}\BibitemShut
  {NoStop}%
\bibitem [{\citenamefont {Tabis}\ \emph {et~al.}(2017)\citenamefont {Tabis},
  \citenamefont {Yu}, \citenamefont {Bialo}, \citenamefont {Bluschke},
  \citenamefont {Kolodziej}, \citenamefont {Kozlowski}, \citenamefont
  {Blackburn}, \citenamefont {Sen}, \citenamefont {Forgan}, \citenamefont
  {Zimmermann}, \citenamefont {Tang}, \citenamefont {Weschke}, \citenamefont
  {Vignolle}, \citenamefont {Hepting}, \citenamefont {Gretarsson},
  \citenamefont {Sutarto}, \citenamefont {He}, \citenamefont {Le~Tacon},
  \citenamefont {Barišić}, \citenamefont {Yu},\ and\ \citenamefont
  {Greven}}]{Tabis:SynchrotronXrayStudyCDW}%
  \BibitemOpen
  \bibfield  {author} {\bibinfo {author} {\bibfnamefont {W.}~\bibnamefont
  {Tabis}}, \bibinfo {author} {\bibfnamefont {B.}~\bibnamefont {Yu}}, \bibinfo
  {author} {\bibfnamefont {I.}~\bibnamefont {Bialo}}, \bibinfo {author}
  {\bibfnamefont {M.}~\bibnamefont {Bluschke}}, \bibinfo {author}
  {\bibfnamefont {T.}~\bibnamefont {Kolodziej}}, \bibinfo {author}
  {\bibfnamefont {A.}~\bibnamefont {Kozlowski}}, \bibinfo {author}
  {\bibfnamefont {E.}~\bibnamefont {Blackburn}}, \bibinfo {author}
  {\bibfnamefont {K.}~\bibnamefont {Sen}}, \bibinfo {author} {\bibfnamefont
  {E.~M.}\ \bibnamefont {Forgan}}, \bibinfo {author} {\bibfnamefont {M.~v.}\
  \bibnamefont {Zimmermann}}, \bibinfo {author} {\bibfnamefont
  {Y.}~\bibnamefont {Tang}}, \bibinfo {author} {\bibfnamefont {E.}~\bibnamefont
  {Weschke}}, \bibinfo {author} {\bibfnamefont {B.}~\bibnamefont {Vignolle}},
  \bibinfo {author} {\bibfnamefont {M.}~\bibnamefont {Hepting}}, \bibinfo
  {author} {\bibfnamefont {H.}~\bibnamefont {Gretarsson}}, \bibinfo {author}
  {\bibfnamefont {R.}~\bibnamefont {Sutarto}}, \bibinfo {author} {\bibfnamefont
  {F.}~\bibnamefont {He}}, \bibinfo {author} {\bibfnamefont {M.}~\bibnamefont
  {Le~Tacon}}, \bibinfo {author} {\bibfnamefont {N.}~\bibnamefont {Barišić}},
  \bibinfo {author} {\bibfnamefont {G.}~\bibnamefont {Yu}},\ and\ \bibinfo
  {author} {\bibfnamefont {M.}~\bibnamefont {Greven}},\ }\bibfield  {title}
  {\bibinfo {title} {Synchrotron x-ray scattering study of charge-density-wave
  order in \text{Hg}\text{Ba}$_2$\text{Cu}\text{O}$_{4+\delta}$},\ }\href
  {https://doi.org/10.1103/PhysRevB.96.134510} {\bibfield  {journal} {\bibinfo
  {journal} {Physical Review B}\ }\textbf {\bibinfo {volume} {96}},\ \bibinfo
  {pages} {134510} (\bibinfo {year} {2017})}\BibitemShut {NoStop}%
\bibitem [{\citenamefont {Li}\ \emph {et~al.}(2013)\citenamefont {Li},
  \citenamefont {Le~Tacon}, \citenamefont {Matiks}, \citenamefont {Boris},
  \citenamefont {Loew}, \citenamefont {Lin}, \citenamefont {Chen},
  \citenamefont {Chan}, \citenamefont {Dorow}, \citenamefont {Ji},
  \citenamefont {Barišić}, \citenamefont {Zhao}, \citenamefont {Greven},\
  and\ \citenamefont {Keimer}}]{Li:DopingDepPhotonScattering}%
  \BibitemOpen
  \bibfield  {author} {\bibinfo {author} {\bibfnamefont {Y.}~\bibnamefont
  {Li}}, \bibinfo {author} {\bibfnamefont {M.}~\bibnamefont {Le~Tacon}},
  \bibinfo {author} {\bibfnamefont {Y.}~\bibnamefont {Matiks}}, \bibinfo
  {author} {\bibfnamefont {A.~V.}\ \bibnamefont {Boris}}, \bibinfo {author}
  {\bibfnamefont {T.}~\bibnamefont {Loew}}, \bibinfo {author} {\bibfnamefont
  {C.~T.}\ \bibnamefont {Lin}}, \bibinfo {author} {\bibfnamefont
  {L.}~\bibnamefont {Chen}}, \bibinfo {author} {\bibfnamefont {M.~K.}\
  \bibnamefont {Chan}}, \bibinfo {author} {\bibfnamefont {C.}~\bibnamefont
  {Dorow}}, \bibinfo {author} {\bibfnamefont {L.}~\bibnamefont {Ji}}, \bibinfo
  {author} {\bibfnamefont {N.}~\bibnamefont {Barišić}}, \bibinfo {author}
  {\bibfnamefont {X.}~\bibnamefont {Zhao}}, \bibinfo {author} {\bibfnamefont
  {M.}~\bibnamefont {Greven}},\ and\ \bibinfo {author} {\bibfnamefont
  {B.}~\bibnamefont {Keimer}},\ }\bibfield  {title} {\bibinfo {title}
  {Doping-dependent photon scattering resonance in the model high-temperature
  superconductor \text{Hg}\text{Ba}$_2$\text{Cu}\text{O}$_{4+\delta}$ revealed
  by raman scattering and optical ellipsometry},\ }\href
  {https://doi.org/10.1103/PhysRevLett.111.187001} {\bibfield  {journal}
  {\bibinfo  {journal} {Physical Review Letters}\ }\textbf {\bibinfo {volume}
  {111}},\ \bibinfo {pages} {187001} (\bibinfo {year} {2013})}\BibitemShut
  {NoStop}%
\bibitem [{\citenamefont {Vishik}\ \emph {et~al.}(2014)\citenamefont {Vishik},
  \citenamefont {Bari\ifmmode \check{s}\else \v{s}\fi{}i\ifmmode~\acute{c}\else
  \'{c}\fi{}}, \citenamefont {Chan}, \citenamefont {Li}, \citenamefont {Xia},
  \citenamefont {Yu}, \citenamefont {Zhao}, \citenamefont {Lee}, \citenamefont
  {Meevasana}, \citenamefont {Devereaux}, \citenamefont {Greven},\ and\
  \citenamefont {Shen}}]{Vishik:Hg1201_PRB2014}%
  \BibitemOpen
  \bibfield  {author} {\bibinfo {author} {\bibfnamefont {I.~M.}\ \bibnamefont
  {Vishik}}, \bibinfo {author} {\bibfnamefont {N.}~\bibnamefont {Bari\ifmmode
  \check{s}\else \v{s}\fi{}i\ifmmode~\acute{c}\else \'{c}\fi{}}}, \bibinfo
  {author} {\bibfnamefont {M.~K.}\ \bibnamefont {Chan}}, \bibinfo {author}
  {\bibfnamefont {Y.}~\bibnamefont {Li}}, \bibinfo {author} {\bibfnamefont
  {D.~D.}\ \bibnamefont {Xia}}, \bibinfo {author} {\bibfnamefont
  {G.}~\bibnamefont {Yu}}, \bibinfo {author} {\bibfnamefont {X.}~\bibnamefont
  {Zhao}}, \bibinfo {author} {\bibfnamefont {W.~S.}\ \bibnamefont {Lee}},
  \bibinfo {author} {\bibfnamefont {W.}~\bibnamefont {Meevasana}}, \bibinfo
  {author} {\bibfnamefont {T.~P.}\ \bibnamefont {Devereaux}}, \bibinfo {author}
  {\bibfnamefont {M.}~\bibnamefont {Greven}},\ and\ \bibinfo {author}
  {\bibfnamefont {Z.-X.}\ \bibnamefont {Shen}},\ }\bibfield  {title} {\bibinfo
  {title} {Angle-resolved photoemission spectroscopy study of
  $\mathrm{Hg}\mathrm{Ba}_{2}\mathrm{Cu}{\mathrm{o}}_{4+\ensuremath{\delta}}$},\
  }\href {https://doi.org/10.1103/PhysRevB.89.195141} {\bibfield  {journal}
  {\bibinfo  {journal} {Phys. Rev. B}\ }\textbf {\bibinfo {volume} {89}},\
  \bibinfo {pages} {195141} (\bibinfo {year} {2014})}\BibitemShut {NoStop}%
\bibitem [{\citenamefont {Rameau}\ \emph {et~al.}(2009)\citenamefont {Rameau},
  \citenamefont {Yang}, \citenamefont {Gu},\ and\ \citenamefont
  {Johnson}}]{Rameau:LEKink}%
  \BibitemOpen
  \bibfield  {author} {\bibinfo {author} {\bibfnamefont {J.~D.}\ \bibnamefont
  {Rameau}}, \bibinfo {author} {\bibfnamefont {H.-B.}\ \bibnamefont {Yang}},
  \bibinfo {author} {\bibfnamefont {G.~D.}\ \bibnamefont {Gu}},\ and\ \bibinfo
  {author} {\bibfnamefont {P.~D.}\ \bibnamefont {Johnson}},\ }\bibfield
  {title} {\bibinfo {title} {Coupling of low-energy electrons in the optimally
  doped \text{Bi}$_{2}$\text{Sr}$_{2}$\text{CaCu}$_{2}$\text{O}$_{8+\delta}$
  superconductor to an optical phonon mode},\ }\href
  {https://doi.org/10.1103/PhysRevB.80.184513} {\bibfield  {journal} {\bibinfo
  {journal} {Phys. Rev. B}\ }\textbf {\bibinfo {volume} {80}},\ \bibinfo
  {pages} {184513} (\bibinfo {year} {2009})}\BibitemShut {NoStop}%
\bibitem [{\citenamefont {Vishik}\ \emph {et~al.}(2010)\citenamefont {Vishik},
  \citenamefont {Lee}, \citenamefont {Schmitt}, \citenamefont {Moritz},
  \citenamefont {Sasagawa}, \citenamefont {Uchida}, \citenamefont {Fujita},
  \citenamefont {Ishida}, \citenamefont {Zhang}, \citenamefont {Devereaux},\
  and\ \citenamefont {Shen}}]{Vishik:LowEnergyKink}%
  \BibitemOpen
  \bibfield  {author} {\bibinfo {author} {\bibfnamefont {I.~M.}\ \bibnamefont
  {Vishik}}, \bibinfo {author} {\bibfnamefont {W.~S.}\ \bibnamefont {Lee}},
  \bibinfo {author} {\bibfnamefont {F.}~\bibnamefont {Schmitt}}, \bibinfo
  {author} {\bibfnamefont {B.}~\bibnamefont {Moritz}}, \bibinfo {author}
  {\bibfnamefont {T.}~\bibnamefont {Sasagawa}}, \bibinfo {author}
  {\bibfnamefont {S.}~\bibnamefont {Uchida}}, \bibinfo {author} {\bibfnamefont
  {K.}~\bibnamefont {Fujita}}, \bibinfo {author} {\bibfnamefont
  {S.}~\bibnamefont {Ishida}}, \bibinfo {author} {\bibfnamefont
  {C.}~\bibnamefont {Zhang}}, \bibinfo {author} {\bibfnamefont {T.~P.}\
  \bibnamefont {Devereaux}},\ and\ \bibinfo {author} {\bibfnamefont {Z.~X.}\
  \bibnamefont {Shen}},\ }\bibfield  {title} {\bibinfo {title}
  {Doping-dependent nodal fermi velocity of the high-temperature superconductor
  \text{Bi}$_{2}$\text{Sr}$_{2}$\text{CaCu}$_{2}$\text{O}$_{8+\delta}$ revealed
  using high-resolution angle-resolved photoemission spectroscopy},\ }\href
  {https://doi.org/10.1103/PhysRevLett.104.207002} {\bibfield  {journal}
  {\bibinfo  {journal} {Phys. Rev. Lett.}\ }\textbf {\bibinfo {volume} {104}},\
  \bibinfo {pages} {207002} (\bibinfo {year} {2010})}\BibitemShut {NoStop}%
\bibitem [{\citenamefont {Kondo}\ \emph {et~al.}(2013)\citenamefont {Kondo},
  \citenamefont {Nakashima}, \citenamefont {Malaeb}, \citenamefont {Ishida},
  \citenamefont {Hamaya}, \citenamefont {Takeuchi},\ and\ \citenamefont
  {Shin}}]{Kondo:LEKink}%
  \BibitemOpen
  \bibfield  {author} {\bibinfo {author} {\bibfnamefont {T.}~\bibnamefont
  {Kondo}}, \bibinfo {author} {\bibfnamefont {Y.}~\bibnamefont {Nakashima}},
  \bibinfo {author} {\bibfnamefont {W.}~\bibnamefont {Malaeb}}, \bibinfo
  {author} {\bibfnamefont {Y.}~\bibnamefont {Ishida}}, \bibinfo {author}
  {\bibfnamefont {Y.}~\bibnamefont {Hamaya}}, \bibinfo {author} {\bibfnamefont
  {T.}~\bibnamefont {Takeuchi}},\ and\ \bibinfo {author} {\bibfnamefont
  {S.}~\bibnamefont {Shin}},\ }\bibfield  {title} {\bibinfo {title} {Anomalous
  doping variation of the nodal low-energy feature of superconducting
  $(\mathrm{Bi},\mathrm{Pb}{)}_{2}(\mathrm{Sr},\mathrm{La}{)}_{2}\mathrm{CuO}_{6+\ensuremath{\delta}}$
  crystals revealed by laser-based angle-resolved photoemission spectroscopy},\
  }\href {https://doi.org/10.1103/PhysRevLett.110.217006} {\bibfield  {journal}
  {\bibinfo  {journal} {Phys. Rev. Lett.}\ }\textbf {\bibinfo {volume} {110}},\
  \bibinfo {pages} {217006} (\bibinfo {year} {2013})}\BibitemShut {NoStop}%
\bibitem [{\citenamefont {Plumb}\ \emph {et~al.}(2010)\citenamefont {Plumb},
  \citenamefont {Reber}, \citenamefont {Koralek}, \citenamefont {Sun},
  \citenamefont {Douglas}, \citenamefont {Aiura}, \citenamefont {Oka},
  \citenamefont {Eisaki},\ and\ \citenamefont {Dessau}}]{Plumb:LEKink}%
  \BibitemOpen
  \bibfield  {author} {\bibinfo {author} {\bibfnamefont {N.~C.}\ \bibnamefont
  {Plumb}}, \bibinfo {author} {\bibfnamefont {T.~J.}\ \bibnamefont {Reber}},
  \bibinfo {author} {\bibfnamefont {J.~D.}\ \bibnamefont {Koralek}}, \bibinfo
  {author} {\bibfnamefont {Z.}~\bibnamefont {Sun}}, \bibinfo {author}
  {\bibfnamefont {J.~F.}\ \bibnamefont {Douglas}}, \bibinfo {author}
  {\bibfnamefont {Y.}~\bibnamefont {Aiura}}, \bibinfo {author} {\bibfnamefont
  {K.}~\bibnamefont {Oka}}, \bibinfo {author} {\bibfnamefont {H.}~\bibnamefont
  {Eisaki}},\ and\ \bibinfo {author} {\bibfnamefont {D.~S.}\ \bibnamefont
  {Dessau}},\ }\bibfield  {title} {\bibinfo {title} {Low-energy ($<$ 10 mev)
  feature in the nodal electron self-energy and strong temperature dependence
  of the fermi velocity in
  \text{Bi}$_{2}$\text{Sr}$_{2}$\text{CaCu}$_{2}$\text{O}$_{8+\ensuremath{\delta}}$},\
  }\href {https://doi.org/10.1103/PhysRevLett.105.046402} {\bibfield  {journal}
  {\bibinfo  {journal} {Phys. Rev. Lett.}\ }\textbf {\bibinfo {volume} {105}},\
  \bibinfo {pages} {046402} (\bibinfo {year} {2010})}\BibitemShut {NoStop}%
\bibitem [{\citenamefont {Johnston}\ \emph {et~al.}(2012)\citenamefont
  {Johnston}, \citenamefont {Vishik}, \citenamefont {Lee}, \citenamefont
  {Schmitt}, \citenamefont {Uchida}, \citenamefont {Fujita}, \citenamefont
  {Ishida}, \citenamefont {Nagaosa}, \citenamefont {Shen},\ and\ \citenamefont
  {Devereaux}}]{Johnston:ForwardScattering}%
  \BibitemOpen
  \bibfield  {author} {\bibinfo {author} {\bibfnamefont {S.}~\bibnamefont
  {Johnston}}, \bibinfo {author} {\bibfnamefont {I.~M.}\ \bibnamefont
  {Vishik}}, \bibinfo {author} {\bibfnamefont {W.~S.}\ \bibnamefont {Lee}},
  \bibinfo {author} {\bibfnamefont {F.}~\bibnamefont {Schmitt}}, \bibinfo
  {author} {\bibfnamefont {S.}~\bibnamefont {Uchida}}, \bibinfo {author}
  {\bibfnamefont {K.}~\bibnamefont {Fujita}}, \bibinfo {author} {\bibfnamefont
  {S.}~\bibnamefont {Ishida}}, \bibinfo {author} {\bibfnamefont
  {N.}~\bibnamefont {Nagaosa}}, \bibinfo {author} {\bibfnamefont {Z.~X.}\
  \bibnamefont {Shen}},\ and\ \bibinfo {author} {\bibfnamefont {T.~P.}\
  \bibnamefont {Devereaux}},\ }\bibfield  {title} {\bibinfo {title} {Evidence
  for the importance of extended coulomb interactions and forward scattering in
  cuprate superconductors},\ }\href
  {https://doi.org/10.1103/PhysRevLett.108.166404} {\bibfield  {journal}
  {\bibinfo  {journal} {Phys. Rev. Lett.}\ }\textbf {\bibinfo {volume} {108}},\
  \bibinfo {pages} {166404} (\bibinfo {year} {2012})}\BibitemShut {NoStop}%
\bibitem [{\citenamefont {Zhao}\ \emph {et~al.}(2006)\citenamefont {Zhao},
  \citenamefont {Yu}, \citenamefont {Cho}, \citenamefont {Chabot-Couture},
  \citenamefont {Bari\ifmmode \check{s}\else \v{s}\fi{}i\ifmmode~\acute{c}\else
  \'{c}\fi{}}, \citenamefont {Bourges}, \citenamefont {Kaneko}, \citenamefont
  {Li}, \citenamefont {Lu}, \citenamefont {Motoyama}, \citenamefont {Vajk},\
  and\ \citenamefont {Greven}}]{Zhao:GrowthCharacterizationModelHg1201}%
  \BibitemOpen
  \bibfield  {author} {\bibinfo {author} {\bibfnamefont {X.}~\bibnamefont
  {Zhao}}, \bibinfo {author} {\bibfnamefont {G.}~\bibnamefont {Yu}}, \bibinfo
  {author} {\bibfnamefont {Y.-C.}\ \bibnamefont {Cho}}, \bibinfo {author}
  {\bibfnamefont {G.}~\bibnamefont {Chabot-Couture}}, \bibinfo {author}
  {\bibfnamefont {N.}~\bibnamefont {Bari\ifmmode \check{s}\else
  \v{s}\fi{}i\ifmmode~\acute{c}\else \'{c}\fi{}}}, \bibinfo {author}
  {\bibfnamefont {P.}~\bibnamefont {Bourges}}, \bibinfo {author} {\bibfnamefont
  {N.}~\bibnamefont {Kaneko}}, \bibinfo {author} {\bibfnamefont
  {Y.}~\bibnamefont {Li}}, \bibinfo {author} {\bibfnamefont {L.}~\bibnamefont
  {Lu}}, \bibinfo {author} {\bibfnamefont {E.}~\bibnamefont {Motoyama}},
  \bibinfo {author} {\bibfnamefont {O.}~\bibnamefont {Vajk}},\ and\ \bibinfo
  {author} {\bibfnamefont {M.}~\bibnamefont {Greven}},\ }\bibfield  {title}
  {\bibinfo {title} {Crystal growth and characterization of the model
  high-temperature superconductor \text{HgBa}$_2$\text{CuO}$_{4+\delta}$},\
  }\href {https://doi.org/10.1002/adma.200600931} {\bibfield  {journal}
  {\bibinfo  {journal} {Advanced Materials}\ }\textbf {\bibinfo {volume}
  {18}},\ \bibinfo {pages} {3243} (\bibinfo {year} {2006})}\BibitemShut
  {NoStop}%
\bibitem [{\citenamefont {Sreedhar}\ \emph {et~al.}(2020)\citenamefont
  {Sreedhar}, \citenamefont {Rossi}, \citenamefont {Nayak}, \citenamefont
  {Anderson}, \citenamefont {Tang}, \citenamefont {Gregory}, \citenamefont
  {Hashimoto}, \citenamefont {Lu}, \citenamefont {Rotenberg}, \citenamefont
  {Birgeneau}, \citenamefont {Greven}, , \citenamefont {Yi},\ and\
  \citenamefont {Vishik}}]{supplementaryMaterials}%
  \BibitemOpen
  \bibfield  {author} {\bibinfo {author} {\bibfnamefont {S.~A.}\ \bibnamefont
  {Sreedhar}}, \bibinfo {author} {\bibfnamefont {A.}~\bibnamefont {Rossi}},
  \bibinfo {author} {\bibfnamefont {J.}~\bibnamefont {Nayak}}, \bibinfo
  {author} {\bibfnamefont {Z.}~\bibnamefont {Anderson}}, \bibinfo {author}
  {\bibfnamefont {Y.}~\bibnamefont {Tang}}, \bibinfo {author} {\bibfnamefont
  {B.}~\bibnamefont {Gregory}}, \bibinfo {author} {\bibfnamefont
  {M.}~\bibnamefont {Hashimoto}}, \bibinfo {author} {\bibfnamefont
  {D.}~\bibnamefont {Lu}}, \bibinfo {author} {\bibfnamefont {E.}~\bibnamefont
  {Rotenberg}}, \bibinfo {author} {\bibfnamefont {R.~J.}\ \bibnamefont
  {Birgeneau}}, \bibinfo {author} {\bibfnamefont {M.}~\bibnamefont {Greven}}, ,
  \bibinfo {author} {\bibfnamefont {M.}~\bibnamefont {Yi}},\ and\ \bibinfo
  {author} {\bibfnamefont {I.~M.}\ \bibnamefont {Vishik}},\ }\bibfield  {title}
  {\bibinfo {title} {Supplementary materials: Angle-resolved x-ray
  photoemission spectroscopy},\ }\href@noop {} {\  (\bibinfo {year}
  {2020})}\BibitemShut {NoStop}%
\bibitem [{\citenamefont {Yamamoto}\ \emph {et~al.}(2000)\citenamefont
  {Yamamoto}, \citenamefont {Hu},\ and\ \citenamefont {Tajima}}]{Yamamoto2000}%
  \BibitemOpen
  \bibfield  {author} {\bibinfo {author} {\bibfnamefont {A.}~\bibnamefont
  {Yamamoto}}, \bibinfo {author} {\bibfnamefont {W.-Z.}\ \bibnamefont {Hu}},\
  and\ \bibinfo {author} {\bibfnamefont {S.}~\bibnamefont {Tajima}},\
  }\bibfield  {title} {\bibinfo {title} {Thermoelectric power and resistivity
  of ${\mathrm{hgba}}_{2}{\mathrm{cuo}}_{4+\ensuremath{\delta}}$ over a wide
  doping range},\ }\href {https://doi.org/10.1103/PhysRevB.63.024504}
  {\bibfield  {journal} {\bibinfo  {journal} {Phys. Rev. B}\ }\textbf {\bibinfo
  {volume} {63}},\ \bibinfo {pages} {024504} (\bibinfo {year}
  {2000})}\BibitemShut {NoStop}%
\bibitem [{\citenamefont {Das}(2012)}]{Das:Hg1201_TB_params}%
  \BibitemOpen
  \bibfield  {author} {\bibinfo {author} {\bibfnamefont {T.}~\bibnamefont
  {Das}},\ }\bibfield  {title} {\bibinfo {title} {$\mathbf{Q}=0$ collective
  modes originating from the low-lying hg-o band in superconducting
  hgba${}_{2}$cuo${}_{4+\ensuremath{\delta}}$},\ }\href
  {https://doi.org/10.1103/PhysRevB.86.054518} {\bibfield  {journal} {\bibinfo
  {journal} {Phys. Rev. B}\ }\textbf {\bibinfo {volume} {86}},\ \bibinfo
  {pages} {054518} (\bibinfo {year} {2012})}\BibitemShut {NoStop}%
\bibitem [{TB:()}]{TB:params}%
  \BibitemOpen
  \href@noop {} {}\bibinfo {note} {$E_k=-2t[\cos(k_x a)+\cos(k_y
  a)]-4t'[\cos(k_x a)\cos(k_y a)]-2t''[\cos(2k_x a)+\cos(2k_y
  a)]-4t'''[\cos(2k_x a)\cos(k_y a)+\cos(k_x a)\cos(2k_y a)]-\mu$ where
  $(t,t',t'',t''')=(0.46, − 0.105,0.08, − 0.02)$ and $\mu$ is adjusted to
  nominal doping}\BibitemShut {NoStop}%
\bibitem [{\citenamefont {Johnson}\ and\ \citenamefont
  {Valla}(2007)}]{johnson2007photoemission}%
  \BibitemOpen
  \bibfield  {author} {\bibinfo {author} {\bibfnamefont {P.~D.}\ \bibnamefont
  {Johnson}}\ and\ \bibinfo {author} {\bibfnamefont {T.}~\bibnamefont
  {Valla}},\ }\bibfield  {title} {\bibinfo {title} {Photoemission as a probe of
  the collective excitations in condensed matter systems},\ }in\ \href@noop {}
  {\emph {\bibinfo {booktitle} {Very High Resolution Photoelectron
  Spectroscopy}}}\ (\bibinfo  {publisher} {Springer},\ \bibinfo {year} {2007})\
  pp.\ \bibinfo {pages} {55--84}\BibitemShut {NoStop}%
\bibitem [{\citenamefont {Peng}\ \emph {et~al.}(2013)\citenamefont {Peng},
  \citenamefont {Meng}, \citenamefont {Mou}, \citenamefont {He}, \citenamefont
  {Zhao}, \citenamefont {Wu}, \citenamefont {Liu}, \citenamefont {Dong},
  \citenamefont {He}, \citenamefont {Zhang}, \citenamefont {Wang},
  \citenamefont {Peng}, \citenamefont {Wang}, \citenamefont {Zhang},
  \citenamefont {Yang}, \citenamefont {Chen}, \citenamefont {Xu}, \citenamefont
  {Lee},\ and\ \citenamefont {Zhou}}]{Peng:Bi2201_nodalPDH}%
  \BibitemOpen
  \bibfield  {author} {\bibinfo {author} {\bibfnamefont {Y.}~\bibnamefont
  {Peng}}, \bibinfo {author} {\bibfnamefont {J.}~\bibnamefont {Meng}}, \bibinfo
  {author} {\bibfnamefont {D.}~\bibnamefont {Mou}}, \bibinfo {author}
  {\bibfnamefont {J.}~\bibnamefont {He}}, \bibinfo {author} {\bibfnamefont
  {L.}~\bibnamefont {Zhao}}, \bibinfo {author} {\bibfnamefont {Y.}~\bibnamefont
  {Wu}}, \bibinfo {author} {\bibfnamefont {G.}~\bibnamefont {Liu}}, \bibinfo
  {author} {\bibfnamefont {X.}~\bibnamefont {Dong}}, \bibinfo {author}
  {\bibfnamefont {S.}~\bibnamefont {He}}, \bibinfo {author} {\bibfnamefont
  {J.}~\bibnamefont {Zhang}}, \bibinfo {author} {\bibfnamefont
  {X.}~\bibnamefont {Wang}}, \bibinfo {author} {\bibfnamefont {Q.}~\bibnamefont
  {Peng}}, \bibinfo {author} {\bibfnamefont {Z.}~\bibnamefont {Wang}}, \bibinfo
  {author} {\bibfnamefont {S.}~\bibnamefont {Zhang}}, \bibinfo {author}
  {\bibfnamefont {F.}~\bibnamefont {Yang}}, \bibinfo {author} {\bibfnamefont
  {C.}~\bibnamefont {Chen}}, \bibinfo {author} {\bibfnamefont {Z.}~\bibnamefont
  {Xu}}, \bibinfo {author} {\bibfnamefont {T.~K.}\ \bibnamefont {Lee}},\ and\
  \bibinfo {author} {\bibfnamefont {X.~J.}\ \bibnamefont {Zhou}},\ }\bibfield
  {title} {\bibinfo {title} {Disappearance of nodal gap across the
  insulator–superconductor transition in a copper-oxide superconductor},\
  }\href {https://doi.org/10.1038/ncomms3459} {\bibfield  {journal} {\bibinfo
  {journal} {Nature Communications}\ }\textbf {\bibinfo {volume} {4}},\
  \bibinfo {pages} {2459} (\bibinfo {year} {2013})}\BibitemShut {NoStop}%
\bibitem [{\citenamefont {d'Astuto}\ \emph {et~al.}(2013)\citenamefont
  {d'Astuto}, \citenamefont {Yamada}, \citenamefont {Giura}, \citenamefont
  {Paulatto}, \citenamefont {Gauzzi}, \citenamefont {Hoesch}, \citenamefont
  {Krisch}, \citenamefont {Azuma},\ and\ \citenamefont
  {Takano}}]{dAstuto:phonon_NaCCOC}%
  \BibitemOpen
  \bibfield  {author} {\bibinfo {author} {\bibfnamefont {M.}~\bibnamefont
  {d'Astuto}}, \bibinfo {author} {\bibfnamefont {I.}~\bibnamefont {Yamada}},
  \bibinfo {author} {\bibfnamefont {P.}~\bibnamefont {Giura}}, \bibinfo
  {author} {\bibfnamefont {L.}~\bibnamefont {Paulatto}}, \bibinfo {author}
  {\bibfnamefont {A.}~\bibnamefont {Gauzzi}}, \bibinfo {author} {\bibfnamefont
  {M.}~\bibnamefont {Hoesch}}, \bibinfo {author} {\bibfnamefont
  {M.}~\bibnamefont {Krisch}}, \bibinfo {author} {\bibfnamefont
  {M.}~\bibnamefont {Azuma}},\ and\ \bibinfo {author} {\bibfnamefont
  {M.}~\bibnamefont {Takano}},\ }\bibfield  {title} {\bibinfo {title} {Phonon
  anomalies and lattice dynamics in the superconducting oxychlorides
  \text{Ca}${}_{2\ensuremath{-}x}$\text{CuO}${}_{2}$\text{Cl}${}_{2}$},\ }\href
  {https://doi.org/10.1103/PhysRevB.88.014522} {\bibfield  {journal} {\bibinfo
  {journal} {Phys. Rev. B}\ }\textbf {\bibinfo {volume} {88}},\ \bibinfo
  {pages} {014522} (\bibinfo {year} {2013})}\BibitemShut {NoStop}%
\bibitem [{\citenamefont {Pintschovius}\ \emph {et~al.}(2006)\citenamefont
  {Pintschovius}, \citenamefont {Reznik},\ and\ \citenamefont
  {Yamada}}]{Pintschovius:BondStretchDopingDep}%
  \BibitemOpen
  \bibfield  {author} {\bibinfo {author} {\bibfnamefont {L.}~\bibnamefont
  {Pintschovius}}, \bibinfo {author} {\bibfnamefont {D.}~\bibnamefont
  {Reznik}},\ and\ \bibinfo {author} {\bibfnamefont {K.}~\bibnamefont
  {Yamada}},\ }\bibfield  {title} {\bibinfo {title} {Oxygen phonon branches in
  overdoped \text{La}$_{1.7}$\text{Sr}$_{0.3}$\text{Cu}$_{3}$\text{O}$_{4}$},\
  }\href {https://doi.org/10.1103/PhysRevB.74.174514} {\bibfield  {journal}
  {\bibinfo  {journal} {Phys. Rev. B}\ }\textbf {\bibinfo {volume} {74}},\
  \bibinfo {pages} {174514} (\bibinfo {year} {2006})}\BibitemShut {NoStop}%
\bibitem [{\citenamefont {Graf}\ \emph {et~al.}(2008)\citenamefont {Graf},
  \citenamefont {d'Astuto}, \citenamefont {Jozwiak}, \citenamefont {Garcia},
  \citenamefont {Saini}, \citenamefont {Krisch}, \citenamefont {Ikeuchi},
  \citenamefont {Baron}, \citenamefont {Eisaki},\ and\ \citenamefont
  {Lanzara}}]{graf:momentumDepKink_Bi2201_OP}%
  \BibitemOpen
  \bibfield  {author} {\bibinfo {author} {\bibfnamefont {J.}~\bibnamefont
  {Graf}}, \bibinfo {author} {\bibfnamefont {M.}~\bibnamefont {d'Astuto}},
  \bibinfo {author} {\bibfnamefont {C.}~\bibnamefont {Jozwiak}}, \bibinfo
  {author} {\bibfnamefont {D.~R.}\ \bibnamefont {Garcia}}, \bibinfo {author}
  {\bibfnamefont {N.~L.}\ \bibnamefont {Saini}}, \bibinfo {author}
  {\bibfnamefont {M.}~\bibnamefont {Krisch}}, \bibinfo {author} {\bibfnamefont
  {K.}~\bibnamefont {Ikeuchi}}, \bibinfo {author} {\bibfnamefont {A.~Q.~R.}\
  \bibnamefont {Baron}}, \bibinfo {author} {\bibfnamefont {H.}~\bibnamefont
  {Eisaki}},\ and\ \bibinfo {author} {\bibfnamefont {A.}~\bibnamefont
  {Lanzara}},\ }\bibfield  {title} {\bibinfo {title} {Bond stretching phonon
  softening and kinks in the angle-resolved photoemission spectra of optimally
  doped
  \text{Bi}$_{2}$\text{Sr}$_{1.6}$\text{La}$_{0.4}$\text{CaCu}$_{2}$\text{O}$_{6+\delta}$
  superconductors},\ }\href {https://doi.org/10.1103/PhysRevLett.100.227002}
  {\bibfield  {journal} {\bibinfo  {journal} {Phys. Rev. Lett.}\ }\textbf
  {\bibinfo {volume} {100}},\ \bibinfo {pages} {227002} (\bibinfo {year}
  {2008})}\BibitemShut {NoStop}%
\bibitem [{\citenamefont {Uchiyama}\ \emph {et~al.}(2004)\citenamefont
  {Uchiyama}, \citenamefont {Baron}, \citenamefont {Tsutsui}, \citenamefont
  {Tanaka}, \citenamefont {Hu}, \citenamefont {Yamamoto}, \citenamefont
  {Tajima},\ and\ \citenamefont {Endoh}}]{Uchiyama:PhononsH1201_2004}%
  \BibitemOpen
  \bibfield  {author} {\bibinfo {author} {\bibfnamefont {H.}~\bibnamefont
  {Uchiyama}}, \bibinfo {author} {\bibfnamefont {A.~Q.~R.}\ \bibnamefont
  {Baron}}, \bibinfo {author} {\bibfnamefont {S.}~\bibnamefont {Tsutsui}},
  \bibinfo {author} {\bibfnamefont {Y.}~\bibnamefont {Tanaka}}, \bibinfo
  {author} {\bibfnamefont {W.-Z.}\ \bibnamefont {Hu}}, \bibinfo {author}
  {\bibfnamefont {A.}~\bibnamefont {Yamamoto}}, \bibinfo {author}
  {\bibfnamefont {S.}~\bibnamefont {Tajima}},\ and\ \bibinfo {author}
  {\bibfnamefont {Y.}~\bibnamefont {Endoh}},\ }\bibfield  {title} {\bibinfo
  {title} {Softening of \text{Cu-O} bond stretching phonons in tetragonal
  \text{HgBa}$_{2}$\text{CuO}$_{4+\ensuremath{\delta}}$},\ }\href
  {https://doi.org/10.1103/PhysRevLett.92.197005} {\bibfield  {journal}
  {\bibinfo  {journal} {Phys. Rev. Lett.}\ }\textbf {\bibinfo {volume} {92}},\
  \bibinfo {pages} {197005} (\bibinfo {year} {2004})}\BibitemShut {NoStop}%
\bibitem [{\citenamefont {Ino}\ \emph {et~al.}(2002)\citenamefont {Ino},
  \citenamefont {Kim}, \citenamefont {Nakamura}, \citenamefont {Yoshida},
  \citenamefont {Mizokawa}, \citenamefont {Fujimori}, \citenamefont {Shen},
  \citenamefont {Kakeshita}, \citenamefont {Eisaki},\ and\ \citenamefont
  {Uchida}}]{Ino:DopingDependenceLSCO}%
  \BibitemOpen
  \bibfield  {author} {\bibinfo {author} {\bibfnamefont {A.}~\bibnamefont
  {Ino}}, \bibinfo {author} {\bibfnamefont {C.}~\bibnamefont {Kim}}, \bibinfo
  {author} {\bibfnamefont {M.}~\bibnamefont {Nakamura}}, \bibinfo {author}
  {\bibfnamefont {T.}~\bibnamefont {Yoshida}}, \bibinfo {author} {\bibfnamefont
  {T.}~\bibnamefont {Mizokawa}}, \bibinfo {author} {\bibfnamefont
  {A.}~\bibnamefont {Fujimori}}, \bibinfo {author} {\bibfnamefont {Z.-X.}\
  \bibnamefont {Shen}}, \bibinfo {author} {\bibfnamefont {T.}~\bibnamefont
  {Kakeshita}}, \bibinfo {author} {\bibfnamefont {H.}~\bibnamefont {Eisaki}},\
  and\ \bibinfo {author} {\bibfnamefont {S.}~\bibnamefont {Uchida}},\
  }\bibfield  {title} {\bibinfo {title} {Doping-dependent evolution of the
  electronic structure of
  \text{La}$_{2\ensuremath{-}x}$\text{Sr}$_{x}$\text{CuO}$_{4}$ in the
  superconducting and metallic phases},\ }\href
  {https://doi.org/10.1103/PhysRevB.65.094504} {\bibfield  {journal} {\bibinfo
  {journal} {Phys. Rev. B}\ }\textbf {\bibinfo {volume} {65}},\ \bibinfo
  {pages} {094504} (\bibinfo {year} {2002})}\BibitemShut {NoStop}%
\bibitem [{\citenamefont {Shen}\ \emph {et~al.}(2004)\citenamefont {Shen},
  \citenamefont {Ronning}, \citenamefont {Lu}, \citenamefont {Lee},
  \citenamefont {Ingle}, \citenamefont {Meevasana}, \citenamefont {Baumberger},
  \citenamefont {Damascelli}, \citenamefont {Armitage}, \citenamefont {Miller},
  \citenamefont {Kohsaka}, \citenamefont {Azuma}, \citenamefont {Takano},
  \citenamefont {Takagi},\ and\ \citenamefont {Shen}}]{Shen:DopingDepCCOC}%
  \BibitemOpen
  \bibfield  {author} {\bibinfo {author} {\bibfnamefont {K.~M.}\ \bibnamefont
  {Shen}}, \bibinfo {author} {\bibfnamefont {F.}~\bibnamefont {Ronning}},
  \bibinfo {author} {\bibfnamefont {D.~H.}\ \bibnamefont {Lu}}, \bibinfo
  {author} {\bibfnamefont {W.~S.}\ \bibnamefont {Lee}}, \bibinfo {author}
  {\bibfnamefont {N.~J.~C.}\ \bibnamefont {Ingle}}, \bibinfo {author}
  {\bibfnamefont {W.}~\bibnamefont {Meevasana}}, \bibinfo {author}
  {\bibfnamefont {F.}~\bibnamefont {Baumberger}}, \bibinfo {author}
  {\bibfnamefont {A.}~\bibnamefont {Damascelli}}, \bibinfo {author}
  {\bibfnamefont {N.~P.}\ \bibnamefont {Armitage}}, \bibinfo {author}
  {\bibfnamefont {L.~L.}\ \bibnamefont {Miller}}, \bibinfo {author}
  {\bibfnamefont {Y.}~\bibnamefont {Kohsaka}}, \bibinfo {author} {\bibfnamefont
  {M.}~\bibnamefont {Azuma}}, \bibinfo {author} {\bibfnamefont
  {M.}~\bibnamefont {Takano}}, \bibinfo {author} {\bibfnamefont
  {H.}~\bibnamefont {Takagi}},\ and\ \bibinfo {author} {\bibfnamefont {Z.-X.}\
  \bibnamefont {Shen}},\ }\bibfield  {title} {\bibinfo {title} {Missing
  quasiparticles and the chemical potential puzzle in the doping evolution of
  the cuprate superconductors},\ }\href
  {https://doi.org/10.1103/PhysRevLett.93.267002} {\bibfield  {journal}
  {\bibinfo  {journal} {Phys. Rev. Lett.}\ }\textbf {\bibinfo {volume} {93}},\
  \bibinfo {pages} {267002} (\bibinfo {year} {2004})}\BibitemShut {NoStop}%
\bibitem [{\citenamefont {Tanaka}\ \emph {et~al.}(2010)\citenamefont {Tanaka},
  \citenamefont {Yoshida}, \citenamefont {Shen}, \citenamefont {Lu},
  \citenamefont {Lee}, \citenamefont {Yagi}, \citenamefont {Fujimori},
  \citenamefont {Shen}, \citenamefont {Risdiana}, \citenamefont {Fujii},\ and\
  \citenamefont {Terasaki}}]{Tanaka:EvolutionElectronicStructureBi2212}%
  \BibitemOpen
  \bibfield  {author} {\bibinfo {author} {\bibfnamefont {K.}~\bibnamefont
  {Tanaka}}, \bibinfo {author} {\bibfnamefont {T.}~\bibnamefont {Yoshida}},
  \bibinfo {author} {\bibfnamefont {K.~M.}\ \bibnamefont {Shen}}, \bibinfo
  {author} {\bibfnamefont {D.~H.}\ \bibnamefont {Lu}}, \bibinfo {author}
  {\bibfnamefont {W.~S.}\ \bibnamefont {Lee}}, \bibinfo {author} {\bibfnamefont
  {H.}~\bibnamefont {Yagi}}, \bibinfo {author} {\bibfnamefont {A.}~\bibnamefont
  {Fujimori}}, \bibinfo {author} {\bibfnamefont {Z.-X.}\ \bibnamefont {Shen}},
  \bibinfo {author} {\bibnamefont {Risdiana}}, \bibinfo {author} {\bibfnamefont
  {T.}~\bibnamefont {Fujii}},\ and\ \bibinfo {author} {\bibfnamefont
  {I.}~\bibnamefont {Terasaki}},\ }\bibfield  {title} {\bibinfo {title}
  {Evolution of electronic structure from insulator to superconductor in
  \text{Bi}$_{2}$\text{Sr}$_{2\ensuremath{-}x}$\text{La}$_{x}$(\text{Ca},\text{Y})\text{Cu}$_{2}$\text{O}$_{8+\ensuremath{\delta}}$},\
  }\href {https://doi.org/10.1103/PhysRevB.81.125115} {\bibfield  {journal}
  {\bibinfo  {journal} {Phys. Rev. B}\ }\textbf {\bibinfo {volume} {81}},\
  \bibinfo {pages} {125115} (\bibinfo {year} {2010})}\BibitemShut {NoStop}%
\bibitem [{\citenamefont {Zhou}\ \emph {et~al.}(2003)\citenamefont {Zhou},
  \citenamefont {Yoshida}, \citenamefont {Lanzara}, \citenamefont {Bogdanov},
  \citenamefont {Kellar}, \citenamefont {Shen}, \citenamefont {Yang},
  \citenamefont {Ronning}, \citenamefont {Sasagawa}, \citenamefont {Kakeshita},
  \citenamefont {Noda}, \citenamefont {Eisaki}, \citenamefont {Uchida},
  \citenamefont {Lin}, \citenamefont {Zhou}, \citenamefont {Xiong},
  \citenamefont {Ti}, \citenamefont {Zhao}, \citenamefont {Fujimori},
  \citenamefont {Hussain},\ and\ \citenamefont
  {Shen}}]{Zhou:UniveralNodalFermiVelocity}%
  \BibitemOpen
  \bibfield  {author} {\bibinfo {author} {\bibfnamefont {X.~J.}\ \bibnamefont
  {Zhou}}, \bibinfo {author} {\bibfnamefont {T.}~\bibnamefont {Yoshida}},
  \bibinfo {author} {\bibfnamefont {A.}~\bibnamefont {Lanzara}}, \bibinfo
  {author} {\bibfnamefont {P.~V.}\ \bibnamefont {Bogdanov}}, \bibinfo {author}
  {\bibfnamefont {S.~A.}\ \bibnamefont {Kellar}}, \bibinfo {author}
  {\bibfnamefont {K.~M.}\ \bibnamefont {Shen}}, \bibinfo {author}
  {\bibfnamefont {W.~L.}\ \bibnamefont {Yang}}, \bibinfo {author}
  {\bibfnamefont {F.}~\bibnamefont {Ronning}}, \bibinfo {author} {\bibfnamefont
  {T.}~\bibnamefont {Sasagawa}}, \bibinfo {author} {\bibfnamefont
  {T.}~\bibnamefont {Kakeshita}}, \bibinfo {author} {\bibfnamefont
  {T.}~\bibnamefont {Noda}}, \bibinfo {author} {\bibfnamefont {H.}~\bibnamefont
  {Eisaki}}, \bibinfo {author} {\bibfnamefont {S.}~\bibnamefont {Uchida}},
  \bibinfo {author} {\bibfnamefont {C.~T.}\ \bibnamefont {Lin}}, \bibinfo
  {author} {\bibfnamefont {F.}~\bibnamefont {Zhou}}, \bibinfo {author}
  {\bibfnamefont {J.~W.}\ \bibnamefont {Xiong}}, \bibinfo {author}
  {\bibfnamefont {W.~X.}\ \bibnamefont {Ti}}, \bibinfo {author} {\bibfnamefont
  {Z.~X.}\ \bibnamefont {Zhao}}, \bibinfo {author} {\bibfnamefont
  {A.}~\bibnamefont {Fujimori}}, \bibinfo {author} {\bibfnamefont
  {Z.}~\bibnamefont {Hussain}},\ and\ \bibinfo {author} {\bibfnamefont {Z.~X.}\
  \bibnamefont {Shen}},\ }\bibfield  {title} {\bibinfo {title} {Universal nodal
  fermi velocity},\ }\href {https://doi.org/10.1038/423398a
  https://www.nature.com/articles/423398a#supplementary-information} {\bibfield
   {journal} {\bibinfo  {journal} {Nature}\ }\textbf {\bibinfo {volume}
  {423}},\ \bibinfo {pages} {398} (\bibinfo {year} {2003})}\BibitemShut
  {NoStop}%
\bibitem [{\citenamefont {Sato}\ \emph {et~al.}(2009)\citenamefont {Sato},
  \citenamefont {Iwasawa}, \citenamefont {Plumb}, \citenamefont {Masui},
  \citenamefont {Yoshida}, \citenamefont {Eisaki}, \citenamefont {Bando},
  \citenamefont {Ino}, \citenamefont {Arita}, \citenamefont {Shimada},
  \citenamefont {Namatame}, \citenamefont {Taniguchi}, \citenamefont {Tajima},
  \citenamefont {Nishihara}, \citenamefont {Dessau},\ and\ \citenamefont
  {Aiura}}]{Sato:Bi2201_2009}%
  \BibitemOpen
  \bibfield  {author} {\bibinfo {author} {\bibfnamefont {K.}~\bibnamefont
  {Sato}}, \bibinfo {author} {\bibfnamefont {H.}~\bibnamefont {Iwasawa}},
  \bibinfo {author} {\bibfnamefont {N.~C.}\ \bibnamefont {Plumb}}, \bibinfo
  {author} {\bibfnamefont {T.}~\bibnamefont {Masui}}, \bibinfo {author}
  {\bibfnamefont {Y.}~\bibnamefont {Yoshida}}, \bibinfo {author} {\bibfnamefont
  {H.}~\bibnamefont {Eisaki}}, \bibinfo {author} {\bibfnamefont
  {H.}~\bibnamefont {Bando}}, \bibinfo {author} {\bibfnamefont
  {A.}~\bibnamefont {Ino}}, \bibinfo {author} {\bibfnamefont {M.}~\bibnamefont
  {Arita}}, \bibinfo {author} {\bibfnamefont {K.}~\bibnamefont {Shimada}},
  \bibinfo {author} {\bibfnamefont {H.}~\bibnamefont {Namatame}}, \bibinfo
  {author} {\bibfnamefont {M.}~\bibnamefont {Taniguchi}}, \bibinfo {author}
  {\bibfnamefont {S.}~\bibnamefont {Tajima}}, \bibinfo {author} {\bibfnamefont
  {Y.}~\bibnamefont {Nishihara}}, \bibinfo {author} {\bibfnamefont {D.~S.}\
  \bibnamefont {Dessau}},\ and\ \bibinfo {author} {\bibfnamefont
  {Y.}~\bibnamefont {Aiura}},\ }\bibfield  {title} {\bibinfo {title}
  {Enhancement of oxygen isotope effect due to out-of-plane disorder in
  \text{Bi}$_{2}$\text{Sr}$_{2}$\text{Ln}$_{0.4}$\text{CuO}$_{6+\ensuremath{\delta}}$
  superconductors},\ }\href {https://doi.org/10.1103/PhysRevB.80.212501}
  {\bibfield  {journal} {\bibinfo  {journal} {Phys. Rev. B}\ }\textbf {\bibinfo
  {volume} {80}},\ \bibinfo {pages} {212501} (\bibinfo {year}
  {2009})}\BibitemShut {NoStop}%
\bibitem [{\citenamefont {d’Astuto}\ \emph {et~al.}(2003)\citenamefont
  {d’Astuto}, \citenamefont {Mirone}, \citenamefont {Giura}, \citenamefont
  {Colson}, \citenamefont {Forget},\ and\ \citenamefont
  {Krisch}}]{dAstuto:Hg1201_phonon2003}%
  \BibitemOpen
  \bibfield  {author} {\bibinfo {author} {\bibfnamefont {M.}~\bibnamefont
  {d’Astuto}}, \bibinfo {author} {\bibfnamefont {A.}~\bibnamefont {Mirone}},
  \bibinfo {author} {\bibfnamefont {P.}~\bibnamefont {Giura}}, \bibinfo
  {author} {\bibfnamefont {D.}~\bibnamefont {Colson}}, \bibinfo {author}
  {\bibfnamefont {A.}~\bibnamefont {Forget}},\ and\ \bibinfo {author}
  {\bibfnamefont {M.}~\bibnamefont {Krisch}},\ }\bibfield  {title} {\bibinfo
  {title} {Phonon dispersion in the one-layer cuprate
  \text{HgBa}$_2$\text{CuO}$_{4+\delta}$},\ }\href
  {http://stacks.iop.org/0953-8984/15/i=50/a=014} {\bibfield  {journal}
  {\bibinfo  {journal} {Journal of Physics: Condensed Matter}\ }\textbf
  {\bibinfo {volume} {15}},\ \bibinfo {pages} {8827} (\bibinfo {year}
  {2003})}\BibitemShut {NoStop}%
\bibitem [{\citenamefont {Zhou}\ \emph {et~al.}(1996)\citenamefont {Zhou},
  \citenamefont {Cardona}, \citenamefont {Chu}, \citenamefont {Lin},
  \citenamefont {Loureiro},\ and\ \citenamefont
  {Marezio}}]{Zhou:RamanHgCuprate1996}%
  \BibitemOpen
  \bibfield  {author} {\bibinfo {author} {\bibfnamefont {X.}~\bibnamefont
  {Zhou}}, \bibinfo {author} {\bibfnamefont {M.}~\bibnamefont {Cardona}},
  \bibinfo {author} {\bibfnamefont {C.}~\bibnamefont {Chu}}, \bibinfo {author}
  {\bibfnamefont {Q.}~\bibnamefont {Lin}}, \bibinfo {author} {\bibfnamefont
  {S.}~\bibnamefont {Loureiro}},\ and\ \bibinfo {author} {\bibfnamefont
  {M.}~\bibnamefont {Marezio}},\ }\bibfield  {title} {\bibinfo {title} {Raman
  study of \text{HgBa$_2$Ca$_{n−1}$Cu$_n$O$_{2n+2+\delta}$ (n$=$1,2,3,4 and
  5)} superconductors},\ }\href
  {https://doi.org/https://doi.org/10.1016/S0921-4534(96)00514-X} {\bibfield
  {journal} {\bibinfo  {journal} {Physica C: Superconductivity}\ }\textbf
  {\bibinfo {volume} {270}},\ \bibinfo {pages} {193 } (\bibinfo {year}
  {1996})}\BibitemShut {NoStop}%
\bibitem [{\citenamefont {Lanzara}\ \emph {et~al.}(2001)\citenamefont
  {Lanzara}, \citenamefont {Bogdanov}, \citenamefont {Zhou}, \citenamefont
  {Kellar}, \citenamefont {Feng}, \citenamefont {Lu}, \citenamefont {Yoshida},
  \citenamefont {Eisaki}, \citenamefont {Fujimori}, \citenamefont {Kishio},
  \citenamefont {Shimoyama}, \citenamefont {Noda}, \citenamefont {Uchida},
  \citenamefont {Hussain},\ and\ \citenamefont
  {Shen}}]{Lanzara:EvidenceUbiquitous}%
  \BibitemOpen
  \bibfield  {author} {\bibinfo {author} {\bibfnamefont {A.}~\bibnamefont
  {Lanzara}}, \bibinfo {author} {\bibfnamefont {P.~V.}\ \bibnamefont
  {Bogdanov}}, \bibinfo {author} {\bibfnamefont {X.~J.}\ \bibnamefont {Zhou}},
  \bibinfo {author} {\bibfnamefont {S.~A.}\ \bibnamefont {Kellar}}, \bibinfo
  {author} {\bibfnamefont {D.~L.}\ \bibnamefont {Feng}}, \bibinfo {author}
  {\bibfnamefont {E.~D.}\ \bibnamefont {Lu}}, \bibinfo {author} {\bibfnamefont
  {T.}~\bibnamefont {Yoshida}}, \bibinfo {author} {\bibfnamefont
  {H.}~\bibnamefont {Eisaki}}, \bibinfo {author} {\bibfnamefont
  {A.}~\bibnamefont {Fujimori}}, \bibinfo {author} {\bibfnamefont
  {K.}~\bibnamefont {Kishio}}, \bibinfo {author} {\bibfnamefont {J.~I.}\
  \bibnamefont {Shimoyama}}, \bibinfo {author} {\bibfnamefont {T.}~\bibnamefont
  {Noda}}, \bibinfo {author} {\bibfnamefont {S.}~\bibnamefont {Uchida}},
  \bibinfo {author} {\bibfnamefont {Z.}~\bibnamefont {Hussain}},\ and\ \bibinfo
  {author} {\bibfnamefont {Z.~X.}\ \bibnamefont {Shen}},\ }\bibfield  {title}
  {\bibinfo {title} {Evidence for ubiquitous strong electron–phonon coupling
  in high-temperature superconductors},\ }\href
  {https://doi.org/10.1038/35087518} {\bibfield  {journal} {\bibinfo  {journal}
  {Nature}\ }\textbf {\bibinfo {volume} {412}},\ \bibinfo {pages} {510}
  (\bibinfo {year} {2001})}\BibitemShut {NoStop}%
\bibitem [{\citenamefont {Johnston}\ \emph
  {et~al.}(2010{\natexlab{a}})\citenamefont {Johnston}, \citenamefont {Lee},
  \citenamefont {Chen}, \citenamefont {Nowadnick}, \citenamefont {Moritz},
  \citenamefont {Shen},\ and\ \citenamefont
  {Devereaux}}]{Johnston:MaterialDopingDepkink}%
  \BibitemOpen
  \bibfield  {author} {\bibinfo {author} {\bibfnamefont {S.}~\bibnamefont
  {Johnston}}, \bibinfo {author} {\bibfnamefont {W.~S.}\ \bibnamefont {Lee}},
  \bibinfo {author} {\bibfnamefont {Y.}~\bibnamefont {Chen}}, \bibinfo {author}
  {\bibfnamefont {E.~A.}\ \bibnamefont {Nowadnick}}, \bibinfo {author}
  {\bibfnamefont {B.}~\bibnamefont {Moritz}}, \bibinfo {author} {\bibfnamefont
  {Z.-X.}\ \bibnamefont {Shen}},\ and\ \bibinfo {author} {\bibfnamefont
  {T.~P.}\ \bibnamefont {Devereaux}},\ }\bibfield  {title} {\bibinfo {title}
  {Material and doping dependence of the nodal and antinodal dispersion
  renormalizations in single- and multilayer cuprates},\ }\href
  {https://doi.org/10.1155/2010/968304} {\bibfield  {journal} {\bibinfo
  {journal} {J Advances in Condensed Matter Physics}\ }\textbf {\bibinfo
  {volume} {2010}},\ \bibinfo {pages} {13} (\bibinfo {year}
  {2010}{\natexlab{a}})}\BibitemShut {NoStop}%
\bibitem [{\citenamefont {Chubukov}\ and\ \citenamefont
  {Norman}(2004)}]{Chubukov:DispersionAnomalyS-shape}%
  \BibitemOpen
  \bibfield  {author} {\bibinfo {author} {\bibfnamefont {A.~V.}\ \bibnamefont
  {Chubukov}}\ and\ \bibinfo {author} {\bibfnamefont {M.~R.}\ \bibnamefont
  {Norman}},\ }\bibfield  {title} {\bibinfo {title} {Dispersion anomalies in
  cuprate superconductors},\ }\href
  {https://doi.org/10.1103/PhysRevB.70.174505} {\bibfield  {journal} {\bibinfo
  {journal} {Phys. Rev. B}\ }\textbf {\bibinfo {volume} {70}},\ \bibinfo
  {pages} {174505} (\bibinfo {year} {2004})}\BibitemShut {NoStop}%
\bibitem [{\citenamefont {Matsuyama}\ \emph {et~al.}(2017)\citenamefont
  {Matsuyama}, \citenamefont {Perepelitsky},\ and\ \citenamefont
  {Shastry}}]{Matsuyama:OriginKinksNoMode}%
  \BibitemOpen
  \bibfield  {author} {\bibinfo {author} {\bibfnamefont {K.}~\bibnamefont
  {Matsuyama}}, \bibinfo {author} {\bibfnamefont {E.}~\bibnamefont
  {Perepelitsky}},\ and\ \bibinfo {author} {\bibfnamefont {B.~S.}\ \bibnamefont
  {Shastry}},\ }\bibfield  {title} {\bibinfo {title} {Origin of kinks in the
  energy dispersion of strongly correlated matter},\ }\href
  {https://doi.org/10.1103/PhysRevB.95.165435} {\bibfield  {journal} {\bibinfo
  {journal} {Phys. Rev. B}\ }\textbf {\bibinfo {volume} {95}},\ \bibinfo
  {pages} {165435} (\bibinfo {year} {2017})}\BibitemShut {NoStop}%
\bibitem [{\citenamefont {Li}\ \emph {et~al.}(2012)\citenamefont {Li},
  \citenamefont {Yu}, \citenamefont {Chan}, \citenamefont {Balédent},
  \citenamefont {Li}, \citenamefont {Barišić}, \citenamefont {Zhao},
  \citenamefont {Hradil}, \citenamefont {Mole}, \citenamefont {Sidis},
  \citenamefont {Steffens}, \citenamefont {Bourges},\ and\ \citenamefont
  {Greven}}]{Li:IsingHg1201}%
  \BibitemOpen
  \bibfield  {author} {\bibinfo {author} {\bibfnamefont {Y.}~\bibnamefont
  {Li}}, \bibinfo {author} {\bibfnamefont {G.}~\bibnamefont {Yu}}, \bibinfo
  {author} {\bibfnamefont {M.~K.}\ \bibnamefont {Chan}}, \bibinfo {author}
  {\bibfnamefont {V.}~\bibnamefont {Balédent}}, \bibinfo {author}
  {\bibfnamefont {Y.}~\bibnamefont {Li}}, \bibinfo {author} {\bibfnamefont
  {N.}~\bibnamefont {Barišić}}, \bibinfo {author} {\bibfnamefont
  {X.}~\bibnamefont {Zhao}}, \bibinfo {author} {\bibfnamefont {K.}~\bibnamefont
  {Hradil}}, \bibinfo {author} {\bibfnamefont {R.~A.}\ \bibnamefont {Mole}},
  \bibinfo {author} {\bibfnamefont {Y.}~\bibnamefont {Sidis}}, \bibinfo
  {author} {\bibfnamefont {P.}~\bibnamefont {Steffens}}, \bibinfo {author}
  {\bibfnamefont {P.}~\bibnamefont {Bourges}},\ and\ \bibinfo {author}
  {\bibfnamefont {M.}~\bibnamefont {Greven}},\ }\bibfield  {title} {\bibinfo
  {title} {Two ising-like magnetic excitations in a single-layer cuprate
  superconductor},\ }\href {https://doi.org/10.1038/nphys2271
  https://www.nature.com/articles/nphys2271#supplementary-information}
  {\bibfield  {journal} {\bibinfo  {journal} {Nature Physics}\ }\textbf
  {\bibinfo {volume} {8}},\ \bibinfo {pages} {404} (\bibinfo {year}
  {2012})}\BibitemShut {NoStop}%
\bibitem [{\citenamefont {Yu}\ \emph {et~al.}(2019)\citenamefont {Yu},
  \citenamefont {Tabis}, \citenamefont {Bialo}, \citenamefont {Yakhou},
  \citenamefont {Brookes}, \citenamefont {Anderson}, \citenamefont {Tang},
  \citenamefont {Yu},\ and\ \citenamefont {Greven}}]{yu2019unusual}%
  \BibitemOpen
  \bibfield  {author} {\bibinfo {author} {\bibfnamefont {B.}~\bibnamefont
  {Yu}}, \bibinfo {author} {\bibfnamefont {W.}~\bibnamefont {Tabis}}, \bibinfo
  {author} {\bibfnamefont {I.}~\bibnamefont {Bialo}}, \bibinfo {author}
  {\bibfnamefont {F.}~\bibnamefont {Yakhou}}, \bibinfo {author} {\bibfnamefont
  {N.}~\bibnamefont {Brookes}}, \bibinfo {author} {\bibfnamefont
  {Z.}~\bibnamefont {Anderson}}, \bibinfo {author} {\bibfnamefont
  {Y.}~\bibnamefont {Tang}}, \bibinfo {author} {\bibfnamefont {G.}~\bibnamefont
  {Yu}},\ and\ \bibinfo {author} {\bibfnamefont {M.}~\bibnamefont {Greven}},\
  }\bibfield  {title} {\bibinfo {title} {Unusual dynamic charge-density-wave
  correlations in \text{HgBa$_2$CuO$_{4+\delta}$}},\ }\href@noop {} {\bibfield
  {journal} {\bibinfo  {journal} {arXiv preprint arXiv:1907.10047}\ } (\bibinfo
  {year} {2019})}\BibitemShut {NoStop}%
\bibitem [{\citenamefont {Yu}\ \emph {et~al.}(2009)\citenamefont {Yu},
  \citenamefont {Li}, \citenamefont {Motoyama},\ and\ \citenamefont
  {Greven}}]{Yu:UniversalRelationMagResDelta}%
  \BibitemOpen
  \bibfield  {author} {\bibinfo {author} {\bibfnamefont {G.}~\bibnamefont
  {Yu}}, \bibinfo {author} {\bibfnamefont {Y.}~\bibnamefont {Li}}, \bibinfo
  {author} {\bibfnamefont {E.~M.}\ \bibnamefont {Motoyama}},\ and\ \bibinfo
  {author} {\bibfnamefont {M.}~\bibnamefont {Greven}},\ }\bibfield  {title}
  {\bibinfo {title} {A universal relationship between magnetic resonance and
  superconducting gap in unconventional superconductors},\ }\href
  {https://doi.org/10.1038/nphys1426} {\bibfield  {journal} {\bibinfo
  {journal} {Nature Physics}\ }\textbf {\bibinfo {volume} {5}},\ \bibinfo
  {pages} {873} (\bibinfo {year} {2009})}\BibitemShut {NoStop}%
\bibitem [{\citenamefont {Li}\ \emph {et~al.}(2010)\citenamefont {Li},
  \citenamefont {Balédent}, \citenamefont {Yu}, \citenamefont {Barišić},
  \citenamefont {Hradil}, \citenamefont {Mole}, \citenamefont {Sidis},
  \citenamefont {Steffens}, \citenamefont {Zhao}, \citenamefont {Bourges},\
  and\ \citenamefont {Greven}}]{Li:HiddenMagneticExcitations}%
  \BibitemOpen
  \bibfield  {author} {\bibinfo {author} {\bibfnamefont {Y.}~\bibnamefont
  {Li}}, \bibinfo {author} {\bibfnamefont {V.}~\bibnamefont {Balédent}},
  \bibinfo {author} {\bibfnamefont {G.}~\bibnamefont {Yu}}, \bibinfo {author}
  {\bibfnamefont {N.}~\bibnamefont {Barišić}}, \bibinfo {author}
  {\bibfnamefont {K.}~\bibnamefont {Hradil}}, \bibinfo {author} {\bibfnamefont
  {R.~A.}\ \bibnamefont {Mole}}, \bibinfo {author} {\bibfnamefont
  {Y.}~\bibnamefont {Sidis}}, \bibinfo {author} {\bibfnamefont
  {P.}~\bibnamefont {Steffens}}, \bibinfo {author} {\bibfnamefont
  {X.}~\bibnamefont {Zhao}}, \bibinfo {author} {\bibfnamefont {P.}~\bibnamefont
  {Bourges}},\ and\ \bibinfo {author} {\bibfnamefont {M.}~\bibnamefont
  {Greven}},\ }\bibfield  {title} {\bibinfo {title} {Hidden magnetic excitation
  in the pseudogap phase of a high-$t_c$ superconductor},\ }\href
  {https://doi.org/10.1038/nature09477} {\bibfield  {journal} {\bibinfo
  {journal} {Nature}\ }\textbf {\bibinfo {volume} {468}},\ \bibinfo {pages}
  {283} (\bibinfo {year} {2010})}\BibitemShut {NoStop}%
\bibitem [{\citenamefont {Pintschovius}\ and\ \citenamefont
  {Braden}(1999)}]{Pintschovius:AnomalousDisp}%
  \BibitemOpen
  \bibfield  {author} {\bibinfo {author} {\bibfnamefont {L.}~\bibnamefont
  {Pintschovius}}\ and\ \bibinfo {author} {\bibfnamefont {M.}~\bibnamefont
  {Braden}},\ }\bibfield  {title} {\bibinfo {title} {Anomalous dispersion of lo
  phonons in \text{La}$_{1.85}$\text{Sr}$_{0.15}$\text{CuO}$_{4}$},\ }\href
  {https://doi.org/10.1103/PhysRevB.60.R15039} {\bibfield  {journal} {\bibinfo
  {journal} {Phys. Rev. B}\ }\textbf {\bibinfo {volume} {60}},\ \bibinfo
  {pages} {R15039} (\bibinfo {year} {1999})}\BibitemShut {NoStop}%
\bibitem [{\citenamefont {Ahmadova}\ \emph {et~al.}(2020)\citenamefont
  {Ahmadova}, \citenamefont {Sterling}, \citenamefont {Sokolik}, \citenamefont
  {Abernathy}, \citenamefont {Greven},\ and\ \citenamefont
  {Reznik}}]{ahmadova2020phonon}%
  \BibitemOpen
  \bibfield  {author} {\bibinfo {author} {\bibfnamefont {I.}~\bibnamefont
  {Ahmadova}}, \bibinfo {author} {\bibfnamefont {T.}~\bibnamefont {Sterling}},
  \bibinfo {author} {\bibfnamefont {A.}~\bibnamefont {Sokolik}}, \bibinfo
  {author} {\bibfnamefont {D.}~\bibnamefont {Abernathy}}, \bibinfo {author}
  {\bibfnamefont {M.}~\bibnamefont {Greven}},\ and\ \bibinfo {author}
  {\bibfnamefont {D.}~\bibnamefont {Reznik}},\ }\bibfield  {title} {\bibinfo
  {title} {Phonon spectrum of underdoped
  \text{HgBa}$_{2}$\text{CuO}$_{4+\ensuremath{\delta}}$ investigated by neutron
  scattering},\ }\href@noop {} {\bibfield  {journal} {\bibinfo  {journal}
  {arXiv preprint arXiv:2002.02593}\ } (\bibinfo {year} {2020})}\BibitemShut
  {NoStop}%
\bibitem [{\citenamefont {Sandvik}\ \emph {et~al.}(2004)\citenamefont
  {Sandvik}, \citenamefont {Scalapino},\ and\ \citenamefont
  {Bickers}}]{Sandvik:EffectsElectronPhononInteration}%
  \BibitemOpen
  \bibfield  {author} {\bibinfo {author} {\bibfnamefont {A.~W.}\ \bibnamefont
  {Sandvik}}, \bibinfo {author} {\bibfnamefont {D.~J.}\ \bibnamefont
  {Scalapino}},\ and\ \bibinfo {author} {\bibfnamefont {N.~E.}\ \bibnamefont
  {Bickers}},\ }\bibfield  {title} {\bibinfo {title} {Effect of an
  electron-phonon interaction on the one-electron spectral weight of a d-wave
  superconductor},\ }\href {https://doi.org/10.1103/PhysRevB.69.094523}
  {\bibfield  {journal} {\bibinfo  {journal} {Phys. Rev. B}\ }\textbf {\bibinfo
  {volume} {69}},\ \bibinfo {pages} {094523} (\bibinfo {year}
  {2004})}\BibitemShut {NoStop}%
\bibitem [{\citenamefont {He}\ \emph {et~al.}(2013)\citenamefont {He},
  \citenamefont {Zhang}, \citenamefont {Bok}, \citenamefont {Mou},
  \citenamefont {Zhao}, \citenamefont {Peng}, \citenamefont {He}, \citenamefont
  {Liu}, \citenamefont {Dong}, \citenamefont {Zhang}, \citenamefont {Wen},
  \citenamefont {Xu}, \citenamefont {Gu}, \citenamefont {Wang}, \citenamefont
  {Peng}, \citenamefont {Wang}, \citenamefont {Zhang}, \citenamefont {Yang},
  \citenamefont {Chen}, \citenamefont {Xu}, \citenamefont {Choi}, \citenamefont
  {Varma},\ and\ \citenamefont {Zhou}}]{He:Bi2212KinkAcrossTc}%
  \BibitemOpen
  \bibfield  {author} {\bibinfo {author} {\bibfnamefont {J.}~\bibnamefont
  {He}}, \bibinfo {author} {\bibfnamefont {W.}~\bibnamefont {Zhang}}, \bibinfo
  {author} {\bibfnamefont {J.~M.}\ \bibnamefont {Bok}}, \bibinfo {author}
  {\bibfnamefont {D.}~\bibnamefont {Mou}}, \bibinfo {author} {\bibfnamefont
  {L.}~\bibnamefont {Zhao}}, \bibinfo {author} {\bibfnamefont {Y.}~\bibnamefont
  {Peng}}, \bibinfo {author} {\bibfnamefont {S.}~\bibnamefont {He}}, \bibinfo
  {author} {\bibfnamefont {G.}~\bibnamefont {Liu}}, \bibinfo {author}
  {\bibfnamefont {X.}~\bibnamefont {Dong}}, \bibinfo {author} {\bibfnamefont
  {J.}~\bibnamefont {Zhang}}, \bibinfo {author} {\bibfnamefont {J.~S.}\
  \bibnamefont {Wen}}, \bibinfo {author} {\bibfnamefont {Z.~J.}\ \bibnamefont
  {Xu}}, \bibinfo {author} {\bibfnamefont {G.~D.}\ \bibnamefont {Gu}}, \bibinfo
  {author} {\bibfnamefont {X.}~\bibnamefont {Wang}}, \bibinfo {author}
  {\bibfnamefont {Q.}~\bibnamefont {Peng}}, \bibinfo {author} {\bibfnamefont
  {Z.}~\bibnamefont {Wang}}, \bibinfo {author} {\bibfnamefont {S.}~\bibnamefont
  {Zhang}}, \bibinfo {author} {\bibfnamefont {F.}~\bibnamefont {Yang}},
  \bibinfo {author} {\bibfnamefont {C.}~\bibnamefont {Chen}}, \bibinfo {author}
  {\bibfnamefont {Z.}~\bibnamefont {Xu}}, \bibinfo {author} {\bibfnamefont
  {H.-Y.}\ \bibnamefont {Choi}}, \bibinfo {author} {\bibfnamefont {C.~M.}\
  \bibnamefont {Varma}},\ and\ \bibinfo {author} {\bibfnamefont {X.~J.}\
  \bibnamefont {Zhou}},\ }\bibfield  {title} {\bibinfo {title} {Coexistence of
  two sharp-mode couplings and their unusual momentum dependence in the
  superconducting state of
  \text{${\mathrm{Bi}}_{2}{\mathrm{Sr}}_{2}{\mathrm{CaCu}}_{2}{\mathbf{O}}_{8\mathbf{+}\ensuremath{\delta}}$}
  revealed by laser-based angle-resolved photoemission},\ }\href
  {https://doi.org/10.1103/PhysRevLett.111.107005} {\bibfield  {journal}
  {\bibinfo  {journal} {Phys. Rev. Lett.}\ }\textbf {\bibinfo {volume} {111}},\
  \bibinfo {pages} {107005} (\bibinfo {year} {2013})}\BibitemShut {NoStop}%
\bibitem [{\citenamefont {Vishik}(2013)}]{Vishik:Thesis}%
  \BibitemOpen
  \bibfield  {author} {\bibinfo {author} {\bibfnamefont {I.}~\bibnamefont
  {Vishik}},\ }\emph {\bibinfo {title} {Low Energy Excitations in Cuprate High
  Temperature Superconductors: Angle-resolved Photoemission Spectroscopy
  Studies}},\ \href@noop {} {Ph.D. thesis},\ \bibinfo  {school} {Stanford
  University} (\bibinfo {year} {2013})\BibitemShut {NoStop}%
\bibitem [{\citenamefont {Lanzara}\ \emph {et~al.}(2006)\citenamefont
  {Lanzara}, \citenamefont {Bogdanov}, \citenamefont {Zhou}, \citenamefont
  {Kaneko}, \citenamefont {Eisaki}, \citenamefont {Greven}, \citenamefont
  {Hussain},\ and\ \citenamefont {Shen}}]{Lanzara:temperatureDepKinkBi2201}%
  \BibitemOpen
  \bibfield  {author} {\bibinfo {author} {\bibfnamefont {A.}~\bibnamefont
  {Lanzara}}, \bibinfo {author} {\bibfnamefont {P.}~\bibnamefont {Bogdanov}},
  \bibinfo {author} {\bibfnamefont {X.}~\bibnamefont {Zhou}}, \bibinfo {author}
  {\bibfnamefont {N.}~\bibnamefont {Kaneko}}, \bibinfo {author} {\bibfnamefont
  {H.}~\bibnamefont {Eisaki}}, \bibinfo {author} {\bibfnamefont
  {M.}~\bibnamefont {Greven}}, \bibinfo {author} {\bibfnamefont
  {Z.}~\bibnamefont {Hussain}},\ and\ \bibinfo {author} {\bibfnamefont
  {Z.}~\bibnamefont {Shen}},\ }\bibfield  {title} {\bibinfo {title} {Normal
  state spectral lineshapes of nodal quasiparticles in single layer bi2201
  superconductor},\ }\href
  {https://doi.org/https://doi.org/10.1016/j.jpcs.2005.10.129} {\bibfield
  {journal} {\bibinfo  {journal} {Journal of Physics and Chemistry of Solids}\
  }\textbf {\bibinfo {volume} {67}},\ \bibinfo {pages} {239 } (\bibinfo {year}
  {2006})}\BibitemShut {NoStop}%
\bibitem [{\citenamefont {Zhao}\ \emph {et~al.}(2011)\citenamefont {Zhao},
  \citenamefont {Wang}, \citenamefont {Shi}, \citenamefont {Zhang},
  \citenamefont {Liu}, \citenamefont {Meng}, \citenamefont {Liu}, \citenamefont
  {Dong}, \citenamefont {Zhang}, \citenamefont {Lu}, \citenamefont {Wang},
  \citenamefont {Zhu}, \citenamefont {Wang}, \citenamefont {Peng},
  \citenamefont {Wang}, \citenamefont {Zhang}, \citenamefont {Yang},
  \citenamefont {Chen}, \citenamefont {Xu},\ and\ \citenamefont
  {Zhou}}]{Zhao:QuantitativeBi2201_Tdep}%
  \BibitemOpen
  \bibfield  {author} {\bibinfo {author} {\bibfnamefont {L.}~\bibnamefont
  {Zhao}}, \bibinfo {author} {\bibfnamefont {J.}~\bibnamefont {Wang}}, \bibinfo
  {author} {\bibfnamefont {J.}~\bibnamefont {Shi}}, \bibinfo {author}
  {\bibfnamefont {W.}~\bibnamefont {Zhang}}, \bibinfo {author} {\bibfnamefont
  {H.}~\bibnamefont {Liu}}, \bibinfo {author} {\bibfnamefont {J.}~\bibnamefont
  {Meng}}, \bibinfo {author} {\bibfnamefont {G.}~\bibnamefont {Liu}}, \bibinfo
  {author} {\bibfnamefont {X.}~\bibnamefont {Dong}}, \bibinfo {author}
  {\bibfnamefont {J.}~\bibnamefont {Zhang}}, \bibinfo {author} {\bibfnamefont
  {W.}~\bibnamefont {Lu}}, \bibinfo {author} {\bibfnamefont {G.}~\bibnamefont
  {Wang}}, \bibinfo {author} {\bibfnamefont {Y.}~\bibnamefont {Zhu}}, \bibinfo
  {author} {\bibfnamefont {X.}~\bibnamefont {Wang}}, \bibinfo {author}
  {\bibfnamefont {Q.}~\bibnamefont {Peng}}, \bibinfo {author} {\bibfnamefont
  {Z.}~\bibnamefont {Wang}}, \bibinfo {author} {\bibfnamefont {S.}~\bibnamefont
  {Zhang}}, \bibinfo {author} {\bibfnamefont {F.}~\bibnamefont {Yang}},
  \bibinfo {author} {\bibfnamefont {C.}~\bibnamefont {Chen}}, \bibinfo {author}
  {\bibfnamefont {Z.}~\bibnamefont {Xu}},\ and\ \bibinfo {author}
  {\bibfnamefont {X.~J.}\ \bibnamefont {Zhou}},\ }\bibfield  {title} {\bibinfo
  {title} {Quantitative determination of \text{Eliashberg} function and
  evidence of strong electron coupling with multiple phonon modes in heavily
  overdoped \text{(Bi,Pb)${}_{2}$Sr${}_{2}$CuO${}_{6+\ensuremath{\delta}}$}},\
  }\href {https://doi.org/10.1103/PhysRevB.83.184515} {\bibfield  {journal}
  {\bibinfo  {journal} {Phys. Rev. B}\ }\textbf {\bibinfo {volume} {83}},\
  \bibinfo {pages} {184515} (\bibinfo {year} {2011})}\BibitemShut {NoStop}%
\bibitem [{\citenamefont {Devereaux}\ \emph {et~al.}(2004)\citenamefont
  {Devereaux}, \citenamefont {Cuk}, \citenamefont {Shen},\ and\ \citenamefont
  {Nagaosa}}]{Devereaux:AnisotropicElectronPhononCoupling2004}%
  \BibitemOpen
  \bibfield  {author} {\bibinfo {author} {\bibfnamefont {T.~P.}\ \bibnamefont
  {Devereaux}}, \bibinfo {author} {\bibfnamefont {T.}~\bibnamefont {Cuk}},
  \bibinfo {author} {\bibfnamefont {Z.-X.}\ \bibnamefont {Shen}},\ and\
  \bibinfo {author} {\bibfnamefont {N.}~\bibnamefont {Nagaosa}},\ }\bibfield
  {title} {\bibinfo {title} {Anisotropic electron-phonon interaction in the
  cuprates},\ }\href {https://doi.org/10.1103/PhysRevLett.93.117004} {\bibfield
   {journal} {\bibinfo  {journal} {Phys. Rev. Lett.}\ }\textbf {\bibinfo
  {volume} {93}},\ \bibinfo {pages} {117004} (\bibinfo {year}
  {2004})}\BibitemShut {NoStop}%
\bibitem [{\citenamefont {Ishihara}\ and\ \citenamefont
  {Nagaosa}(2004)}]{Ishihara:BondStretchEnhance}%
  \BibitemOpen
  \bibfield  {author} {\bibinfo {author} {\bibfnamefont {S.}~\bibnamefont
  {Ishihara}}\ and\ \bibinfo {author} {\bibfnamefont {N.}~\bibnamefont
  {Nagaosa}},\ }\bibfield  {title} {\bibinfo {title} {Interplay of
  electron-phonon interaction and electron correlation in high-temperature
  superconductivity},\ }\href {https://doi.org/10.1103/PhysRevB.69.144520}
  {\bibfield  {journal} {\bibinfo  {journal} {Phys. Rev. B}\ }\textbf {\bibinfo
  {volume} {69}},\ \bibinfo {pages} {144520} (\bibinfo {year}
  {2004})}\BibitemShut {NoStop}%
\bibitem [{\citenamefont {Sakai}\ \emph {et~al.}(1997)\citenamefont {Sakai},
  \citenamefont {Poilblanc},\ and\ \citenamefont
  {Scalapino}}]{Sakai:Pairing_tJ_model}%
  \BibitemOpen
  \bibfield  {author} {\bibinfo {author} {\bibfnamefont {T.}~\bibnamefont
  {Sakai}}, \bibinfo {author} {\bibfnamefont {D.}~\bibnamefont {Poilblanc}},\
  and\ \bibinfo {author} {\bibfnamefont {D.~J.}\ \bibnamefont {Scalapino}},\
  }\bibfield  {title} {\bibinfo {title} {Hole pairing and phonon dynamics in
  generalized two-dimensional t-j holstein models},\ }\href
  {https://doi.org/10.1103/PhysRevB.55.8445} {\bibfield  {journal} {\bibinfo
  {journal} {Phys. Rev. B}\ }\textbf {\bibinfo {volume} {55}},\ \bibinfo
  {pages} {8445} (\bibinfo {year} {1997})}\BibitemShut {NoStop}%
\bibitem [{\citenamefont {Johnston}\ \emph
  {et~al.}(2010{\natexlab{b}})\citenamefont {Johnston}, \citenamefont {Vernay},
  \citenamefont {Moritz}, \citenamefont {Shen}, \citenamefont {Nagaosa},
  \citenamefont {Zaanen},\ and\ \citenamefont
  {Devereaux}}]{Johnston:SystematicStudyEPC}%
  \BibitemOpen
  \bibfield  {author} {\bibinfo {author} {\bibfnamefont {S.}~\bibnamefont
  {Johnston}}, \bibinfo {author} {\bibfnamefont {F.}~\bibnamefont {Vernay}},
  \bibinfo {author} {\bibfnamefont {B.}~\bibnamefont {Moritz}}, \bibinfo
  {author} {\bibfnamefont {Z.-X.}\ \bibnamefont {Shen}}, \bibinfo {author}
  {\bibfnamefont {N.}~\bibnamefont {Nagaosa}}, \bibinfo {author} {\bibfnamefont
  {J.}~\bibnamefont {Zaanen}},\ and\ \bibinfo {author} {\bibfnamefont {T.~P.}\
  \bibnamefont {Devereaux}},\ }\bibfield  {title} {\bibinfo {title} {Systematic
  study of electron-phonon coupling to oxygen modes across the cuprates},\
  }\href {https://doi.org/10.1103/PhysRevB.82.064513} {\bibfield  {journal}
  {\bibinfo  {journal} {Phys. Rev. B}\ }\textbf {\bibinfo {volume} {82}},\
  \bibinfo {pages} {064513} (\bibinfo {year} {2010}{\natexlab{b}})}\BibitemShut
  {NoStop}%
\bibitem [{\citenamefont {Park}\ \emph {et~al.}(2013)\citenamefont {Park},
  \citenamefont {Cao}, \citenamefont {Wang}, \citenamefont {Fujita},
  \citenamefont {Yamada}, \citenamefont {Mo}, \citenamefont {Dessau},\ and\
  \citenamefont {Reznik}}]{Park:BrokenRelationship}%
  \BibitemOpen
  \bibfield  {author} {\bibinfo {author} {\bibfnamefont {S.~R.}\ \bibnamefont
  {Park}}, \bibinfo {author} {\bibfnamefont {Y.}~\bibnamefont {Cao}}, \bibinfo
  {author} {\bibfnamefont {Q.}~\bibnamefont {Wang}}, \bibinfo {author}
  {\bibfnamefont {M.}~\bibnamefont {Fujita}}, \bibinfo {author} {\bibfnamefont
  {K.}~\bibnamefont {Yamada}}, \bibinfo {author} {\bibfnamefont {S.-K.}\
  \bibnamefont {Mo}}, \bibinfo {author} {\bibfnamefont {D.~S.}\ \bibnamefont
  {Dessau}},\ and\ \bibinfo {author} {\bibfnamefont {D.}~\bibnamefont
  {Reznik}},\ }\bibfield  {title} {\bibinfo {title} {Broken relationship
  between superconducting pairing interaction and electronic dispersion kinks
  in \text{La${}_{2\ensuremath{-}x}$Sr${}_{x}$CuO${}_{4}$} measured by
  angle-resolved photoemission},\ }\href
  {https://doi.org/10.1103/PhysRevB.88.220503} {\bibfield  {journal} {\bibinfo
  {journal} {Phys. Rev. B}\ }\textbf {\bibinfo {volume} {88}},\ \bibinfo
  {pages} {220503} (\bibinfo {year} {2013})}\BibitemShut {NoStop}%
\bibitem [{\citenamefont {Lee}\ \emph {et~al.}(2009)\citenamefont {Lee},
  \citenamefont {Tanaka}, \citenamefont {Vishik}, \citenamefont {Lu},
  \citenamefont {Moore}, \citenamefont {Eisaki}, \citenamefont {Iyo},
  \citenamefont {Devereaux},\ and\ \citenamefont {Shen}}]{Lee:LayerNumberDep}%
  \BibitemOpen
  \bibfield  {author} {\bibinfo {author} {\bibfnamefont {W.~S.}\ \bibnamefont
  {Lee}}, \bibinfo {author} {\bibfnamefont {K.}~\bibnamefont {Tanaka}},
  \bibinfo {author} {\bibfnamefont {I.~M.}\ \bibnamefont {Vishik}}, \bibinfo
  {author} {\bibfnamefont {D.~H.}\ \bibnamefont {Lu}}, \bibinfo {author}
  {\bibfnamefont {R.~G.}\ \bibnamefont {Moore}}, \bibinfo {author}
  {\bibfnamefont {H.}~\bibnamefont {Eisaki}}, \bibinfo {author} {\bibfnamefont
  {A.}~\bibnamefont {Iyo}}, \bibinfo {author} {\bibfnamefont {T.~P.}\
  \bibnamefont {Devereaux}},\ and\ \bibinfo {author} {\bibfnamefont {Z.~X.}\
  \bibnamefont {Shen}},\ }\bibfield  {title} {\bibinfo {title} {Dependence of
  band-renormalization effects on the number of copper oxide layers in tl-based
  copper oxide superconductors revealed by angle-resolved photoemission
  spectroscopy},\ }\href {https://doi.org/10.1103/PhysRevLett.103.067003}
  {\bibfield  {journal} {\bibinfo  {journal} {Phys. Rev. Lett.}\ }\textbf
  {\bibinfo {volume} {103}},\ \bibinfo {pages} {067003} (\bibinfo {year}
  {2009})}\BibitemShut {NoStop}%
\bibitem [{\citenamefont {Wei}\ \emph {et~al.}(2008)\citenamefont {Wei},
  \citenamefont {Zhang}, \citenamefont {Ou}, \citenamefont {Xie}, \citenamefont
  {Shen}, \citenamefont {Zhao}, \citenamefont {Yang}, \citenamefont {Arita},
  \citenamefont {Shimada}, \citenamefont {Namatame}, \citenamefont {Taniguchi},
  \citenamefont {Yoshida}, \citenamefont {Eisaki},\ and\ \citenamefont
  {Feng}}]{Wei:Bi2201_OP_lineshape}%
  \BibitemOpen
  \bibfield  {author} {\bibinfo {author} {\bibfnamefont {J.}~\bibnamefont
  {Wei}}, \bibinfo {author} {\bibfnamefont {Y.}~\bibnamefont {Zhang}}, \bibinfo
  {author} {\bibfnamefont {H.~W.}\ \bibnamefont {Ou}}, \bibinfo {author}
  {\bibfnamefont {B.~P.}\ \bibnamefont {Xie}}, \bibinfo {author} {\bibfnamefont
  {D.~W.}\ \bibnamefont {Shen}}, \bibinfo {author} {\bibfnamefont {J.~F.}\
  \bibnamefont {Zhao}}, \bibinfo {author} {\bibfnamefont {L.~X.}\ \bibnamefont
  {Yang}}, \bibinfo {author} {\bibfnamefont {M.}~\bibnamefont {Arita}},
  \bibinfo {author} {\bibfnamefont {K.}~\bibnamefont {Shimada}}, \bibinfo
  {author} {\bibfnamefont {H.}~\bibnamefont {Namatame}}, \bibinfo {author}
  {\bibfnamefont {M.}~\bibnamefont {Taniguchi}}, \bibinfo {author}
  {\bibfnamefont {Y.}~\bibnamefont {Yoshida}}, \bibinfo {author} {\bibfnamefont
  {H.}~\bibnamefont {Eisaki}},\ and\ \bibinfo {author} {\bibfnamefont {D.~L.}\
  \bibnamefont {Feng}},\ }\bibfield  {title} {\bibinfo {title} {Superconducting
  coherence peak in the electronic excitations of a single-layer
  \text{${\mathrm{Bi}}_{2}{\mathrm{Sr}}_{1.6}{\mathrm{La}}_{0.4}{\mathrm{CuO}}_{6+\ensuremath{\delta}}$}
  cuprate superconductor},\ }\href
  {https://doi.org/10.1103/PhysRevLett.101.097005} {\bibfield  {journal}
  {\bibinfo  {journal} {Phys. Rev. Lett.}\ }\textbf {\bibinfo {volume} {101}},\
  \bibinfo {pages} {097005} (\bibinfo {year} {2008})}\BibitemShut {NoStop}%
\bibitem [{\citenamefont {Feng}\ \emph {et~al.}(2002)\citenamefont {Feng},
  \citenamefont {Damascelli}, \citenamefont {Shen}, \citenamefont {Motoyama},
  \citenamefont {Lu}, \citenamefont {Eisaki}, \citenamefont {Shimizu},
  \citenamefont {Shimoyama}, \citenamefont {Kishio}, \citenamefont {Kaneko},
  \citenamefont {Greven}, \citenamefont {Gu}, \citenamefont {Zhou},
  \citenamefont {Kim}, \citenamefont {Ronning}, \citenamefont {Armitage},\ and\
  \citenamefont {Shen}}]{Feng:Bi2223ARPES}%
  \BibitemOpen
  \bibfield  {author} {\bibinfo {author} {\bibfnamefont {D.~L.}\ \bibnamefont
  {Feng}}, \bibinfo {author} {\bibfnamefont {A.}~\bibnamefont {Damascelli}},
  \bibinfo {author} {\bibfnamefont {K.~M.}\ \bibnamefont {Shen}}, \bibinfo
  {author} {\bibfnamefont {N.}~\bibnamefont {Motoyama}}, \bibinfo {author}
  {\bibfnamefont {D.~H.}\ \bibnamefont {Lu}}, \bibinfo {author} {\bibfnamefont
  {H.}~\bibnamefont {Eisaki}}, \bibinfo {author} {\bibfnamefont
  {K.}~\bibnamefont {Shimizu}}, \bibinfo {author} {\bibfnamefont {J.-i.}\
  \bibnamefont {Shimoyama}}, \bibinfo {author} {\bibfnamefont {K.}~\bibnamefont
  {Kishio}}, \bibinfo {author} {\bibfnamefont {N.}~\bibnamefont {Kaneko}},
  \bibinfo {author} {\bibfnamefont {M.}~\bibnamefont {Greven}}, \bibinfo
  {author} {\bibfnamefont {G.~D.}\ \bibnamefont {Gu}}, \bibinfo {author}
  {\bibfnamefont {X.~J.}\ \bibnamefont {Zhou}}, \bibinfo {author}
  {\bibfnamefont {C.}~\bibnamefont {Kim}}, \bibinfo {author} {\bibfnamefont
  {F.}~\bibnamefont {Ronning}}, \bibinfo {author} {\bibfnamefont {N.~P.}\
  \bibnamefont {Armitage}},\ and\ \bibinfo {author} {\bibfnamefont {Z.-X.}\
  \bibnamefont {Shen}},\ }\bibfield  {title} {\bibinfo {title} {Electronic
  structure of the trilayer cuprate superconductor
  \text{${\mathrm{Bi}}_{2}{\mathrm{Sr}}_{2}{\mathrm{Ca}}_{2}{\mathrm{Cu}}_{3}{O}_{10+\mathit{\ensuremath{\delta}}}$}},\
  }\href {https://doi.org/10.1103/PhysRevLett.88.107001} {\bibfield  {journal}
  {\bibinfo  {journal} {Phys. Rev. Lett.}\ }\textbf {\bibinfo {volume} {88}},\
  \bibinfo {pages} {107001} (\bibinfo {year} {2002})}\BibitemShut {NoStop}%
\bibitem [{\citenamefont {Campuzano}\ \emph {et~al.}(1999)\citenamefont
  {Campuzano}, \citenamefont {Ding}, \citenamefont {Norman}, \citenamefont
  {Fretwell}, \citenamefont {Randeria}, \citenamefont {Kaminski}, \citenamefont
  {Mesot}, \citenamefont {Takeuchi}, \citenamefont {Sato}, \citenamefont
  {Yokoya}, \citenamefont {Takahashi}, \citenamefont {Mochiku}, \citenamefont
  {Kadowaki}, \citenamefont {Guptasarma}, \citenamefont {Hinks}, \citenamefont
  {Konstantinovic}, \citenamefont {Li},\ and\ \citenamefont
  {Raffy}}]{Campuzano:Resonance}%
  \BibitemOpen
  \bibfield  {author} {\bibinfo {author} {\bibfnamefont {J.~C.}\ \bibnamefont
  {Campuzano}}, \bibinfo {author} {\bibfnamefont {H.}~\bibnamefont {Ding}},
  \bibinfo {author} {\bibfnamefont {M.~R.}\ \bibnamefont {Norman}}, \bibinfo
  {author} {\bibfnamefont {H.~M.}\ \bibnamefont {Fretwell}}, \bibinfo {author}
  {\bibfnamefont {M.}~\bibnamefont {Randeria}}, \bibinfo {author}
  {\bibfnamefont {A.}~\bibnamefont {Kaminski}}, \bibinfo {author}
  {\bibfnamefont {J.}~\bibnamefont {Mesot}}, \bibinfo {author} {\bibfnamefont
  {T.}~\bibnamefont {Takeuchi}}, \bibinfo {author} {\bibfnamefont
  {T.}~\bibnamefont {Sato}}, \bibinfo {author} {\bibfnamefont {T.}~\bibnamefont
  {Yokoya}}, \bibinfo {author} {\bibfnamefont {T.}~\bibnamefont {Takahashi}},
  \bibinfo {author} {\bibfnamefont {T.}~\bibnamefont {Mochiku}}, \bibinfo
  {author} {\bibfnamefont {K.}~\bibnamefont {Kadowaki}}, \bibinfo {author}
  {\bibfnamefont {P.}~\bibnamefont {Guptasarma}}, \bibinfo {author}
  {\bibfnamefont {D.~G.}\ \bibnamefont {Hinks}}, \bibinfo {author}
  {\bibfnamefont {Z.}~\bibnamefont {Konstantinovic}}, \bibinfo {author}
  {\bibfnamefont {Z.~Z.}\ \bibnamefont {Li}},\ and\ \bibinfo {author}
  {\bibfnamefont {H.}~\bibnamefont {Raffy}},\ }\bibfield  {title} {\bibinfo
  {title} {Electronic spectra and their relation to the
  $(\mathit{\ensuremath{\pi}},\mathit{\ensuremath{\pi}})$ collective mode in
  high- ${T}_{c}$ superconductors},\ }\href
  {https://doi.org/10.1103/PhysRevLett.83.3709} {\bibfield  {journal} {\bibinfo
   {journal} {Phys. Rev. Lett.}\ }\textbf {\bibinfo {volume} {83}},\ \bibinfo
  {pages} {3709} (\bibinfo {year} {1999})}\BibitemShut {NoStop}%
\bibitem [{\citenamefont {Cuk}\ \emph {et~al.}(2004)\citenamefont {Cuk},
  \citenamefont {Baumberger}, \citenamefont {Lu}, \citenamefont {Ingle},
  \citenamefont {Zhou}, \citenamefont {Eisaki}, \citenamefont {Kaneko},
  \citenamefont {Hussain}, \citenamefont {Devereaux}, \citenamefont {Nagaosa},\
  and\ \citenamefont {Shen}}]{Cuk:B1g_phononAntinode}%
  \BibitemOpen
  \bibfield  {author} {\bibinfo {author} {\bibfnamefont {T.}~\bibnamefont
  {Cuk}}, \bibinfo {author} {\bibfnamefont {F.}~\bibnamefont {Baumberger}},
  \bibinfo {author} {\bibfnamefont {D.~H.}\ \bibnamefont {Lu}}, \bibinfo
  {author} {\bibfnamefont {N.}~\bibnamefont {Ingle}}, \bibinfo {author}
  {\bibfnamefont {X.~J.}\ \bibnamefont {Zhou}}, \bibinfo {author}
  {\bibfnamefont {H.}~\bibnamefont {Eisaki}}, \bibinfo {author} {\bibfnamefont
  {N.}~\bibnamefont {Kaneko}}, \bibinfo {author} {\bibfnamefont
  {Z.}~\bibnamefont {Hussain}}, \bibinfo {author} {\bibfnamefont {T.~P.}\
  \bibnamefont {Devereaux}}, \bibinfo {author} {\bibfnamefont {N.}~\bibnamefont
  {Nagaosa}},\ and\ \bibinfo {author} {\bibfnamefont {Z.-X.}\ \bibnamefont
  {Shen}},\ }\bibfield  {title} {\bibinfo {title} {Coupling of the ${B}_{1g}$
  phonon to the antinodal electronic states of
  \text{${\mathrm{B}\mathrm{i}}_{2}{\mathrm{S}\mathrm{r}}_{2}{\mathrm{C}\mathrm{a}}_{0.92}{\mathrm{Y}}_{0.08}{\mathrm{C}\mathrm{u}}_{2}{\mathrm{O}}_{8+\ensuremath{\delta}}$}},\
  }\href {https://doi.org/10.1103/PhysRevLett.93.117003} {\bibfield  {journal}
  {\bibinfo  {journal} {Phys. Rev. Lett.}\ }\textbf {\bibinfo {volume} {93}},\
  \bibinfo {pages} {117003} (\bibinfo {year} {2004})}\BibitemShut {NoStop}%
\bibitem [{\citenamefont {He}\ \emph {et~al.}(2018)\citenamefont {He},
  \citenamefont {Hashimoto}, \citenamefont {Song}, \citenamefont {Chen},
  \citenamefont {He}, \citenamefont {Vishik}, \citenamefont {Moritz},
  \citenamefont {Lee}, \citenamefont {Nagaosa}, \citenamefont {Zaanen},
  \citenamefont {Devereaux}, \citenamefont {Yoshida}, \citenamefont {Eisaki},
  \citenamefont {Lu},\ and\ \citenamefont {Shen}}]{He:RapidBi2212_Science}%
  \BibitemOpen
  \bibfield  {author} {\bibinfo {author} {\bibfnamefont {Y.}~\bibnamefont
  {He}}, \bibinfo {author} {\bibfnamefont {M.}~\bibnamefont {Hashimoto}},
  \bibinfo {author} {\bibfnamefont {D.}~\bibnamefont {Song}}, \bibinfo {author}
  {\bibfnamefont {S.-D.}\ \bibnamefont {Chen}}, \bibinfo {author}
  {\bibfnamefont {J.}~\bibnamefont {He}}, \bibinfo {author} {\bibfnamefont
  {I.~M.}\ \bibnamefont {Vishik}}, \bibinfo {author} {\bibfnamefont
  {B.}~\bibnamefont {Moritz}}, \bibinfo {author} {\bibfnamefont {D.-H.}\
  \bibnamefont {Lee}}, \bibinfo {author} {\bibfnamefont {N.}~\bibnamefont
  {Nagaosa}}, \bibinfo {author} {\bibfnamefont {J.}~\bibnamefont {Zaanen}},
  \bibinfo {author} {\bibfnamefont {T.~P.}\ \bibnamefont {Devereaux}}, \bibinfo
  {author} {\bibfnamefont {Y.}~\bibnamefont {Yoshida}}, \bibinfo {author}
  {\bibfnamefont {H.}~\bibnamefont {Eisaki}}, \bibinfo {author} {\bibfnamefont
  {D.~H.}\ \bibnamefont {Lu}},\ and\ \bibinfo {author} {\bibfnamefont {Z.-X.}\
  \bibnamefont {Shen}},\ }\bibfield  {title} {\bibinfo {title} {Rapid change of
  superconductivity and electron-phonon coupling through critical doping in
  \text{Bi}-2212},\ }\href {https://doi.org/10.1126/science.aar3394} {\bibfield
   {journal} {\bibinfo  {journal} {Science}\ }\textbf {\bibinfo {volume}
  {362}},\ \bibinfo {pages} {62} (\bibinfo {year} {2018})}\BibitemShut
  {NoStop}%
\bibitem [{\citenamefont {Hashimoto}\ \emph
  {et~al.}(2014{\natexlab{b}})\citenamefont {Hashimoto}, \citenamefont
  {Nowadnick}, \citenamefont {He}, \citenamefont {Vishik}, \citenamefont
  {Moritz}, \citenamefont {He}, \citenamefont {Tanaka}, \citenamefont {Moore},
  \citenamefont {Lu}, \citenamefont {Yoshida}, \citenamefont {Ishikado},
  \citenamefont {Sasagawa}, \citenamefont {Fujita}, \citenamefont {Ishida},
  \citenamefont {Uchida}, \citenamefont {Eisaki}, \citenamefont {Hussain},
  \citenamefont {Devereaux},\ and\ \citenamefont
  {Shen}}]{Hashimoto:PhaseCompetition}%
  \BibitemOpen
  \bibfield  {author} {\bibinfo {author} {\bibfnamefont {M.}~\bibnamefont
  {Hashimoto}}, \bibinfo {author} {\bibfnamefont {E.~A.}\ \bibnamefont
  {Nowadnick}}, \bibinfo {author} {\bibfnamefont {R.-H.}\ \bibnamefont {He}},
  \bibinfo {author} {\bibfnamefont {I.~M.}\ \bibnamefont {Vishik}}, \bibinfo
  {author} {\bibfnamefont {B.}~\bibnamefont {Moritz}}, \bibinfo {author}
  {\bibfnamefont {Y.}~\bibnamefont {He}}, \bibinfo {author} {\bibfnamefont
  {K.}~\bibnamefont {Tanaka}}, \bibinfo {author} {\bibfnamefont {R.~G.}\
  \bibnamefont {Moore}}, \bibinfo {author} {\bibfnamefont {D.}~\bibnamefont
  {Lu}}, \bibinfo {author} {\bibfnamefont {Y.}~\bibnamefont {Yoshida}},
  \bibinfo {author} {\bibfnamefont {M.}~\bibnamefont {Ishikado}}, \bibinfo
  {author} {\bibfnamefont {T.}~\bibnamefont {Sasagawa}}, \bibinfo {author}
  {\bibfnamefont {K.}~\bibnamefont {Fujita}}, \bibinfo {author} {\bibfnamefont
  {S.}~\bibnamefont {Ishida}}, \bibinfo {author} {\bibfnamefont
  {S.}~\bibnamefont {Uchida}}, \bibinfo {author} {\bibfnamefont
  {H.}~\bibnamefont {Eisaki}}, \bibinfo {author} {\bibfnamefont
  {Z.}~\bibnamefont {Hussain}}, \bibinfo {author} {\bibfnamefont {T.~P.}\
  \bibnamefont {Devereaux}},\ and\ \bibinfo {author} {\bibfnamefont {Z.-X.}\
  \bibnamefont {Shen}},\ }\bibfield  {title} {\bibinfo {title} {Direct
  spectroscopic evidence for phase competition between the pseudogap and
  superconductivity in \text{Bi$_2$Sr$_2$CaCu$_2$O$_{8+\delta}$}},\ }\href
  {https://doi.org/10.1038/nmat4116
  https://www.nature.com/articles/nmat4116#supplementary-information}
  {\bibfield  {journal} {\bibinfo  {journal} {Nature Materials}\ }\textbf
  {\bibinfo {volume} {14}},\ \bibinfo {pages} {37} (\bibinfo {year}
  {2014}{\natexlab{b}})}\BibitemShut {NoStop}%
\bibitem [{\citenamefont {Shen}\ \emph {et~al.}(2007)\citenamefont {Shen},
  \citenamefont {Ronning}, \citenamefont {Meevasana}, \citenamefont {Lu},
  \citenamefont {Ingle}, \citenamefont {Baumberger}, \citenamefont {Lee},
  \citenamefont {Miller}, \citenamefont {Kohsaka}, \citenamefont {Azuma},
  \citenamefont {Takano}, \citenamefont {Takagi},\ and\ \citenamefont
  {Shen}}]{Shen:LatticePolaronFormation}%
  \BibitemOpen
  \bibfield  {author} {\bibinfo {author} {\bibfnamefont {K.~M.}\ \bibnamefont
  {Shen}}, \bibinfo {author} {\bibfnamefont {F.}~\bibnamefont {Ronning}},
  \bibinfo {author} {\bibfnamefont {W.}~\bibnamefont {Meevasana}}, \bibinfo
  {author} {\bibfnamefont {D.~H.}\ \bibnamefont {Lu}}, \bibinfo {author}
  {\bibfnamefont {N.~J.~C.}\ \bibnamefont {Ingle}}, \bibinfo {author}
  {\bibfnamefont {F.}~\bibnamefont {Baumberger}}, \bibinfo {author}
  {\bibfnamefont {W.~S.}\ \bibnamefont {Lee}}, \bibinfo {author} {\bibfnamefont
  {L.~L.}\ \bibnamefont {Miller}}, \bibinfo {author} {\bibfnamefont
  {Y.}~\bibnamefont {Kohsaka}}, \bibinfo {author} {\bibfnamefont
  {M.}~\bibnamefont {Azuma}}, \bibinfo {author} {\bibfnamefont
  {M.}~\bibnamefont {Takano}}, \bibinfo {author} {\bibfnamefont
  {H.}~\bibnamefont {Takagi}},\ and\ \bibinfo {author} {\bibfnamefont {Z.-X.}\
  \bibnamefont {Shen}},\ }\bibfield  {title} {\bibinfo {title} {Angle-resolved
  photoemission studies of lattice polaron formation in the cuprate
  \text{${\mathrm{Ca}}_{2}\mathrm{Cu}{\mathrm{O}}_{2}{\mathrm{Cl}}_{2}$}},\
  }\href {https://doi.org/10.1103/PhysRevB.75.075115} {\bibfield  {journal}
  {\bibinfo  {journal} {Phys. Rev. B}\ }\textbf {\bibinfo {volume} {75}},\
  \bibinfo {pages} {075115} (\bibinfo {year} {2007})}\BibitemShut {NoStop}%
\bibitem [{\citenamefont {Vishik}\ \emph {et~al.}(2012)\citenamefont {Vishik},
  \citenamefont {Hashimoto}, \citenamefont {He}, \citenamefont {Lee},
  \citenamefont {Schmitt}, \citenamefont {Lu}, \citenamefont {Moore},
  \citenamefont {Zhang}, \citenamefont {Meevasana}, \citenamefont {Sasagawa},
  \citenamefont {Uchida}, \citenamefont {Fujita}, \citenamefont {Ishida},
  \citenamefont {Ishikado}, \citenamefont {Yoshida}, \citenamefont {Eisaki},
  \citenamefont {Hussain}, \citenamefont {Devereaux},\ and\ \citenamefont
  {Shen}}]{Vishik:PNAS}%
  \BibitemOpen
  \bibfield  {author} {\bibinfo {author} {\bibfnamefont {I.~M.}\ \bibnamefont
  {Vishik}}, \bibinfo {author} {\bibfnamefont {M.}~\bibnamefont {Hashimoto}},
  \bibinfo {author} {\bibfnamefont {R.-H.}\ \bibnamefont {He}}, \bibinfo
  {author} {\bibfnamefont {W.-S.}\ \bibnamefont {Lee}}, \bibinfo {author}
  {\bibfnamefont {F.}~\bibnamefont {Schmitt}}, \bibinfo {author} {\bibfnamefont
  {D.}~\bibnamefont {Lu}}, \bibinfo {author} {\bibfnamefont {R.~G.}\
  \bibnamefont {Moore}}, \bibinfo {author} {\bibfnamefont {C.}~\bibnamefont
  {Zhang}}, \bibinfo {author} {\bibfnamefont {W.}~\bibnamefont {Meevasana}},
  \bibinfo {author} {\bibfnamefont {T.}~\bibnamefont {Sasagawa}}, \bibinfo
  {author} {\bibfnamefont {S.}~\bibnamefont {Uchida}}, \bibinfo {author}
  {\bibfnamefont {K.}~\bibnamefont {Fujita}}, \bibinfo {author} {\bibfnamefont
  {S.}~\bibnamefont {Ishida}}, \bibinfo {author} {\bibfnamefont
  {M.}~\bibnamefont {Ishikado}}, \bibinfo {author} {\bibfnamefont
  {Y.}~\bibnamefont {Yoshida}}, \bibinfo {author} {\bibfnamefont
  {H.}~\bibnamefont {Eisaki}}, \bibinfo {author} {\bibfnamefont
  {Z.}~\bibnamefont {Hussain}}, \bibinfo {author} {\bibfnamefont {T.~P.}\
  \bibnamefont {Devereaux}},\ and\ \bibinfo {author} {\bibfnamefont {Z.-X.}\
  \bibnamefont {Shen}},\ }\bibfield  {title} {\bibinfo {title} {Phase
  competition in trisected superconducting dome},\ }\href
  {https://doi.org/10.1073/pnas.1209471109} {\bibfield  {journal} {\bibinfo
  {journal} {Proceedings of the National Academy of Sciences}\ }\textbf
  {\bibinfo {volume} {109}},\ \bibinfo {pages} {18332} (\bibinfo {year}
  {2012})}\BibitemShut {NoStop}%
\bibitem [{\citenamefont {Drachuck}\ \emph {et~al.}(2014)\citenamefont
  {Drachuck}, \citenamefont {Razzoli}, \citenamefont {Bazalitski},
  \citenamefont {Kanigel}, \citenamefont {Niedermayer}, \citenamefont {Shi},\
  and\ \citenamefont {Keren}}]{Drachuk:NodalGapSpin}%
  \BibitemOpen
  \bibfield  {author} {\bibinfo {author} {\bibfnamefont {G.}~\bibnamefont
  {Drachuck}}, \bibinfo {author} {\bibfnamefont {E.}~\bibnamefont {Razzoli}},
  \bibinfo {author} {\bibfnamefont {G.}~\bibnamefont {Bazalitski}}, \bibinfo
  {author} {\bibfnamefont {A.}~\bibnamefont {Kanigel}}, \bibinfo {author}
  {\bibfnamefont {C.}~\bibnamefont {Niedermayer}}, \bibinfo {author}
  {\bibfnamefont {M.}~\bibnamefont {Shi}},\ and\ \bibinfo {author}
  {\bibfnamefont {A.}~\bibnamefont {Keren}},\ }\bibfield  {title} {\bibinfo
  {title} {Comprehensive study of the spin-charge interplay in
  antiferromagnetic la$_{2−x}$sr$_x$cuo$_4$},\ }\href
  {https://doi.org/10.1038/ncomms4390} {\bibfield  {journal} {\bibinfo
  {journal} {Nature Communications}\ }\textbf {\bibinfo {volume} {5}},\
  \bibinfo {pages} {3390} (\bibinfo {year} {2014})}\BibitemShut {NoStop}%
\bibitem [{\citenamefont {Ino}\ \emph {et~al.}(1997)\citenamefont {Ino},
  \citenamefont {Mizokawa}, \citenamefont {Fujimori}, \citenamefont {Tamasaku},
  \citenamefont {Eisaki}, \citenamefont {Uchida}, \citenamefont {Kimura},
  \citenamefont {Sasagawa},\ and\ \citenamefont
  {Kishio}}]{Ino:ChemicalPotentialShiftLSCO}%
  \BibitemOpen
  \bibfield  {author} {\bibinfo {author} {\bibfnamefont {A.}~\bibnamefont
  {Ino}}, \bibinfo {author} {\bibfnamefont {T.}~\bibnamefont {Mizokawa}},
  \bibinfo {author} {\bibfnamefont {A.}~\bibnamefont {Fujimori}}, \bibinfo
  {author} {\bibfnamefont {K.}~\bibnamefont {Tamasaku}}, \bibinfo {author}
  {\bibfnamefont {H.}~\bibnamefont {Eisaki}}, \bibinfo {author} {\bibfnamefont
  {S.}~\bibnamefont {Uchida}}, \bibinfo {author} {\bibfnamefont
  {T.}~\bibnamefont {Kimura}}, \bibinfo {author} {\bibfnamefont
  {T.}~\bibnamefont {Sasagawa}},\ and\ \bibinfo {author} {\bibfnamefont
  {K.}~\bibnamefont {Kishio}},\ }\bibfield  {title} {\bibinfo {title} {Chemical
  potential shift in overdoped and underdoped
  \text{${\mathrm{La}}_{2\ensuremath{-}\mathit{x}}{\mathrm{Sr}}_{\mathit{x}}{\mathrm{CuO}}_{4}$}},\
  }\href {https://doi.org/10.1103/PhysRevLett.79.2101} {\bibfield  {journal}
  {\bibinfo  {journal} {Phys. Rev. Lett.}\ }\textbf {\bibinfo {volume} {79}},\
  \bibinfo {pages} {2101} (\bibinfo {year} {1997})}\BibitemShut {NoStop}%
\bibitem [{\citenamefont {Michon}\ \emph {et~al.}(2019)\citenamefont {Michon},
  \citenamefont {Girod}, \citenamefont {Badoux}, \citenamefont {Kačmarčík},
  \citenamefont {Ma}, \citenamefont {Dragomir}, \citenamefont {Dabkowska},
  \citenamefont {Gaulin}, \citenamefont {Zhou}, \citenamefont {Pyon},
  \citenamefont {Takayama}, \citenamefont {Takagi}, \citenamefont {Verret},
  \citenamefont {Doiron-Leyraud}, \citenamefont {Marcenat}, \citenamefont
  {Taillefer},\ and\ \citenamefont {Klein}}]{Michon:quantumCriticality}%
  \BibitemOpen
  \bibfield  {author} {\bibinfo {author} {\bibfnamefont {B.}~\bibnamefont
  {Michon}}, \bibinfo {author} {\bibfnamefont {C.}~\bibnamefont {Girod}},
  \bibinfo {author} {\bibfnamefont {S.}~\bibnamefont {Badoux}}, \bibinfo
  {author} {\bibfnamefont {J.}~\bibnamefont {Kačmarčík}}, \bibinfo {author}
  {\bibfnamefont {Q.}~\bibnamefont {Ma}}, \bibinfo {author} {\bibfnamefont
  {M.}~\bibnamefont {Dragomir}}, \bibinfo {author} {\bibfnamefont {H.~A.}\
  \bibnamefont {Dabkowska}}, \bibinfo {author} {\bibfnamefont {B.~D.}\
  \bibnamefont {Gaulin}}, \bibinfo {author} {\bibfnamefont {J.~S.}\
  \bibnamefont {Zhou}}, \bibinfo {author} {\bibfnamefont {S.}~\bibnamefont
  {Pyon}}, \bibinfo {author} {\bibfnamefont {T.}~\bibnamefont {Takayama}},
  \bibinfo {author} {\bibfnamefont {H.}~\bibnamefont {Takagi}}, \bibinfo
  {author} {\bibfnamefont {S.}~\bibnamefont {Verret}}, \bibinfo {author}
  {\bibfnamefont {N.}~\bibnamefont {Doiron-Leyraud}}, \bibinfo {author}
  {\bibfnamefont {C.}~\bibnamefont {Marcenat}}, \bibinfo {author}
  {\bibfnamefont {L.}~\bibnamefont {Taillefer}},\ and\ \bibinfo {author}
  {\bibfnamefont {T.}~\bibnamefont {Klein}},\ }\bibfield  {title} {\bibinfo
  {title} {Thermodynamic signatures of quantum criticality in cuprate
  superconductors},\ }\href {https://doi.org/10.1038/s41586-019-0932-x}
  {\bibfield  {journal} {\bibinfo  {journal} {Nature}\ }\textbf {\bibinfo
  {volume} {567}},\ \bibinfo {pages} {218} (\bibinfo {year}
  {2019})}\BibitemShut {NoStop}%
\bibitem [{\citenamefont {Pelc}\ \emph
  {et~al.}(2019{\natexlab{a}})\citenamefont {Pelc}, \citenamefont {Pop{\v
  c}evi{\'c}}, \citenamefont {Po{\v z}ek}, \citenamefont {Greven},\ and\
  \citenamefont {Bari{\v s}i{\'c}}}]{Pelc:heterogeneousBehavior}%
  \BibitemOpen
  \bibfield  {author} {\bibinfo {author} {\bibfnamefont {D.}~\bibnamefont
  {Pelc}}, \bibinfo {author} {\bibfnamefont {P.}~\bibnamefont {Pop{\v
  c}evi{\'c}}}, \bibinfo {author} {\bibfnamefont {M.}~\bibnamefont {Po{\v
  z}ek}}, \bibinfo {author} {\bibfnamefont {M.}~\bibnamefont {Greven}},\ and\
  \bibinfo {author} {\bibfnamefont {N.}~\bibnamefont {Bari{\v s}i{\'c}}},\
  }\bibfield  {title} {\bibinfo {title} {Unusual behavior of cuprates explained
  by heterogeneous charge localization},\ }\bibfield  {journal} {\bibinfo
  {journal} {Science Advances}\ }\textbf {\bibinfo {volume} {5}},\ \href
  {https://doi.org/10.1126/sciadv.aau4538} {10.1126/sciadv.aau4538} (\bibinfo
  {year} {2019}{\natexlab{a}})\BibitemShut {NoStop}%
\bibitem [{\citenamefont {Pelc}\ \emph
  {et~al.}(2019{\natexlab{b}})\citenamefont {Pelc}, \citenamefont {Veit},
  \citenamefont {Chan}, \citenamefont {Dorow}, \citenamefont {Ge},
  \citenamefont {Bari{\v{s}}i{\'c}},\ and\ \citenamefont
  {Greven}}]{pelc2019resistivity}%
  \BibitemOpen
  \bibfield  {author} {\bibinfo {author} {\bibfnamefont {D.}~\bibnamefont
  {Pelc}}, \bibinfo {author} {\bibfnamefont {M.}~\bibnamefont {Veit}}, \bibinfo
  {author} {\bibfnamefont {M.}~\bibnamefont {Chan}}, \bibinfo {author}
  {\bibfnamefont {C.}~\bibnamefont {Dorow}}, \bibinfo {author} {\bibfnamefont
  {Y.}~\bibnamefont {Ge}}, \bibinfo {author} {\bibfnamefont {N.}~\bibnamefont
  {Bari{\v{s}}i{\'c}}},\ and\ \bibinfo {author} {\bibfnamefont
  {M.}~\bibnamefont {Greven}},\ }\bibfield  {title} {\bibinfo {title} {The
  resistivity phase diagram of cuprates revisited},\ }\href@noop {} {\bibfield
  {journal} {\bibinfo  {journal} {arXiv preprint arXiv:1902.00529}\ } (\bibinfo
  {year} {2019}{\natexlab{b}})}\BibitemShut {NoStop}%
\bibitem [{\citenamefont {Krumhansl}(1992)}]{krumhansl1992fine}%
  \BibitemOpen
  \bibfield  {author} {\bibinfo {author} {\bibfnamefont {J.}~\bibnamefont
  {Krumhansl}},\ }\bibfield  {title} {\bibinfo {title} {Fine scale
  mesostructures in superconducting and other materials},\ }in\ \href@noop {}
  {\emph {\bibinfo {booktitle} {Lattice Effects in High-Tc Superconductors}}}\
  (\bibinfo  {publisher} {World Scientific},\ \bibinfo {year} {1992})\ pp.\
  \bibinfo {pages} {503--516}\BibitemShut {NoStop}%
\bibitem [{\citenamefont {Pelc}\ \emph
  {et~al.}(2019{\natexlab{c}})\citenamefont {Pelc}, \citenamefont {Anderson},
  \citenamefont {Yu}, \citenamefont {Leighton},\ and\ \citenamefont
  {Greven}}]{Pelc:UniversalPrecursor}%
  \BibitemOpen
  \bibfield  {author} {\bibinfo {author} {\bibfnamefont {D.}~\bibnamefont
  {Pelc}}, \bibinfo {author} {\bibfnamefont {Z.}~\bibnamefont {Anderson}},
  \bibinfo {author} {\bibfnamefont {B.}~\bibnamefont {Yu}}, \bibinfo {author}
  {\bibfnamefont {C.}~\bibnamefont {Leighton}},\ and\ \bibinfo {author}
  {\bibfnamefont {M.}~\bibnamefont {Greven}},\ }\bibfield  {title} {\bibinfo
  {title} {Universal superconducting precursor in three classes of
  unconventional superconductors},\ }\href
  {https://doi.org/10.1038/s41467-019-10635-w} {\bibfield  {journal} {\bibinfo
  {journal} {Nature Communications}\ }\textbf {\bibinfo {volume} {10}},\
  \bibinfo {pages} {2729} (\bibinfo {year} {2019}{\natexlab{c}})}\BibitemShut
  {NoStop}%
\bibitem [{\citenamefont {Zaki}\ \emph {et~al.}(2017)\citenamefont {Zaki},
  \citenamefont {Yang}, \citenamefont {Rameau}, \citenamefont {Johnson},
  \citenamefont {Claus},\ and\ \citenamefont
  {Hinks}}]{Zaki:CupratePhaseDiagram}%
  \BibitemOpen
  \bibfield  {author} {\bibinfo {author} {\bibfnamefont {N.}~\bibnamefont
  {Zaki}}, \bibinfo {author} {\bibfnamefont {H.-B.}\ \bibnamefont {Yang}},
  \bibinfo {author} {\bibfnamefont {J.~D.}\ \bibnamefont {Rameau}}, \bibinfo
  {author} {\bibfnamefont {P.~D.}\ \bibnamefont {Johnson}}, \bibinfo {author}
  {\bibfnamefont {H.}~\bibnamefont {Claus}},\ and\ \bibinfo {author}
  {\bibfnamefont {D.~G.}\ \bibnamefont {Hinks}},\ }\bibfield  {title} {\bibinfo
  {title} {Cuprate phase diagram and the influence of nanoscale
  inhomogeneities},\ }\href {https://doi.org/10.1103/PhysRevB.96.195163}
  {\bibfield  {journal} {\bibinfo  {journal} {Phys. Rev. B}\ }\textbf {\bibinfo
  {volume} {96}},\ \bibinfo {pages} {195163} (\bibinfo {year}
  {2017})}\BibitemShut {NoStop}%
\bibitem [{\citenamefont {Chen}\ \emph {et~al.}(2019)\citenamefont {Chen},
  \citenamefont {Dong},\ and\ \citenamefont {Li}}]{Chen:chargeFlux}%
  \BibitemOpen
  \bibfield  {author} {\bibinfo {author} {\bibfnamefont {X.}~\bibnamefont
  {Chen}}, \bibinfo {author} {\bibfnamefont {J.}~\bibnamefont {Dong}},\ and\
  \bibinfo {author} {\bibfnamefont {X.}~\bibnamefont {Li}},\ }\bibfield
  {title} {\bibinfo {title} {A picture of pseudogap phase related to charge
  fluxes},\ }\href@noop {} {\bibfield  {journal} {\bibinfo  {journal} {arXiv
  preprint arXiv:1912.11222}\ } (\bibinfo {year} {2019})}\BibitemShut {NoStop}%
\bibitem [{\citenamefont {Dal~Conte}\ \emph {et~al.}(2012)\citenamefont
  {Dal~Conte}, \citenamefont {Giannetti}, \citenamefont {Coslovich},
  \citenamefont {Cilento}, \citenamefont {Bossini}, \citenamefont {Abebaw},
  \citenamefont {Banfi}, \citenamefont {Ferrini}, \citenamefont {Eisaki},
  \citenamefont {Greven}, \citenamefont {Damascelli}, \citenamefont {van~der
  Marel},\ and\ \citenamefont
  {Parmigiani}}]{DalConte:DisentanglingElectronicPhononic}%
  \BibitemOpen
  \bibfield  {author} {\bibinfo {author} {\bibfnamefont {S.}~\bibnamefont
  {Dal~Conte}}, \bibinfo {author} {\bibfnamefont {C.}~\bibnamefont
  {Giannetti}}, \bibinfo {author} {\bibfnamefont {G.}~\bibnamefont
  {Coslovich}}, \bibinfo {author} {\bibfnamefont {F.}~\bibnamefont {Cilento}},
  \bibinfo {author} {\bibfnamefont {D.}~\bibnamefont {Bossini}}, \bibinfo
  {author} {\bibfnamefont {T.}~\bibnamefont {Abebaw}}, \bibinfo {author}
  {\bibfnamefont {F.}~\bibnamefont {Banfi}}, \bibinfo {author} {\bibfnamefont
  {G.}~\bibnamefont {Ferrini}}, \bibinfo {author} {\bibfnamefont
  {H.}~\bibnamefont {Eisaki}}, \bibinfo {author} {\bibfnamefont
  {M.}~\bibnamefont {Greven}}, \bibinfo {author} {\bibfnamefont
  {A.}~\bibnamefont {Damascelli}}, \bibinfo {author} {\bibfnamefont
  {D.}~\bibnamefont {van~der Marel}},\ and\ \bibinfo {author} {\bibfnamefont
  {F.}~\bibnamefont {Parmigiani}},\ }\bibfield  {title} {\bibinfo {title}
  {Disentangling the electronic and phononic glue in a high-t$_c$
  superconductor},\ }\href {https://doi.org/10.1126/science.1216765} {\bibfield
   {journal} {\bibinfo  {journal} {Science}\ }\textbf {\bibinfo {volume}
  {335}},\ \bibinfo {pages} {1600} (\bibinfo {year} {2012})}\BibitemShut
  {NoStop}%
\end{thebibliography}%


\begin{thebibliography}{13}%
\makeatletter
\providecommand \@ifxundefined [1]{%
 \@ifx{#1\undefined}
}%
\providecommand \@ifnum [1]{%
 \ifnum #1\expandafter \@firstoftwo
 \else \expandafter \@secondoftwo
 \fi
}%
\providecommand \@ifx [1]{%
 \ifx #1\expandafter \@firstoftwo
 \else \expandafter \@secondoftwo
 \fi
}%
\providecommand \natexlab [1]{#1}%
\providecommand \enquote  [1]{``#1''}%
\providecommand \bibnamefont  [1]{#1}%
\providecommand \bibfnamefont [1]{#1}%
\providecommand \citenamefont [1]{#1}%
\providecommand \href@noop [0]{\@secondoftwo}%
\providecommand \href [0]{\begingroup \@sanitize@url \@href}%
\providecommand \@href[1]{\@@startlink{#1}\@@href}%
\providecommand \@@href[1]{\endgroup#1\@@endlink}%
\providecommand \@sanitize@url [0]{\catcode `\\12\catcode `\$12\catcode
  `\&12\catcode `\#12\catcode `\^12\catcode `\_12\catcode `\%12\relax}%
\providecommand \@@startlink[1]{}%
\providecommand \@@endlink[0]{}%
\providecommand \url  [0]{\begingroup\@sanitize@url \@url }%
\providecommand \@url [1]{\endgroup\@href {#1}{\urlprefix }}%
\providecommand \urlprefix  [0]{URL }%
\providecommand \Eprint [0]{\href }%
\providecommand \doibase [0]{http://dx.doi.org/}%
\providecommand \selectlanguage [0]{\@gobble}%
\providecommand \bibinfo  [0]{\@secondoftwo}%
\providecommand \bibfield  [0]{\@secondoftwo}%
\providecommand \translation [1]{[#1]}%
\providecommand \BibitemOpen [0]{}%
\providecommand \bibitemStop [0]{}%
\providecommand \bibitemNoStop [0]{.\EOS\space}%
\providecommand \EOS [0]{\spacefactor3000\relax}%
\providecommand \BibitemShut  [1]{\csname bibitem#1\endcsname}%
\let\auto@bib@innerbib\@empty
\bibitem [{\citenamefont {Knox}\ \emph {et~al.}(2011)\citenamefont {Knox},
  \citenamefont {Locatelli}, \citenamefont {Yilmaz}, \citenamefont {Cvetko},
  \citenamefont {Mente\ifmmode~\mbox{\c{s}}\else \c{s}\fi{}}, \citenamefont
  {Ni\~no}, \citenamefont {Kim}, \citenamefont {Morgante},\ and\ \citenamefont
  {Osgood}}]{KRKnox:SuspendedGraphene}%
  \BibitemOpen
  \bibfield  {author} {\bibinfo {author} {\bibfnamefont {K.~R.}\ \bibnamefont
  {Knox}}, \bibinfo {author} {\bibfnamefont {A.}~\bibnamefont {Locatelli}},
  \bibinfo {author} {\bibfnamefont {M.~B.}\ \bibnamefont {Yilmaz}}, \bibinfo
  {author} {\bibfnamefont {D.}~\bibnamefont {Cvetko}}, \bibinfo {author}
  {\bibfnamefont {T.~O.}\ \bibnamefont {Mente\ifmmode~\mbox{\c{s}}\else
  \c{s}\fi{}}}, \bibinfo {author} {\bibfnamefont {M.~A.}\ \bibnamefont
  {Ni\~no}}, \bibinfo {author} {\bibfnamefont {P.}~\bibnamefont {Kim}},
  \bibinfo {author} {\bibfnamefont {A.}~\bibnamefont {Morgante}}, \ and\
  \bibinfo {author} {\bibfnamefont {R.~M.}\ \bibnamefont {Osgood}},\ }\href
  {\doibase 10.1103/PhysRevB.84.115401} {\bibfield  {journal} {\bibinfo
  {journal} {Phys. Rev. B}\ }\textbf {\bibinfo {volume} {84}},\ \bibinfo
  {pages} {115401} (\bibinfo {year} {2011})}\BibitemShut {NoStop}%
\bibitem [{\citenamefont {Fauster}\ \emph {et~al.}(2000)\citenamefont
  {Fauster}, \citenamefont {Reu\ss{}}, \citenamefont {Shumay}, \citenamefont
  {Weinelt}, \citenamefont {Theilmann},\ and\ \citenamefont
  {Goldmann}}]{Fauster:MorphologyCu}%
  \BibitemOpen
  \bibfield  {author} {\bibinfo {author} {\bibfnamefont {T.}~\bibnamefont
  {Fauster}}, \bibinfo {author} {\bibfnamefont {C.}~\bibnamefont {Reu\ss{}}},
  \bibinfo {author} {\bibfnamefont {I.~L.}\ \bibnamefont {Shumay}}, \bibinfo
  {author} {\bibfnamefont {M.}~\bibnamefont {Weinelt}}, \bibinfo {author}
  {\bibfnamefont {F.}~\bibnamefont {Theilmann}}, \ and\ \bibinfo {author}
  {\bibfnamefont {A.}~\bibnamefont {Goldmann}},\ }\href {\doibase
  10.1103/PhysRevB.61.16168} {\bibfield  {journal} {\bibinfo  {journal} {Phys.
  Rev. B}\ }\textbf {\bibinfo {volume} {61}},\ \bibinfo {pages} {16168}
  (\bibinfo {year} {2000})}\BibitemShut {NoStop}%
\bibitem [{\citenamefont {Bari\ifmmode \check{s}\else
  \v{s}\fi{}i\ifmmode~\acute{c}\else \'{c}\fi{}}\ \emph
  {et~al.}(2008)\citenamefont {Bari\ifmmode \check{s}\else
  \v{s}\fi{}i\ifmmode~\acute{c}\else \'{c}\fi{}}, \citenamefont {Li},
  \citenamefont {Zhao}, \citenamefont {Cho}, \citenamefont {Chabot-Couture},
  \citenamefont {Yu},\ and\ \citenamefont
  {Greven}}]{Barisic:DemonstratingModelNature}%
  \BibitemOpen
  \bibfield  {author} {\bibinfo {author} {\bibfnamefont {N.}~\bibnamefont
  {Bari\ifmmode \check{s}\else \v{s}\fi{}i\ifmmode~\acute{c}\else \'{c}\fi{}}},
  \bibinfo {author} {\bibfnamefont {Y.}~\bibnamefont {Li}}, \bibinfo {author}
  {\bibfnamefont {X.}~\bibnamefont {Zhao}}, \bibinfo {author} {\bibfnamefont
  {Y.-C.}\ \bibnamefont {Cho}}, \bibinfo {author} {\bibfnamefont
  {G.}~\bibnamefont {Chabot-Couture}}, \bibinfo {author} {\bibfnamefont
  {G.}~\bibnamefont {Yu}}, \ and\ \bibinfo {author} {\bibfnamefont
  {M.}~\bibnamefont {Greven}},\ }\href {\doibase 10.1103/PhysRevB.78.054518}
  {\bibfield  {journal} {\bibinfo  {journal} {Phys. Rev. B}\ }\textbf {\bibinfo
  {volume} {78}},\ \bibinfo {pages} {054518} (\bibinfo {year}
  {2008})}\BibitemShut {NoStop}%
\bibitem [{\citenamefont {Seah}\ and\ \citenamefont
  {Dench}(1979)}]{UniversalCurve}%
  \BibitemOpen
  \bibfield  {author} {\bibinfo {author} {\bibfnamefont {M.~P.}\ \bibnamefont
  {Seah}}\ and\ \bibinfo {author} {\bibfnamefont {W.~A.}\ \bibnamefont
  {Dench}},\ }\href {\doibase 10.1002/sia.740010103} {\bibfield  {journal}
  {\bibinfo  {journal} {Surface and Interface Analysis}\ }\textbf {\bibinfo
  {volume} {1}},\ \bibinfo {pages} {2} (\bibinfo {year} {1979})}\BibitemShut
  {NoStop}%
\bibitem [{\citenamefont {Merzlikin}\ \emph {et~al.}(2008)\citenamefont
  {Merzlikin}, \citenamefont {Tolkachev}, \citenamefont {Strunskus},
  \citenamefont {Witte}, \citenamefont {{Glogowski}}, \citenamefont
  {W{\"o}ll},\ and\ \citenamefont
  {Gr{\"u}nert}}]{Merzlikin:ResolvingDepthCoordinate}%
  \BibitemOpen
  \bibfield  {author} {\bibinfo {author} {\bibfnamefont {S.~V.}\ \bibnamefont
  {Merzlikin}}, \bibinfo {author} {\bibfnamefont {N.~N.}\ \bibnamefont
  {Tolkachev}}, \bibinfo {author} {\bibfnamefont {T.}~\bibnamefont
  {Strunskus}}, \bibinfo {author} {\bibfnamefont {G.}~\bibnamefont {Witte}},
  \bibinfo {author} {\bibfnamefont {T.}~\bibnamefont {{Glogowski}}}, \bibinfo
  {author} {\bibfnamefont {C.}~\bibnamefont {W{\"o}ll}}, \ and\ \bibinfo
  {author} {\bibfnamefont {W.}~\bibnamefont {Gr{\"u}nert}},\ }\href {\doibase
  10.1016/j.susc.2007.12.005} {\bibfield  {journal} {\bibinfo  {journal}
  {Surface Science}\ }\textbf {\bibinfo {volume} {602}},\ \bibinfo {pages}
  {755} (\bibinfo {year} {2008})}\BibitemShut {NoStop}%
\bibitem [{\citenamefont {Matthew}(2004)}]{matthew2004surface}%
  \BibitemOpen
  \bibfield  {author} {\bibinfo {author} {\bibfnamefont {J.}~\bibnamefont
  {Matthew}},\ }\href@noop {} {\bibfield  {journal} {\bibinfo  {journal}
  {Surface and Interface Analysis: An International Journal devoted to the
  development and application of techniques for the analysis of surfaces,
  interfaces and thin films}\ }\textbf {\bibinfo {volume} {36}},\ \bibinfo
  {pages} {1647} (\bibinfo {year} {2004})}\BibitemShut {NoStop}%
\bibitem [{\citenamefont {Vishik}\ \emph {et~al.}(2014)\citenamefont {Vishik},
  \citenamefont {Bari\ifmmode \check{s}\else \v{s}\fi{}i\ifmmode~\acute{c}\else
  \'{c}\fi{}}, \citenamefont {Chan}, \citenamefont {Li}, \citenamefont {Xia},
  \citenamefont {Yu}, \citenamefont {Zhao}, \citenamefont {Lee}, \citenamefont
  {Meevasana}, \citenamefont {Devereaux}, \citenamefont {Greven},\ and\
  \citenamefont {Shen}}]{Inna:PRBHg1201}%
  \BibitemOpen
  \bibfield  {author} {\bibinfo {author} {\bibfnamefont {I.~M.}\ \bibnamefont
  {Vishik}}, \bibinfo {author} {\bibfnamefont {N.}~\bibnamefont {Bari\ifmmode
  \check{s}\else \v{s}\fi{}i\ifmmode~\acute{c}\else \'{c}\fi{}}}, \bibinfo
  {author} {\bibfnamefont {M.~K.}\ \bibnamefont {Chan}}, \bibinfo {author}
  {\bibfnamefont {Y.}~\bibnamefont {Li}}, \bibinfo {author} {\bibfnamefont
  {D.~D.}\ \bibnamefont {Xia}}, \bibinfo {author} {\bibfnamefont
  {G.}~\bibnamefont {Yu}}, \bibinfo {author} {\bibfnamefont {X.}~\bibnamefont
  {Zhao}}, \bibinfo {author} {\bibfnamefont {W.~S.}\ \bibnamefont {Lee}},
  \bibinfo {author} {\bibfnamefont {W.}~\bibnamefont {Meevasana}}, \bibinfo
  {author} {\bibfnamefont {T.~P.}\ \bibnamefont {Devereaux}}, \bibinfo {author}
  {\bibfnamefont {M.}~\bibnamefont {Greven}}, \ and\ \bibinfo {author}
  {\bibfnamefont {Z.-X.}\ \bibnamefont {Shen}},\ }\href {\doibase
  10.1103/PhysRevB.89.195141} {\bibfield  {journal} {\bibinfo  {journal} {Phys.
  Rev. B}\ }\textbf {\bibinfo {volume} {89}},\ \bibinfo {pages} {195141}
  (\bibinfo {year} {2014})}\BibitemShut {NoStop}%
\bibitem [{\citenamefont {Hossain}\ \emph {et~al.}(2008)\citenamefont
  {Hossain}, \citenamefont {Mottershead}, \citenamefont {Fournier},
  \citenamefont {Bostwick}, \citenamefont {McChesney}, \citenamefont
  {Rotenberg}, \citenamefont {Liang}, \citenamefont {Hardy}, \citenamefont
  {Sawatzky}, \citenamefont {Elfimov}, \citenamefont {Bonn},\ and\
  \citenamefont {Damascelli}}]{Hossain:DopingControl}%
  \BibitemOpen
  \bibfield  {author} {\bibinfo {author} {\bibfnamefont {M.~A.}\ \bibnamefont
  {Hossain}}, \bibinfo {author} {\bibfnamefont {J.~D.~F.}\ \bibnamefont
  {Mottershead}}, \bibinfo {author} {\bibfnamefont {D.}~\bibnamefont
  {Fournier}}, \bibinfo {author} {\bibfnamefont {A.}~\bibnamefont {Bostwick}},
  \bibinfo {author} {\bibfnamefont {J.~L.}\ \bibnamefont {McChesney}}, \bibinfo
  {author} {\bibfnamefont {E.}~\bibnamefont {Rotenberg}}, \bibinfo {author}
  {\bibfnamefont {R.}~\bibnamefont {Liang}}, \bibinfo {author} {\bibfnamefont
  {W.~N.}\ \bibnamefont {Hardy}}, \bibinfo {author} {\bibfnamefont {G.~A.}\
  \bibnamefont {Sawatzky}}, \bibinfo {author} {\bibfnamefont {I.~S.}\
  \bibnamefont {Elfimov}}, \bibinfo {author} {\bibfnamefont {D.~A.}\
  \bibnamefont {Bonn}}, \ and\ \bibinfo {author} {\bibfnamefont
  {A.}~\bibnamefont {Damascelli}},\ }\href {\doibase 10.1038/nphys998}
  {\bibfield  {journal} {\bibinfo  {journal} {Nature Physics}\ }\textbf
  {\bibinfo {volume} {4}},\ \bibinfo {pages} {527} (\bibinfo {year}
  {2008})}\BibitemShut {NoStop}%
\bibitem [{\citenamefont {Das}(2012)}]{Das:Hg1201_TB_params}%
  \BibitemOpen
  \bibfield  {author} {\bibinfo {author} {\bibfnamefont {T.}~\bibnamefont
  {Das}},\ }\href {\doibase 10.1103/PhysRevB.86.054518} {\bibfield  {journal}
  {\bibinfo  {journal} {Phys. Rev. B}\ }\textbf {\bibinfo {volume} {86}},\
  \bibinfo {pages} {054518} (\bibinfo {year} {2012})}\BibitemShut {NoStop}%
\bibitem [{\citenamefont {Uchiyama}\ \emph {et~al.}(2004)\citenamefont
  {Uchiyama}, \citenamefont {Baron}, \citenamefont {Tsutsui}, \citenamefont
  {Tanaka}, \citenamefont {Hu}, \citenamefont {Yamamoto}, \citenamefont
  {Tajima},\ and\ \citenamefont {Endoh}}]{Uchiyama:PhononsH1201_2004}%
  \BibitemOpen
  \bibfield  {author} {\bibinfo {author} {\bibfnamefont {H.}~\bibnamefont
  {Uchiyama}}, \bibinfo {author} {\bibfnamefont {A.~Q.~R.}\ \bibnamefont
  {Baron}}, \bibinfo {author} {\bibfnamefont {S.}~\bibnamefont {Tsutsui}},
  \bibinfo {author} {\bibfnamefont {Y.}~\bibnamefont {Tanaka}}, \bibinfo
  {author} {\bibfnamefont {W.-Z.}\ \bibnamefont {Hu}}, \bibinfo {author}
  {\bibfnamefont {A.}~\bibnamefont {Yamamoto}}, \bibinfo {author}
  {\bibfnamefont {S.}~\bibnamefont {Tajima}}, \ and\ \bibinfo {author}
  {\bibfnamefont {Y.}~\bibnamefont {Endoh}},\ }\href {\doibase
  10.1103/PhysRevLett.92.197005} {\bibfield  {journal} {\bibinfo  {journal}
  {Phys. Rev. Lett.}\ }\textbf {\bibinfo {volume} {92}},\ \bibinfo {pages}
  {197005} (\bibinfo {year} {2004})}\BibitemShut {NoStop}%
\bibitem [{\citenamefont {van Heumen}\ \emph {et~al.}(2009)\citenamefont {van
  Heumen}, \citenamefont {Muhlethaler}, \citenamefont {Kuzmenko}, \citenamefont
  {Eisaki}, \citenamefont {Meevasana}, \citenamefont {Greven},\ and\
  \citenamefont {van~der Marel}}]{vanHeumen:OpticalDetermination}%
  \BibitemOpen
  \bibfield  {author} {\bibinfo {author} {\bibfnamefont {E.}~\bibnamefont {van
  Heumen}}, \bibinfo {author} {\bibfnamefont {E.}~\bibnamefont {Muhlethaler}},
  \bibinfo {author} {\bibfnamefont {A.~B.}\ \bibnamefont {Kuzmenko}}, \bibinfo
  {author} {\bibfnamefont {H.}~\bibnamefont {Eisaki}}, \bibinfo {author}
  {\bibfnamefont {W.}~\bibnamefont {Meevasana}}, \bibinfo {author}
  {\bibfnamefont {M.}~\bibnamefont {Greven}}, \ and\ \bibinfo {author}
  {\bibfnamefont {D.}~\bibnamefont {van~der Marel}},\ }\href {\doibase
  10.1103/PhysRevB.79.184512} {\bibfield  {journal} {\bibinfo  {journal} {Phys.
  Rev. B}\ }\textbf {\bibinfo {volume} {79}},\ \bibinfo {pages} {184512}
  (\bibinfo {year} {2009})}\BibitemShut {NoStop}%
\bibitem [{\citenamefont {He}\ \emph {et~al.}(2018)\citenamefont {He},
  \citenamefont {Hashimoto}, \citenamefont {Song}, \citenamefont {Chen},
  \citenamefont {He}, \citenamefont {Vishik}, \citenamefont {Moritz},
  \citenamefont {Lee}, \citenamefont {Nagaosa}, \citenamefont {Zaanen},
  \citenamefont {Devereaux}, \citenamefont {Yoshida}, \citenamefont {Eisaki},
  \citenamefont {Lu},\ and\ \citenamefont {Shen}}]{He:RapidBi2212_Science}%
  \BibitemOpen
  \bibfield  {author} {\bibinfo {author} {\bibfnamefont {Y.}~\bibnamefont
  {He}}, \bibinfo {author} {\bibfnamefont {M.}~\bibnamefont {Hashimoto}},
  \bibinfo {author} {\bibfnamefont {D.}~\bibnamefont {Song}}, \bibinfo {author}
  {\bibfnamefont {S.-D.}\ \bibnamefont {Chen}}, \bibinfo {author}
  {\bibfnamefont {J.}~\bibnamefont {He}}, \bibinfo {author} {\bibfnamefont
  {I.~M.}\ \bibnamefont {Vishik}}, \bibinfo {author} {\bibfnamefont
  {B.}~\bibnamefont {Moritz}}, \bibinfo {author} {\bibfnamefont {D.-H.}\
  \bibnamefont {Lee}}, \bibinfo {author} {\bibfnamefont {N.}~\bibnamefont
  {Nagaosa}}, \bibinfo {author} {\bibfnamefont {J.}~\bibnamefont {Zaanen}},
  \bibinfo {author} {\bibfnamefont {T.~P.}\ \bibnamefont {Devereaux}}, \bibinfo
  {author} {\bibfnamefont {Y.}~\bibnamefont {Yoshida}}, \bibinfo {author}
  {\bibfnamefont {H.}~\bibnamefont {Eisaki}}, \bibinfo {author} {\bibfnamefont
  {D.~H.}\ \bibnamefont {Lu}}, \ and\ \bibinfo {author} {\bibfnamefont {Z.-X.}\
  \bibnamefont {Shen}},\ }\href {\doibase 10.1126/science.aar3394} {\bibfield
  {journal} {\bibinfo  {journal} {Science}\ }\textbf {\bibinfo {volume}
  {362}},\ \bibinfo {pages} {62} (\bibinfo {year} {2018})}\BibitemShut
  {NoStop}%
\bibitem [{\citenamefont {Hashimoto}\ \emph {et~al.}(2014)\citenamefont
  {Hashimoto}, \citenamefont {Nowadnick}, \citenamefont {He}, \citenamefont
  {Vishik}, \citenamefont {Moritz}, \citenamefont {He}, \citenamefont {Tanaka},
  \citenamefont {Moore}, \citenamefont {Lu}, \citenamefont {Yoshida},
  \citenamefont {Ishikado}, \citenamefont {Sasagawa}, \citenamefont {Fujita},
  \citenamefont {Ishida}, \citenamefont {Uchida}, \citenamefont {Eisaki},
  \citenamefont {Hussain}, \citenamefont {Devereaux},\ and\ \citenamefont
  {Shen}}]{Hashimoto:PhaseCompetition}%
  \BibitemOpen
  \bibfield  {author} {\bibinfo {author} {\bibfnamefont {M.}~\bibnamefont
  {Hashimoto}}, \bibinfo {author} {\bibfnamefont {E.~A.}\ \bibnamefont
  {Nowadnick}}, \bibinfo {author} {\bibfnamefont {R.-H.}\ \bibnamefont {He}},
  \bibinfo {author} {\bibfnamefont {I.~M.}\ \bibnamefont {Vishik}}, \bibinfo
  {author} {\bibfnamefont {B.}~\bibnamefont {Moritz}}, \bibinfo {author}
  {\bibfnamefont {Y.}~\bibnamefont {He}}, \bibinfo {author} {\bibfnamefont
  {K.}~\bibnamefont {Tanaka}}, \bibinfo {author} {\bibfnamefont {R.~G.}\
  \bibnamefont {Moore}}, \bibinfo {author} {\bibfnamefont {D.}~\bibnamefont
  {Lu}}, \bibinfo {author} {\bibfnamefont {Y.}~\bibnamefont {Yoshida}},
  \bibinfo {author} {\bibfnamefont {M.}~\bibnamefont {Ishikado}}, \bibinfo
  {author} {\bibfnamefont {T.}~\bibnamefont {Sasagawa}}, \bibinfo {author}
  {\bibfnamefont {K.}~\bibnamefont {Fujita}}, \bibinfo {author} {\bibfnamefont
  {S.}~\bibnamefont {Ishida}}, \bibinfo {author} {\bibfnamefont
  {S.}~\bibnamefont {Uchida}}, \bibinfo {author} {\bibfnamefont
  {H.}~\bibnamefont {Eisaki}}, \bibinfo {author} {\bibfnamefont
  {Z.}~\bibnamefont {Hussain}}, \bibinfo {author} {\bibfnamefont {T.~P.}\
  \bibnamefont {Devereaux}}, \ and\ \bibinfo {author} {\bibfnamefont {Z.-X.}\
  \bibnamefont {Shen}},\ }\href {\doibase 10.1038/nmat4116
  https://www.nature.com/articles/nmat4116#supplementary-information}
  {\bibfield  {journal} {\bibinfo  {journal} {Nature Materials}\ }\textbf
  {\bibinfo {volume} {14}},\ \bibinfo {pages} {37} (\bibinfo {year}
  {2014})}\BibitemShut {NoStop}%
\end{thebibliography}%
\end{document}


\title{Supplementary materials: Three interaction energy scales in single-layer high-T$_C$ cuprate HgBa$_2$CuO$_{4+\delta}$}

\maketitle
\section{Surface termination}
While ARPES is a technique that allows direct visualisation of the band structure, the observed spectrum shows strong dependence on sample surface structure \cite{KRKnox:SuspendedGraphene,Fauster:MorphologyCu}. Typically, one obtains a chemically pure surface by mechanical cleaving $-$ gluing a top-post on the sample and knocking it off in ultra-high vacuum before measurement. Hg1201, due to the absence of a neutral cleavage plane  \cite{Barisic:DemonstratingModelNature}, has no obvious preferred termination upon cleaving. In the absence of atomic-scale studies of Hg1201 in the literature, surface terminations can be inferred from photoemission studies.

Angle-resolved X-Ray Photoemission Spectroscopy (AR-XPS) is a technique that probes core levels of elements in a sample up to different depths by varying the angle between detector and the surface. Depth sensitivity in photoemission is limited by the mean free path of the electrons after they encounter an optical excitation in the solid \cite{UniversalCurve}, but before they reach and leave the surface \cite{Merzlikin:ResolvingDepthCoordinate}. By varying the angle of the surface normal of the sample with respect to the direction of the detector (the photoemission angle), electrons from deeper inside the sample must travel longer distances through the sample before reaching the sample surface, causing them to be scattered more than electrons located closer to the surface. To a first approximation, the photoemission intensity for a homogeneous sample as a function of depth and photemission angle can be modelled as
\begin{equation}
    I=I_{z=0}\, exp(\frac{-z}{\lambda \, cos\theta})
    \label{eqn1}
\end{equation}
where $z$ is the depth of origin for the electron from the surface, $\lambda$ is the inelastic mean free path, and $\theta$ is the photoemission angle. While the model does not account for multiple scattering, elastic scattering or non-isotropic effects, the principle that the electrons that reach the detector predominantly originate from a depth of the order of $\lambda$ holds \cite{matthew2004surface}. Based on (\ref{eqn1}), atoms located on the surface are less affected by the orientation of the surface, while atoms located deeper into the sample originate fewer photoelectrons at oblique photoemission angles due to higher scattering.

\begin{figure*}[ht]
\includegraphics[width=1\columnwidth]{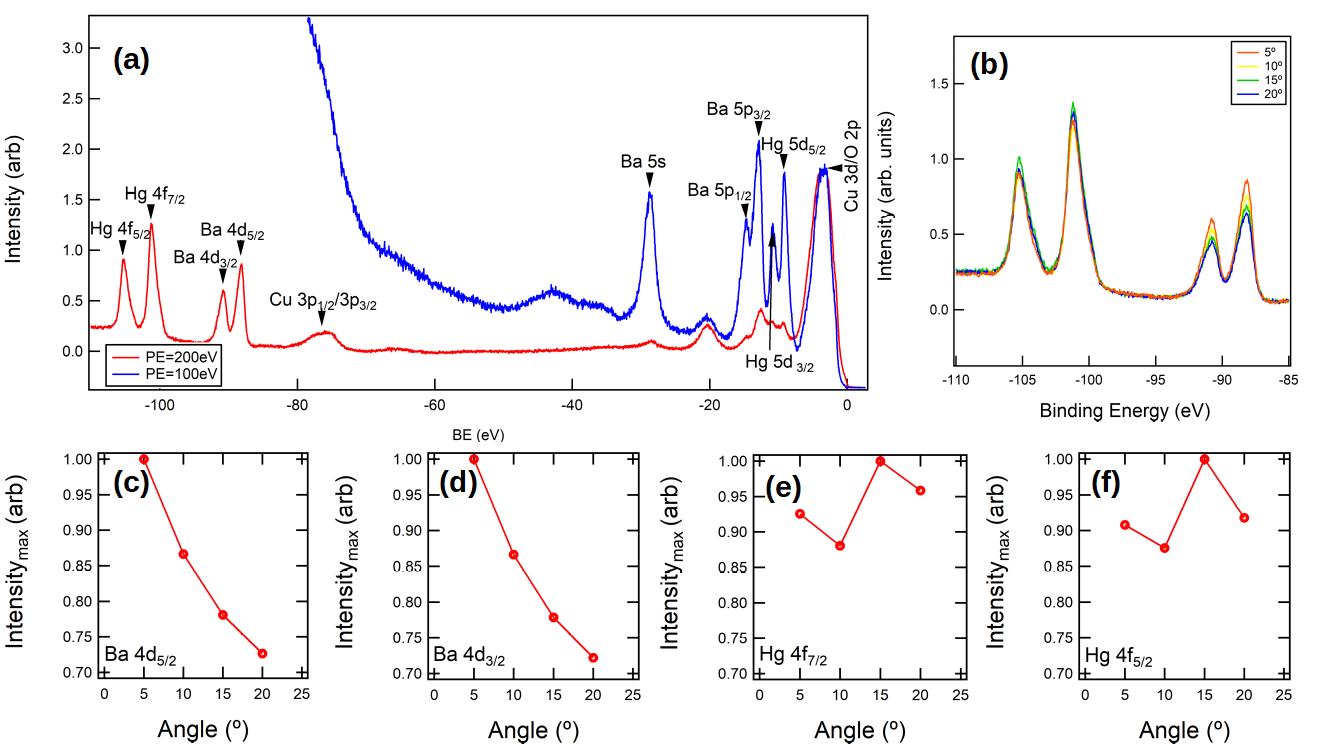}
\captionsetup{justification=raggedright,singlelinecheck=false}
	\caption{ (a) Core Levels of Hg1201 at photon energies of 100 eV and 200 eV taken at normal incidence. (b) Hg 4f and Ba 4d core levels at different polar angles taken with 200eV (c-f) Polar angle-dependence of the peak intensities for various core levels normalized to the maximum in each series.}
	\label{fig:figs1}
\end{figure*}

Fig. \ref{fig:figs1}(a) shows the relevant core levels on Hg1201 obtained at photon energies of 100 eV and  200 eV. Both spectra have been rescaled with respect to the Cu-O valence band peak for comparison. We observe a strong photon energy dependence on the relative intensities of the core levels between the two spectra. The following discussion focuses only on the deeper Hg and Ba core levels(4f and 4d), measured with 200eV photon energy, for an angle-dependent study. Fig \ref{fig:figs1}(b) details the XPS intensities of the Ba 4d and Hg 4f at different photoemission angles between the surface normal and the detector. We observe a systematic decrease in the peak intensities of the Ba core levels with increasing photoemission angle. The maximum intensities are then obtained by fitting Lorentzians to each of the peaks. The intensities are then normalised to the maximum within the angle series and plotted in Fig \ref{fig:figs1}(c-f). The data show a steady decrease in intensities of the peaks for the Ba core levels while showing no monotonic trend for the Hg core levels. This suggests that Ba atoms are located further from the surface than Hg atoms. 

These data are consistent with cleavage close to the Hg plane, and further from the Ba plane, specifically at the Hg-O bonds. Due to the symmetry of the Hg-O bonds, both bonds above and below a Hg atom are equally likely to be broken resulting in a mixed O and Hg surface termination. This effectively gives a charge-neutral cleaved surface, possibly explaining why as-cleaved Hg1201 does not exhibit polar catastrophe as as-cleaved YBCO \cite{Inna:PRBHg1201,Hossain:DopingControl}. Confirmation of the same will require further atomic resolved experiments not yet available in literature, but this work provides a photoemission-based guide.

\section{Momentum dependence of intermediate-energy kink}
\begin{figure*}[ht]
\includegraphics[width=1\columnwidth]{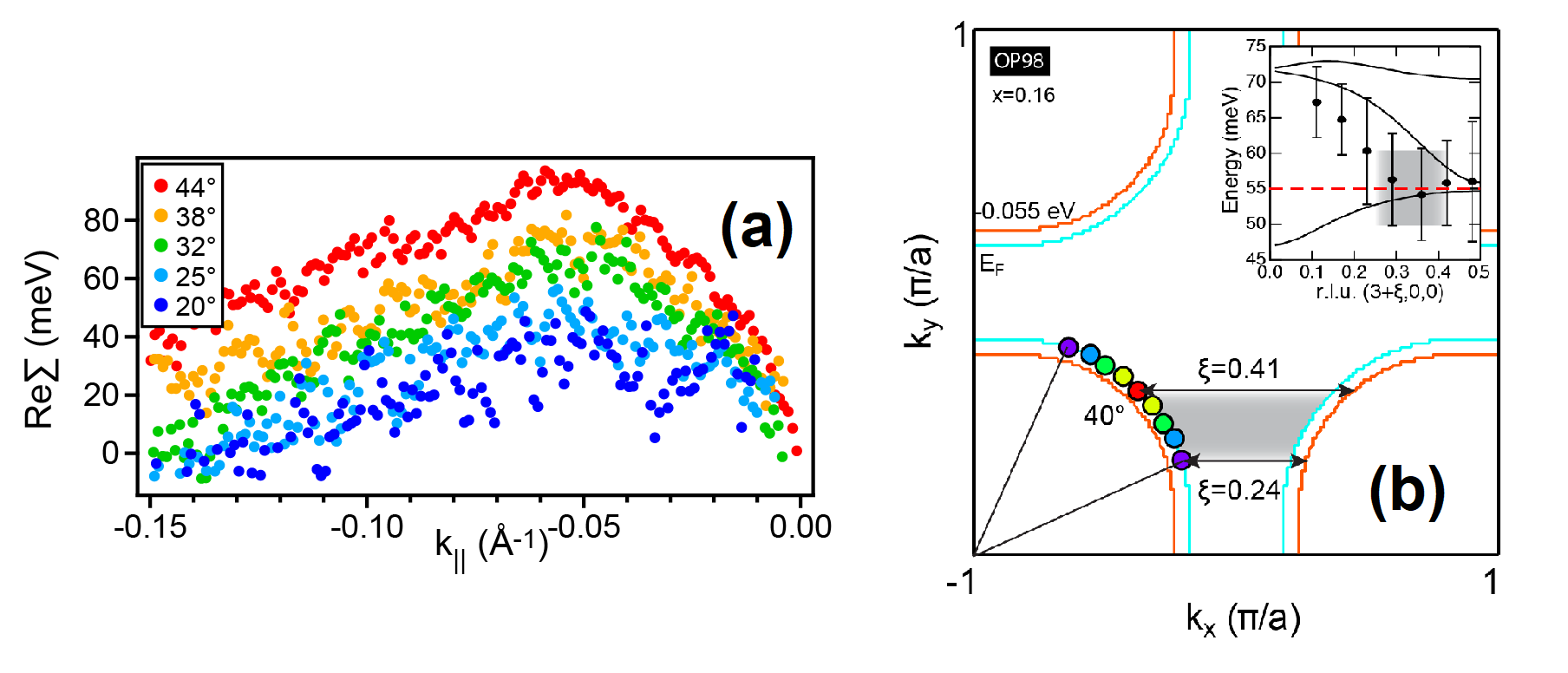}
\captionsetup{justification=raggedright,singlelinecheck=false}
	\caption{
	Momentum dependence of intermediate-energy kink at optimal doping (a) $Re \Sigma$ for different Fermi surface angles.  Data taken at 30K. (b) Cyan: tight-binding Fermi surface for $p=0.16$. Orange: constant energy contour at kink energy (-0.055 eV).  Tight binding parameters from Ref. \onlinecite{Das:Hg1201_TB_params}.  Dots on Fermi surface denote $k_F$ for cuts in (a).  Arrows connect one constant energy contour to another along (q,0,0) starting from node and also starting 20 degrees away from node.  Inset: dispersion of bond-stretching phonon measured by inelastic x-ray scattering from Ref. \onlinecite{Uchiyama:PhononsH1201_2004}. }
	\label{fig:figs2}
\end{figure*}

Figure \ref{fig:figs2} provides further details regarding the interpretation of the momentum dependence of intermediate kink in terms of the bond-stretching phonon at large momentum transfer.  Figure \ref{fig:figs2}(a) shows the real part of the self energy ($Re \Sigma$) at optimal doping ($T_c$=98) as a function of Fermi surface angle, defined in the manuscript.  Tight-binding bare bands were used, yielding a $Re \Sigma$ which decreases by $75\%$ going from the node to approximately halfway to the antinode.

Coupling between a near-nodal fermion and a phonon with energy $\Omega$ and wavevector along (q, 0, 0) will involve scattering from one near-node at $E_F$ to the opposite near-node at $E_F\pm\Omega$ as sketeched in Figure \ref{fig:figs2}(b).  Notably wavevectors that connect a contour at $E_F$ on one side of the Brillouin zone to a contour at -55 meV on the other side coincide with the momentum range where the bond-stretching phonon energy is considerably softened from its value at q=0, consistent with proposals that strong coupling to electrons is the cause of the softening.

\section{Nodal peak-dip-hump: momentum and temperature dependence}
\begin{figure*}[ht]
\includegraphics[width=1\columnwidth]{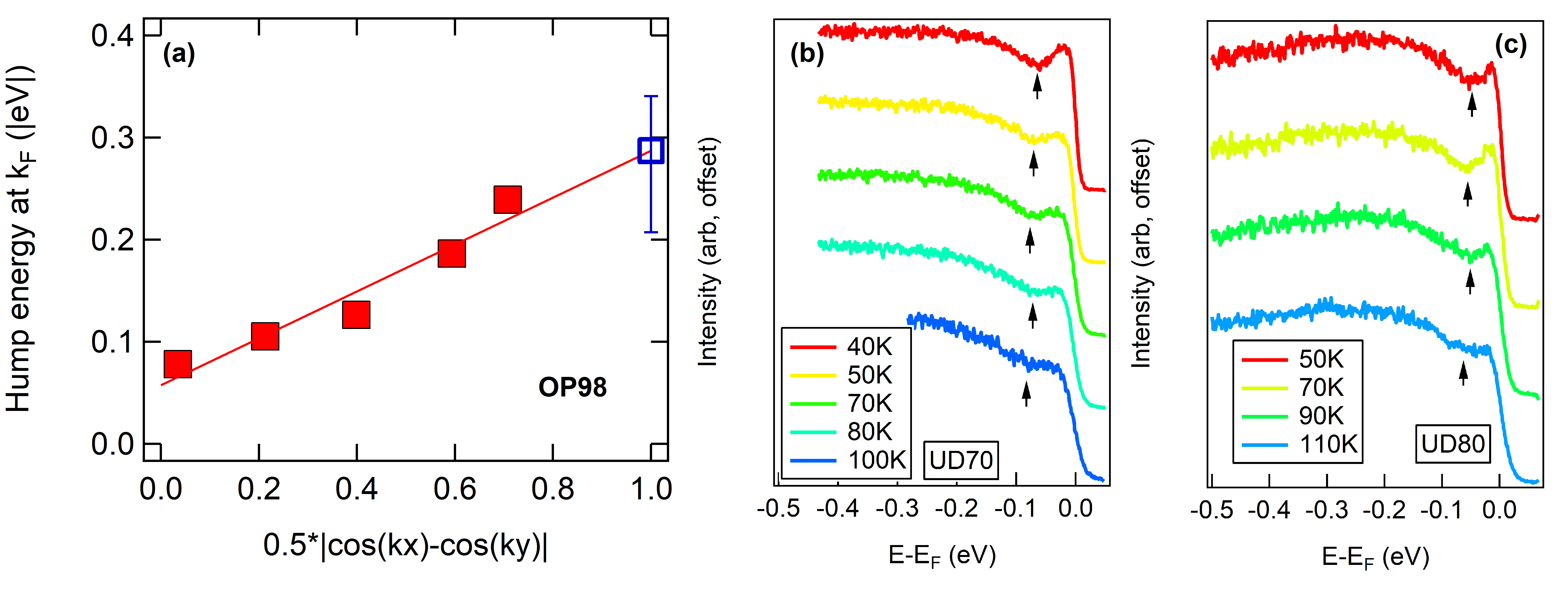}
\captionsetup{justification=raggedright,singlelinecheck=false}
\caption{(a) Momentum dependence of hump position at optimal doping, from Fig. 4 of the manuscript, plotted as a function of a simple \textit{d}-wave form.  Blue symbol is extrapolation to antinode. (b) Temperature dependence of nodal EDC at $k_F$ for UD80 through $T_c$ with arrow pointing to dip energy. (c) same for UD70 }
	\label{fig:figs3}
\end{figure*}

Figure \ref{fig:figs3} shows additional data with regards to the nodal and near-nodal PDH structure.  Figure \ref{fig:figs3}(a) focuses on the momentum dependence of the hump energy scale, as shown in Fig. 4 of the manuscript, at optimal doping, plotted as a function of the simple \textit{d}-wave form, $0.5*|cos(k_x)-cos(k_y)|$.  In the momentum range where the hump is observed, its energy scale roughly follows a simple \textit{d}-wave form, extrapolating to an energy of $\approx0.3 eV$ at the antinode.  Interestingly, this agrees well with the maximum of the broad continuum of excitations derived from optics experiments at the same doping \cite{vanHeumen:OpticalDetermination}, suggesting a common origin of the two if optics is dominated by antinodal states. 

Figure \ref{fig:figs3}(b)-(c) show EDCs at $k_F$ in a temperature range spanning 30K below $T_c$ to 30K above $T_c$ for UD70 and UD80.  The dip feature is the emphasis of these panels, and it clearly persists above $T_c$ for both dopings.  This is distinct from the antinodal PDH feature in multi-CuO$_2$-plane cuprates, where the dip is strongly sensitive to $T_c$ \cite{He:RapidBi2212_Science,Hashimoto:PhaseCompetition}.

\bibliography{Supp.bib}